\documentclass[12pt]{article}
\usepackage{amsmath, amssymb, amsfonts}
\usepackage[dvips]{graphicx}
\usepackage{multirow}

\def\nn{\nonumber}
\def\dfrac{\displaystyle\frac}
\def\numt#1#2{#1 \times 10^{#2}}
\def\etal{{\it et~al.}}

\def\PR#1#2#3{Phys. Rev. {\bf #1}, #2 (#3)}
\def\PRL#1#2#3{Phys. Rev. Lett. {\bf #1}, #2 (#3)}
\def\PL#1#2#3{Phys. Lett. {\bf #1}, #2 (#3)}
\def\NP#1#2#3{Nucl. Phys. {\bf #1}, #2 (#3)}
\def\PTP#1#2#3{Prog. Theor. Phys. {\bf #1}, #2 (#3)}
\def\EPJ#1#2#3{Eur. Phys. J. {\bf #1}, #2 (#3)}

\def\Mod#1#2#3{Mod. Phys. Lett. {\bf #1}, #2 (#3)}
\def\MPL#1#2#3{Mod. Phys. Lett. {\bf #1}, #2 (#3)}
\def\PTP#1#2#3{Prog. Theor. Phys. {\bf #1}, #2 (#3)}
\def\EPJ#1#2#3{Eur. Phys. J. {\bf #1}, #2 (#3)}

\def\JHEP#1#2#3{JHEP{\bf #1}, #2 (#3)}

\def\gsim{~{\rlap{\lower 3.5pt\hbox{$\mathchar\sim$}}\raise 1pt\hbox{$>$}}\,}
\def\lsim{~{\rlap{\lower 3.5pt\hbox{$\mathchar\sim$}}\raise 1pt\hbox{$<$}}\,}
\def\cerenkov{$\check{\rm C}$erenkov~}
\def\sss{\scriptscriptstyle}
\def\atm{\sss{\rm ATM}}
\def\rct{\sss{\rm RCT}}
\def\sun{\sss{\rm SOL}}
\def\dmns{\delta_{\sss{\rm MNS}}}
\def\dsun{\delta m^2_{\sss{\rm SOL}}}
\def\datm{\delta m^2_{\sss{\rm ATM}}}
\def\dm#1#2{\delta m^2_{#1#2}}
\def\ssun#1{\sin^2 #1\theta_{\sss{\rm SOL}}}
\def\satm#1{\sin^2 #1\theta_{\sss{\rm ATM}}}
\def\srct#1{\sin^2 #1\theta_{\sss{\rm RCT}}}
\def\csun#1{\cos^2 #1\theta_{\sss{\rm SOL}}}

\def\sSK{\sss{\rm T2K}}
\def\sOKI{\sss{\rm T2Oki}}
\def\sKR{\sss{\rm T2Kr(1000km)}}
\def\T2Kn{T2K$_{122}$}
\begin{document}
\renewcommand{\thefootnote}{\fnsymbol{footnote}}
\renewcommand{\thefigure}{\arabic{figure}}
\title{ Physics potential of neutrino oscillation experiment \\
with a far detector in Oki Island\\
along the T2K baseline}
\author{
{Kaoru~Hagiwara}$^{1,2}$,
{Takayuki Kiwanami}$^{2}$\thanks{
\scriptsize
Present address~:~PLP Division, 
SHINKO ELECTRIC INDUSTRIES CO., LTD,
Nagano, 381-0103, Japan.
},\\
{Naotoshi Okamura}$^3$\thanks{
\scriptsize
e-mail~:~nokamura@yamanashi.ac.jp}, and
{Ken-ichi Senda}$^1$\thanks{
\scriptsize
e-mail~:~senda@post.kek.jp}\\
{\small \it $^1$ KEK Theory Center, Tsukuba, 305-0801, Japan}\\
{\small \it $^2$ Sokendai, (The Graduate University for Advanced
Studies), Tsukuba, 305-0801, Japan}\\
{\small \it $^3$ Faculty of Engineering, University of Yamanashi,
Kofu, Yamanashi, 400-8511, Japan}}
\date{~}
\maketitle
\vspace{-11.5cm}
\begin{flushright}
KEK-TH-1568
\end{flushright}
\vspace{ 11.5cm}
\vspace{-1.9cm}

\begin{abstract}
Oki Island is located between Japan and Korea 
along the Tokai-To-Kamioka (T2K) baseline.
The distance from J-PARC to Oki Island is about 653km,
which is twice that of the T2K experiment ($L=295$km).
When the off-axis angle of the neutrino beam from J-PARC 
is $3.0^\circ$ ($2.0^\circ$) at Super-Kamiokande (SK),
the off-axis beam (OAB) with $1.4^\circ$ ($0.6^\circ$)
reaches at Oki Island.
We examine physics case of placing a far detector
in Oki Island during the T2K experimental period.
We estimate the matter density profile along the Tokai-to-Oki baseline
by using recent seismological measurements.
For a detector of 100~kton fiducial volume and
$\numt{2.5}{21}$ POT (protons on target) exposure
for both $\nu_\mu$ and $\bar{\nu}_\mu$ beams,
we find that the mass hierarchy pattern can be distinguished
at 3~$\sigma$ level 
if $\srct{2} \equiv 4|U_{e3}|^2(1-|U_{e3}|^2) \gsim 0.09$,
by observing the electron-like CCQE (Charged-Current Quasi Elastic)
events.
The CP phase in the Maki-Nakagawa-Sakata lepton flavor mixing matrix, 
$\dmns$, can be constrained with $\pm 20^\circ$.
As a reference, we repeat the same analysis by placing the same
detector in Korea at $L=1000$~km and OAB=$0.5^\circ$ (T2KK)
and also by placing it at the SK site (\T2Kn).
The Tokai-to-Kamioka-OKI (T2KO)
 sensitivity to the mass hierarchy is about $1/3$
(in $\Delta \chi^2_{\rm min}$) of T2KK,
while the sensitivity to the phase $\dmns$ is similar between T2KO
and T2KK. 
The T2K$_{122}$ option has almost no sensitivity to the mass hierarchy,
and cannot measure the CP phase except
when $\dmns\sim -90^\circ$ ($90^\circ$) for
the normal (inverted) hierarchy.
\end{abstract}

\section{Introduction}
\label{sec:1}

 Since the neutrino oscillation was first observed at
Super-Kamiokande (SK) \cite{SK98},
many experiments have measured the physics parameters
of the neutrino oscillation
\cite{atm}\nocite{K2K,T2K,minos,SKatm,
solar,SNO,KamLAND,
CHOOZ, PaloVerde}-\cite{miniboone}.

 Some of these experiments observe the survival probability of 
$\nu_\mu$ and $\bar{\nu}_\mu$ which are generated in the atmosphere
by the cosmic ray \cite{atm}.
 Accelerator based long baseline experiments \cite{K2K,T2K,minos} 
also measure the $\nu_\mu$ survival probability.
From the combined results of these experiments \cite{atm}-\cite{minos},
the mass-squared difference and the mixing angle are obtained as
\begin{subequations} 
\begin{eqnarray}
\satm{2}&>& 0.90~~~(90\% \mbox{C.L.})\,, 
\label{eq:mixing_atm} \\ 
\left|\datm \right| &=& \numt{\left(2.35^{+0.11}_{-0.08}\right)}{-3}
 {\mbox{eV}}^2\,.
\label{eq:mass_atm} 
\end{eqnarray}
\label{eq:atm_data}
\end{subequations}
$\!\!\!\!$
The sign of the mass-squared difference, eq.~(\ref{eq:mass_atm}),
cannot be determined from these experiments.
The SK collaboration also reported that the atmospheric neutrinos
oscillate into active neutrinos \cite{SKatm}.

The combined results of
the solar neutrino observations \cite{solar, SNO},
which measure the survival probability of $\nu_e$ from the sun,
and
the KamLAND experiment \cite{KamLAND},
which measure the
$\bar{\nu}_e$ flux from the reactors at distances
of a few 100~km, find 
\begin{subequations} 
\begin{eqnarray}
 \ssun{2} &=& 0.852^{+0.024}_{-0.026}\,,
\label{eq:mixing_sun} \\
 \dsun &=& \numt{\left(7.50^{+0.19}_{-0.20}\right)}{-5} \mbox{eV}^2
\label{eq:mass_sun}\,,
\end{eqnarray}
\label{eq:sun_data} 
\end{subequations}
$\!\!\!\!$
where the sign of mass-squared difference has been determined by the matter
effect inside the sun \cite{msw}.
 The SNO experiment determined that $\nu_e$ from the sun
changes into active neutrinos \cite{SNO}.

For the reactor experiments,
which measure the survival probability of $\bar{\nu}_e$ from the
reactor at distances of $L\sim O(1)$~km \cite{CHOOZ,PaloVerde},
the CHOOZ experiment observed no signal,
giving
\begin{subequations} 
\begin{eqnarray}
 \srct{2} &<& 0.17\,,
\label{eq:mixing_rct} \\
\mbox{for } \left|\delta m^2_{\rct} \right| &=& 
\numt{2.35}{-3} {\mbox{eV}}^2\,,
\label{eq:mass_rct}
\end{eqnarray} 
\label{eq:rct_data}
\end{subequations}
$\!\!\!\!$
at the 90\% confidence level \cite{CHOOZ}. 
More recently,
the T2K collaboration reported several
$\nu_\mu \to \nu_e$ appearance candidate events giving
\begin{equation}
 0.03~(0.04) < \srct{2} < 0.28~(0.34)
\label{eq:T2K_s13}
\end{equation}
for $\sin^2 \theta_{23}=0.5$ and $\dmns=0^\circ$
with the normal (inverted) hierarchy at the 90\% confidence level
\cite{T2Knue}.
The MINOS collaboration reported
\begin{equation}
 \srct{2} = 0.041^{+0.047}_{-0.031}~(0.079^{+0.071}_{-0.053})
\label{eq:MINOS_s13}
\end{equation}
also for $\sin^2 \theta_{23}=0.5$ and $\dmns=0^\circ$
with the normal (inverted) hierarchy \cite{MINOSnue}.
The Double CHOOZ collaboration, 
which is one of the new reactor experiments, 
found hints of reactor electron anti-neutrino
disappearance consistent with neutrino oscillation
and reported
\begin{equation}
 \srct{2} = 0.086 \pm 0.041 {\rm~(stat.)~} \pm 0.030 {\rm~(syst.)}
 \label{eq:DCHOOZ_s13}
\end{equation} 
from analyzing both the rate of the prompt positrons
and their energy spectrum \cite{DoubleCHOOZ}.
Recently another new reactor experiment, 
the DayaBay experiment \cite{DayaBay},
announced that they have measured the neutrino mixing angle as
\begin{equation}
 \srct{2} = 0.092 \pm 0.016 {\rm~(stat.)~} \pm 0.005 {\rm~(syst.)}\,,
 \label{eq:DayaBay_s13}
\end{equation}
which is more than $5\sigma$ away from zero.
The RENO collaboration,
which also measure the reactor $\bar{\nu}_e$
survival probability,
shows the evidence of the non-zero mixing angle;
\begin{equation}
 \srct{2} = 0.113 \pm 0.013 {\rm~(stat.)~} \pm 0.019 {\rm~(syst.)}\,,
 \label{eq:RENO_s13}
\end{equation}
from a rate-only analysis,
which is $4.9\sigma$ away from zero.

Since the MiniBooNE experiment \cite{miniboone}
did not confirm the LSND observation of rapid 
$\bar{\nu}_\mu \to \bar{\nu}_e$ oscillation \cite{LSND},
there is no clear indication of experimental data
which suggests more than three neutrinos.
Therefore the $\nu_\mu \to \nu_e$ appearance analysis of
T2K \cite{T2Knue} and MINOS\cite{MINOSnue}
presented above assume the 3 neutrino model,
with the $3 \times 3$ flavor mixing, 
the MNS (Maki-Nakagawa-Sakata) matrix \cite{MNS}
\begin{equation}
\left(
 \begin{array}{c}
 \nu_e \\
 \nu_\mu \\
 \nu_\tau
 \end{array}
\right)
=
\left(
 \begin{array}{ccc}
 U_{e1}     & U_{e2}     & U_{e3} \\
 U_{\mu 1}  & U_{\mu 2}  & U_{\mu 3} \\
 U_{\tau 1} & U_{\tau 2} & U_{\tau 3}
 \end{array}
\right)
\left(
 \begin{array}{c}
 \nu_1 \\
 \nu_2 \\
 \nu_3
 \end{array}
\right)\,,
\label{eq:def_MNS}
\end{equation}
relating the weak interaction eigenstates
$(\nu_e, \nu_\mu, \nu_\tau)$
and the mass eigenstate $\nu_i$
with the mass $m_i$ ($i=1,2,3$).
The mass-squared differences that dictate the neutrino oscillation
phase are then identified as
\begin{subequations} 
\begin{eqnarray}
\datm = \delta m^2_{\rct} & = & m^2_3-m^2_1 \,,\\
\dsun &=& m^2_2 - m^2_1 \,, 
\end{eqnarray}
\end{subequations}
where only the magnitude of the larger mass-squared difference,
$m^2_3 - m^2_1$ is determined in eq.~(\ref{eq:mass_atm}).
The $m^2_3 > m^2_1$ case is called normal,
while $m^2_3 < m^2_1$ case is called inverted neutrino
mass hierarchy.
With a good approximation \cite{3G},
the three independent mixing angles of the MNS matrix
can be related to the oscillation amplitudes in
eqs.~(\ref{eq:mixing_atm}), 
(\ref{eq:mixing_sun}), and
$\srct{2}$ in eqs.~(\ref{eq:mixing_rct}),
(\ref{eq:T2K_s13}) to (\ref{eq:DayaBay_s13}),
with the MNS matrix elements $U_{\alpha i}$ :
\begin{subequations} 
\begin{eqnarray}
 \sin \theta_{\atm} &=& U_{\mu 3} \hspace*{4.3ex}
= ~ \sin \theta_{23} \cos \theta_{13}\,,
\label{eq:def_satm}\\
 \sin \theta_{\rct} &=& \left|U_{e3}\right| \hspace*{3.2ex}
= ~ \sin \theta_{13}\,,
\label{eq:def_srct}\\
 \sin 2\theta_{\sun} &=& 2U_{e1} U_{e2}
= ~ \sin2\theta_{12} \cos^2\theta_{13}\,.
\label{eq:def_ssun}
\end{eqnarray} 
\label{eq:MNS_exp}
\end{subequations}
$\!\!\!\!$
In the last equations, the defining region of
the three mixing angles $\theta_{ij}=\theta_{ji}$ 
can be chosen as $0\leq \theta_{12},\theta_{13}, \theta_{23} \leq \pi/2$
\cite{PDB},
which is consistent with the convention of non-negative
$U_{e1}$ and $U_{e2}$ \cite{HO1}.
The argument of $U_{e3}$ gives the CP phase of the MNS matrix,
\begin{equation}
\dmns = -\mbox{arg} U_{e3}\,.
\label{eq:def_dmns}
\end{equation}

Even after $\theta_{\rct}$, the smallest of the three mixing angles,
is measured by the accelerator based long
baseline experiments \cite{T2Knue,MINOSnue} and the reactor experiments
\cite{DoubleCHOOZ,DayaBay,RENO},
three unknowns remain in the 3 neutrino model,
which are 
the sign of the larger mass-squared differences,
normal ($m^2_3-m^2_1>0$) or inverted ($m^2_3-m^2_1<0$),
the leptonic CP phase ($\dmns$),
and 
the sign of $\satm{}-0.5$ if its magnitude differs from zero
significantly.
The main purpose of the next generation neutrino oscillation
experiments is to determine these three unknown parameters
of the three neutrino model.

The accelerator based long baseline neutrino oscillation experiment
with two-detectors for one-beam,
such as Tokai-to-Kamioka-and-Korea (T2KK) experiment 
\cite{HOSetc}\nocite{Kajita, HOS1, HOS2, T2KKatm, T2KKbg}-\cite{HOSF},
is one of the promising experiments for measuring 
all the three unknowns.
When one measures the neutrino energy ($E_\nu$)
and the magnitude of the $\nu_e$ appearance probability at 
significantly different baseline lengths,
the degeneracy between the sign of the larger mass-squared
difference and the CP phase can be resolved \cite{HOS1}-\cite{HOSF},
since they affect the magnitude of the $\nu_e$ appearance
probability and 
the neutrino energy at the first oscillation maximum
differently at the two baseline lengths.
Once the mass hierarchy and the CP phase are determined,
the sign of $\satm{}-0.5$ can be determined \cite{T2KKatm}
since the $\nu_\mu \to \nu_e$ oscillation probability is
proportional to $\satm{}\srct{}$.

In the previous works \cite{HOS1}-\cite{HOSF},
physics impacts of the T2KK experiment
have been studied systematically,
and the following observations have been made:
If a 100~kton water \cerenkov detector is placed in Korea
at $L = 1000$~km
observing the $0.5^\circ$ off-axis beam (OAB) during 
the T2K exposure time of $\numt{5}{21}$ POT (protons on target),
the mass hierarchy can be resolved at 3$\sigma$ level
for $\srct{2} \gsim 0.05~(0.06)$
when the hierarchy is normal (inverted),
and the CP phase can be constrained uniquely,
by measuring the CCQE (Charged-Current Quasi Elastic) events
\cite{HOS1,HOS2}.
The sign of $\satm{}-0.5$ can also be determined for
$|\satm{}-0.5|=0.1$ with $3\sigma$ level,
if $\srct{2}>0.12$ for the normal hierarchy \cite{T2KKatm}.
When we take into account the smearing of reconstructed neutrino energy
due to finite detector resolution
and the Fermi motion of target nucleons,
resonance production,
and
the neutral current (NC) $\pi^0$ production background to 
the $\nu_e$ appearance signal,
it is found that the mass hierarchy pattern can still be determined
at $3\sigma$ level for $\srct{2} \gsim 0.08~(0.09)$
when the hierarchy is normal (inverted),
but the CP phase can no longer be established at $3\sigma$ level
with $\numt{5}{21}$ POT $\nu_\mu$ exposure \cite{T2KKbg}.
In Ref.\cite{HOSF},
matter distribution profile along the T2K and Tokai-to-Korea baselines
have been studied,
and merits of splitting the total exposure time into half $\nu_\mu$
and half $\bar{\nu}_\mu$ beam have been reported.
Studies on the impacts of the neutrino energy smearing and the NC
$\pi^0$ background for the $\bar{\nu}_\mu \to \bar{\nu}_e$ oscillation
measurements are in progress \cite{T2K5}.

\begin{figure}[t]
\centering
 \includegraphics[scale=0.75]{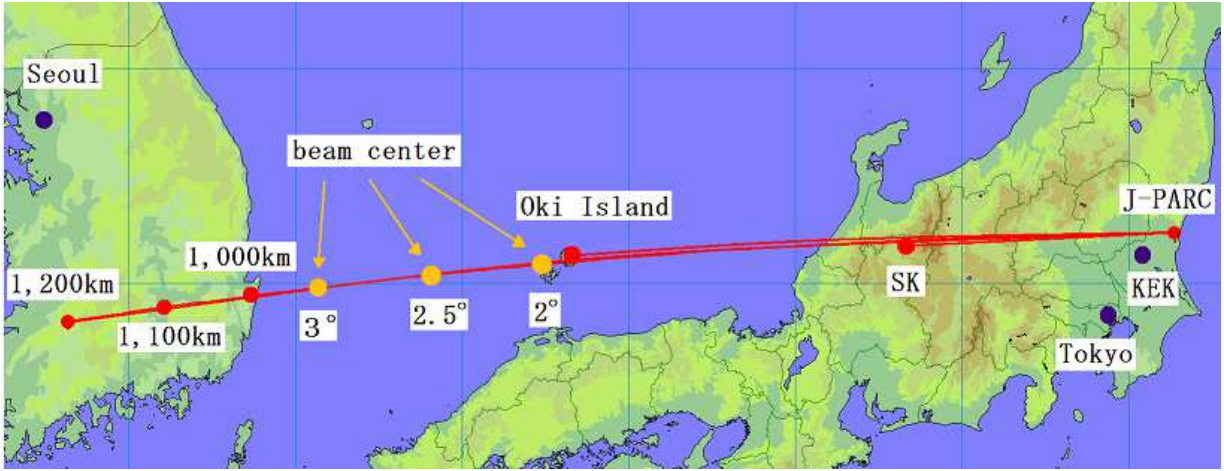}
 \caption{
The surface map of the T2K, T2KO, and T2KK experiment.
The yellow blobs show the center of the neutrino beam for the T2K
experiment at the sea level,
where the number in the white box is the off-axis angle at SK.
} 
\label{fig:T2OKI1}
\end{figure}

Oki Island\footnote{{%
This island is sometimes called ``Oki-no-Shima'',
because ``Shima'' means an ``Island'',
``no'' is ``in'',
and
``Oki'' means the ``Offing'';
an island in the offing, or an offshore island.}}
is placed between Japan and Korea along the
T2K baseline and inside Japanese territory.
The distance from J-PARC to Oki Island is about $L=653$~km
which is about two times longer than that of the T2K experiment
($L=295$~km).
In Fig.~\ref{fig:T2OKI1}, 
we show the surface map of the T2K, T2KO (Tokai-to-Kamioka-and-Oki),
and T2KK experiments,
in which the red lines denote the baselines for each experiment
and the yellow blobs show the center of the neutrino beam 
from J-PARC at the sea level,
when the off-axis angle at SK is $2^\circ$, $2.5^\circ$, 
and $3^\circ$.
The authors of Ref.\cite{OkiOld1} consider the physics performance
of a 100~kton Liquid Argon Time Projection Chamber placed at Oki Island.
In Ref.\cite{OkiOld2}, the authors studied the Oki Island
site from the geological, geographic and infrastructure points of view
for the possibility to construct a large detector.
They conclude that Oki Island is one of the good candidate 
sites for a large detector.

In this paper,
we study the physics potential of a 100~kton
water \cerenkov detector placed in Oki Island
instead of Korea 
during the T2K experimental period,
by using exactly the same setting assumed for the T2KK experiment
in Ref.\cite{HOSF},
except for the location of the detector.
This allows us to compare the physics capability of the two proposals
on the same footing.
Since we do assume in this analysis that the
$\nu_\mu \to \nu_\mu, \nu_e$ CCQE events can be distinguished from the
background,
our results should equally be applied for a more advanced
Liquid Argon detector \cite{OkiOld1}.

 This article is organized as follows.
 In the next section,
 we briefly review neutrino oscillation in the matter and
give useful approximation formula for 
$\nu_\mu \to \nu_\mu, \nu_e$ and
$\bar{\nu}_\mu \to \bar{\nu}_\mu, \bar{\nu}_e$
oscillation probabilities.
 In section \ref{sec:far},
 we show the matter profile between J-PARC and 
Oki Island 
by using recent seismological measurements and
give relations between the off-axis angle
observable at Oki and that at SK.
 The expected numbers of $\nu_\mu \to \nu_\mu, \nu_e$
$(\bar{\nu}_\mu \to \bar{\nu}_\mu, \bar{\nu}_e)$
CCQE events at the far (Oki) and the near (SK) detectors
are shown for typical parameters in section \ref{sec:event}.
 In section \ref{sec:chi},
 we present the $\chi^2$ function with which we estimate
the statistical sensitivity of the T2KO and T2KK experiments on the
 mass hierarchy and the CP phase.
 In section \ref{sec:mass},
 we show the results on the mass hierarchy determination of the T2KO
 experiment and compare it with T2KK, and also with \T2Kn
where the 100~kton detector is placed in the Kamioka site 
making the total fiducial volume 122~kton at
$L=295$~km.
 In section \ref{sec:CP},
 we show the CP phase sensitivity of T2KO, T2KK, and 
 \T2Kn experiments.
 We summarize our findings in the last section.

\section{Neutrino oscillation in the matter}
\label{sec:osc_theory}

In long baseline neutrino oscillation experiments,
neutrinos from the accelerator interact coherently
with electrons and nucleons by charged and neutral current
interactions.
These coherent interactions make an additional potential in the
effective Hamiltonian \cite{msw}.
Because the potential from the neutral current interactions are
flavor-blind, it does not affect the neutrino oscillation probability
in the three neutrino model.
On the other hand,
the charged-current interactions contribute only to the effective
potential of $\nu_e$ and $\bar{\nu}_e$.
The neutrino oscillation probabilities hence depend on
the matter density profile along the baseline.
Detailed studies on the matter distribution dependence of
the oscillation probabilities for
the T2K and T2KK experiments are found in Ref.~\cite{HOSF}.
In this exploratory analysis, we compare the physics
capability of the T2KO, T2KK, and T2K experiments
by using the average matter density along the baselines
which is found to approximate the matter effects rather accurately
\cite{HOSF}.

The Hamiltonian of a neutrino propagating in the matter is 
then expressed as
\begin{subequations} 
\begin{eqnarray}
H&=&\dfrac{1}{2E}
\left[
U^{}
\left(
\begin{array}{ccc}
0&& \\
&\dm12& \\
&&\dm13
\end{array}
\right)
U^\dagger
+
\left(
\begin{array}{ccc}
\bar{a}_0&& \\
&0& \\
&&0
\end{array}
\right)
\right]\,,
\label{eq:Hamiltonian1}\\
&=&
\dfrac{1}{2E}
\tilde{U}^{}
\left(
\begin{array}{ccc}
\lambda_1&& \\
&\lambda_2& \\
&&\lambda_3
\end{array}
\right)
\tilde{U}^\dagger\,,
\label{eq:Hamiltonian2}
\end{eqnarray} 
\label{eq:Hamiltonians}
\end{subequations}
$\!\!\!\!$
on the flavor space $(\nu_e,\nu_\mu,\nu_\tau)^T$,
with $\delta m^2_{ij} \equiv m^2_j -m^2_i$,
where
\begin{equation}
 \bar{a}_0=2\sqrt{2} G_F E_\nu n_e
 \simeq {7.56}{\times}10^{-5}\mbox{eV$^2$} 
\left(\dfrac{\bar{\rho}}{\mbox{g/cm$^3$}}\right)
\left(\dfrac{E_\nu}{\mbox{GeV}}\right)
\label{eq:defA}
\end{equation}
gives the matter effect with the electron number density $n_e$,
which is approximated by the average matter density
$\bar{\rho}$~({g/cm$^3$}) along the baseline.
By using the solution of eq.~(\ref{eq:Hamiltonian2}),
the oscillation probability that
an initial flavor eigenstate $\nu_\alpha$ 
is observed as a flavor eigenstate $\nu_\beta$
after traveling a distance $L$ along the baseline is
\begin{subequations} 
\begin{eqnarray}
 P_{\nu_\alpha \to \nu_\beta}
\!\! &=& \!\!
\left|
\left\langle \nu_\beta \right|
\exp\left({-i\int_0^L H dx}\right)
\left|\nu_{\alpha} \right\rangle
\right|^2 \,,
\label{eq:PPP1}\\
\!\!&=& \!\!
{\delta}_{\alpha\beta}
-4\sum_{i<j}
 {\rm Re}({\tilde U}^{\ast}_{\alpha i}{\tilde U}_{\beta i}^{}
     {\tilde U}_{\alpha j}^{}{\tilde U}^{\ast}_{\beta j})
 \sin^2\dfrac{{\tilde \Delta}_{ij}}{2} 
+2\sum_{i<j}
 {\rm Im}({\tilde U}^{\ast}_{\alpha i}{\tilde U}_{\beta i}^{}
     {\tilde U}_{\alpha j}^{}{\tilde U}^{\ast}_{\beta j})
 \sin{\tilde \Delta}_{ij}\,,~~~~~
\label{eq:PPP2}
\end{eqnarray}
\label{eq:PPPs}
\end{subequations}
$\!\!\!\!$
where
\begin{equation}
 \tilde{\Delta}_{ij}
\equiv
 \dfrac{\lambda_j-\lambda_i}{2E_\nu}L\,.
\end{equation}

All our numerical results are based on eq.~(\ref{eq:PPPs}).
However, we find the following analytic 
approximations \cite{HOS1,HOSF} useful for
understanding the reason 
why and how the one-beam two-detectors 
long baseline experiments such as T2KK \cite{HOS1}-\cite{HOSF}
and T2KO, with a far detector in Oki Island,
can determine the neutrino mass hierarchy and the CP phase
simultaneously.
They are obtained by expanding the oscillation probabilities
in terms of the three small parameters;
the matter effects, which is small at energies below a few GeV
around the earth crust for $\bar{\rho} \sim 3$ g/cm$^3$ and
$L\lsim 1000$~km,
the oscillation phase $\Delta_{12} \equiv \dm12 L/2E_\nu$,
which is also small ($\sim \pi/30$)
near the first oscillation maximum,
$|\Delta_{13}|\sim\pi$,
and the mixing factor $|U_{e3}|^2=\srct{}$, eq.~(\ref{eq:def_srct}),
which is about $1/40$ from the recent reactor measurements
eqs.~(\ref{eq:DayaBay_s13}) and (\ref{eq:RENO_s13}).

The $\nu_\mu$ survival probability
can then be approximated as
\begin{eqnarray}
 P_{\nu_\mu \to \nu_\mu} &=&
 1 - \satm{2}\left( 1 + A^{\mu} \right)
       \sin^2 \left( \dfrac{\Delta_{13}+B^{\mu}}{2}\right)
\,,
\label{eq:Pmm}
\end{eqnarray}
around the first maximum of the main oscillation phase
\begin{equation}
 \Delta_{13} \equiv \dm13\dfrac{L}{2E_\nu} =
 \dfrac{m_3^2-m_1^2}{2E_\nu}L\,,
\end{equation}
with
\begin{subequations}
\label{eq:ABmu}
\begin{eqnarray}
A^{\mu} &=& - \dfrac{a_0 L}{\Delta_{13}E_\nu} 
\left(1-\tan^2\theta_{\atm}\right)
\srct{} 
\label{eq:Amu}\,,\\
B^{\mu} &=& \dfrac{a_0 L}{2E_\nu} 
\left(1-\tan^2\theta_{\atm}\right)
\srct{} 
\label{eq:Bmu}
\\
&&
- \Delta_{12}
\left(\csun{}+ \tan^2 \theta_{\atm}\ssun{}\srct{}
 - \tan \theta_{\atm}\sin 2\theta_{\sun}\sin\theta_{\rct}\cos\dmns
\right) \nn
\end{eqnarray}
\end{subequations}
We find that the above formula reproduce 
the survival probability with 1\%
accuracy throughout the parameter range explored in this analysis,
which covers all the three neutrino model parameters in
the $3\sigma$ allowed range of the present measurements
eqs.~(\ref{eq:atm_data})-(\ref{eq:RENO_s13}),
for the neutrino energies 400~MeV $<E_\nu<$ 4~GeV
(for $\nu_\mu$, $\bar{\nu}_\mu$),
and for the baseline lengths $295$~km $<L<$ 1200~km,
except where the probability is very small,
($P_{\nu_\mu\to\nu_\mu} \lsim 10^{-5}$) \cite{HOS2,HOSF}.
The matter effects are proportional to the term
\begin{equation}
 \dfrac{a_0L}{2E_\nu} \simeq 0.58 
\left(
\dfrac{\bar{\rho}}{3\mbox{{g/cm$^3$}}}
\right)
\left(
\dfrac{L}{1000\mbox{{km}}}
\right)
\end{equation}
and it appears with the opposite sign both in the amplitude
correction term $A^\mu$ eq.~(\ref{eq:Amu})
where $\Delta_{13}=|\Delta_{13}|$ for the normal while
$\Delta_{13}=-|\Delta_{13}|$ for the inverted hierarchy,
and also in the phase shift in eq.~(\ref{eq:Pmm}), 
where $|\Delta_{13}+B^\mu|=\pi+B^\mu$ for the normal,
$\pi-B^\mu$ for the inverted hierarchy around the oscillation
maximum $|\Delta_{13}|\simeq \pi$.
Nevertheless, the effect is too small to be observed in the near
future
because the term $1-\tan^2\theta_{\atm}$ is bounded as
\begin{equation}
 0.48 > 1-\tan^2\theta_{\atm} > -0.92
\end{equation}
at $90\%$ C.L. from eq.~(\ref{eq:mixing_atm}),
and 
because both $A^\mu$ and $B^\mu$ are proportional to
the product of the two small parameters, 
the matter effect term of $a_0 L/2E_\nu$ and $\srct{}$.
The coefficient of $\cos \dmns$ in $B^\mu$ is also small,
only $1\%$ of the main oscillation phase $|\Delta_{13}|$.
In other words, the $\nu_\mu$ survival probability is very
insensitive to the unknown parameters of the three neutrino model,
and hence can be used to measure $\satm{}$ and $|\dm13|$ accurately.

Under the same conditions that give eq.~(\ref{eq:Pmm}) 
for the $\nu_\mu$ survival probability,
the $\nu_e$ appearance probability can be approximated as \cite{HOSF}
\begin{eqnarray}
P_{\nu_{\mu} \to \nu_e} &=& 
4 \satm{} \srct{}
\left\{
 \left( 1 + A^{e} \right)
  \sin^2 \left( \dfrac{\Delta_{13}}{2} \right)
+\dfrac{B^e}{2} \sin \Delta_{13}
\right\}
+C^e
 \,,
\label{eq:Pme}
\end{eqnarray}
where
we retain both linear and quadratic terms of 
$\Delta_{12}$ and $a_0$.
The analytic expressions for the correction terms 
$A^e$, $B^e$ and $C^e$ are found in Ref.\cite{HOSF}.
For our semi-quantitative discussion below, the following
numerical estimates \cite{HOSF} for $\satm{2}=1$ and
$\ssun{2}=0.852$ suffice:
\begin{subequations}
\begin{eqnarray}
A^e &\simeq& 0.37
\dfrac{\bar{\rho}}{3\mbox{g/cm}^3}
\dfrac{L}{1000\mbox{km}}
\dfrac{\pi}{\Delta_{13}}
\left(1-\dfrac{\srct{2}}{2}\right)\nn\\
&&-0.29
\left|\dfrac{\Delta_{13}}{\pi}\right|
\sqrt{\dfrac{0.1}{\srct{2}}}
\left[1+0.18\dfrac{\bar{\rho}}{3\mbox{g/cm}^3}
\dfrac{L}{1000\mbox{km}}
\dfrac{\pi}{\Delta_{13}}
\right]\sin\dmns\,,
\label{eq:AeAprox}\\
B^e &\simeq& -0.58
\dfrac{\bar{\rho}}{3\mbox{g/cm}^3}
\dfrac{L}{1000\mbox{km}}
\left(1-\dfrac{\srct{2}}{2}\right) \nn\\
&&+
0.30\left|\dfrac{\Delta_{13}}{\pi}\right|
\left[
\sqrt{\dfrac{0.1}{\srct{2}}}\cos\dmns
-0.11\right]
\left[
1+0.18
\dfrac{\bar{\rho}}{3\mbox{g/cm}^3}
\dfrac{L}{1000\mbox{km}}
\dfrac{\pi}{\Delta_{13}}
\right]\,.~~~~~~~
\label{eq:BeAprox}
\end{eqnarray}
\label{eq:ABeApprox}
\end{subequations}
$\!\!\!$
The term $C^e$ is relevant only when the $\nu_\mu \to \nu_e$
oscillation probability is very small \cite{HOSF}.

The first term in $A^e$ in eq.~(\ref{eq:AeAprox}) is sensitive
not only to the matter effect
but also to the mass hierarchy pattern,
since $\Delta_{13}\sim\pi$ for the normal 
while $\Delta_{13} \sim -\pi$ for the inverted hierarchy
around the oscillation maximum.
For the normal (inverted) hierarchy, 
the magnitude of the $\nu_\mu \to \nu_e$ transition probability is
enhanced (suppressed) by 
about $10\%$ at Kamioka, 
$24\%$ at Oki Island, 
and $37\%$ at $L\sim 1000$~km in Korea,
around the first oscillation maximum,
$|\Delta_{13}|\sim \pi$.
When $L/E_\nu$ is fixed at $|\Delta_{13}| \sim \pi$,
the difference between the two hierarchy cases grows with $L$,
because the matter effect grows with $E_\nu$.
Within the allowed range of the model parameters,
the difference of the $A^e$ between SK and a far detector at Oki or
Korea becomes
\begin{subequations} 
\begin{eqnarray}
 A^e_{\rm peak}(L=653{\rm km}) -
 A^e_{\rm peak}(L=295{\rm km}) &\simeq&
\pm 0.13\,,\label{eq:Ae_OKI-SK}\\
 A^e_{\rm peak}(L\sim 1000 {\rm km}) -
 A^e_{\rm peak}(L=295 {\rm km}) &\simeq&
\pm0.26\,,\label{eq:Ae_Korea-SK}
\end{eqnarray}
 \label{eq:Ae_diffs}
\end{subequations}
$\!\!\!\!$
where the upper sign corresponds to the normal,
and the lower sign for the inverted hierarchy.
The hierarchy pattern can hence be determined by comparing 
$P_{\nu_\mu\to\nu_e}$ near the oscillation maximum $|\Delta_{13}|\simeq\pi$
at two vastly different baseline lengths~\cite{HOS1}-\cite{HOSF},
{\it independently} of the sign and magnitude of $\sin \dmns$.

In eq.~(\ref{eq:BeAprox}), it is also found that
the first term in $B^e$,
which shifts the oscillation phase from $|\Delta_{13}|$ to
$|\Delta_{13}+B^e|=|\Delta_{13}| \pm B^e$, 
is also sensitive to the mass hierarchy pattern.
As in the case for $A^e$, 
the difference in $B^e$ between SK and a far detectors is found
to be
\begin{subequations} 
\begin{eqnarray}
B^e_{\rm peak}(L=653{\rm km}) -
 B^e_{\rm peak}(L=295{\rm km}) &\simeq&
\mp 0.10\,,\label{eq:Be_OKI-SK}\\
 B^e_{\rm peak}(L\sim 1000 {\rm km}) -
 B^e_{\rm peak}(L=295 {\rm km}) &\simeq&
\mp0.20\,,\label{eq:Be_Korea-SK}
\end{eqnarray}
\label{eq:Be_diffs}
\end{subequations}
$\!\!\!\!$
where the upper sign is for the normal,
and the lower sign for the inverted hierarchy.
This implies that the mass hierarchy pattern can also be discriminated
by measuring the neutrino energy of the first oscillation maximum
at two different baseline lengths.
As in the case of the oscillation amplitude above,
this determination can be made independent of the unknown $\cos \dmns$ 
terms since they cancel in the differences, eq.~(\ref{eq:Be_diffs}).

Once the sign of $\Delta_{13}$ is fixed by the terms linear in $\bar{\rho}$, 
the terms linear in $\Delta_{12}$, 
which appear as those proportional to $|\Delta_{13}|\sim 30 \Delta_{12}$ 
in eq.~(\ref{eq:ABeApprox}),
allow us to constrain $\sin \dmns$ via the amplitude $A^e$, and
$\cos \dmns$ via the phase shift $B^e$.
Therefore, $\dmns$ can be measured uniquely
once the mass hierarchy pattern is determined.
From the above discussions, we understand qualitatively why the mass
hierarchy as well as both $\sin \dmns$ and $\cos \dmns$ can be determined
uniquely by observing the $\nu_\mu \to \nu_e$ oscillation probability
around the first oscillation maximum at two vastly different baseline lengths.
Therefore, in order to take advantage of this very efficient mechanism
to determine all the main unknowns of the three neutrino model,
one should arrange for high neutrino flux both at near and far
detectors around the first oscillation maximum which appears at the
same energy-to-baseline-length ratio,
\begin{equation}
 \left|\dfrac{\Delta_{13}}{\pi}\right| = 
\dfrac{(L/295{\mbox{{km}}})}{(E_\nu/{0.55\mbox{{GeV}}})}
=
\dfrac{(L/653{\mbox{{km}}})}{(E_\nu/{1.22\mbox{{GeV}}})}
=
\dfrac{(L/1000{\mbox{{km}}})}{(E_\nu/{1.86\mbox{{GeV}}})}
\label{eq:EL_ratio}
\end{equation}
for $|\dm13|=\numt{2.35}{-3}$eV$^2$; eq.~(\ref{eq:mass_atm}).
This observation 
led to the T2KK proposal of choosing the $3^\circ$ off-axis beam
($E_\nu^{\rm peak}\sim 0.55$GeV) at SK
and
$\sim 0.5^\circ$ off-axis beam ($E_\nu^{\rm peak}\sim 1.1$GeV) at
$L\sim 1000$km \cite{HOS1,HOS2}.
Since the baseline length of $L \sim 653$km to Oki Island is about
a factor of two longer than $L \sim 295$km for T2K, we may also expect
that a beam with smaller off-axis angle at Oki Island 
will enhance its physics capability;
see, however,
discussions in section \ref{sec:mass}.

The oscillation probabilities for the anti-neutrino,
$P(\bar{\nu}_\alpha \to \bar{\nu}_\beta)$,
are obtained from the expressions for 
$P({\nu}_\alpha \to {\nu}_\beta)$
by reversing the sign of the matter effect term,
($\bar{\rho} \to -\bar{\rho}$),
and that of $\dmns$,
($\dmns \to -\dmns$).
The differences in the shift terms
$A^e$ and $B^e$ for $\nu_\mu \to \nu_e$ and 
$\bar{A}^e$ and $\bar{B}^e$ for $\bar{\nu}_\mu \to \bar{\nu}_e$ 
oscillation probabilities are, respectively,
\begin{subequations} 
\begin{eqnarray}
A^e -\bar{A}^e
&=& \pm 0.74
\dfrac{\bar{\rho}}{3\mbox{g/cm}^3}
\dfrac{L}{1000\mbox{km}}
\left|\dfrac{\pi}{\Delta_{13}}\right|
\left(1-\dfrac{\srct{2}}{2}\right)\nn\\
&&
-0.58
\left|\dfrac{\Delta_{13}}{\pi}\right|
\sqrt{\dfrac{0.1}{\srct{2}}}\sin\dmns\,,
\label{eq:diff_Ae}\\
B^e - \bar{B}^e 
&=&-1.16
\dfrac{\bar{\rho}}{3\mbox{g/cm}^3}
\dfrac{L}{1000\mbox{km}} 
\nn\\
&&\times
\left[
1-\dfrac{\srct{2}}{2}
\mp 0.093
\left(\sqrt{\dfrac{0.1}{\srct{2}}}\cos\dmns-0.11\right)
\right]\,,
\label{eq:diff_Be}
\end{eqnarray}
 \label{eq:diff_Ae_And_Be}
\end{subequations}
$\!\!\!\!$
where the upper sign should be taken for the normal and
the lower sign for the inverted hierarchy.
From eq.~(\ref{eq:diff_Ae}), 
the difference of the amplitude between $\nu_e$ and $\bar{\nu}_e$
appearance probability grows with the baseline length
with the opposite sign for the normal and inverted hierarchies.
The dependence on $\sin \dmns$ is also strong since 
$A^e$ and $\bar{A}^e$ changes in the opposite direction 
when we vary $\sin \dmns$.
The phase-shift term $B^e$ and $\bar{B}^e$ also change sign in the
opposite direction for the normal and inverted hierarchies,
whose sign is independent of $\cos \dmns$ because of the smallness
of the $\cos \dmns$ dependence in eq.~(\ref{eq:diff_Be}).
By the same token,
the use of both $\nu_\mu$ and $\bar{\nu}_\mu$ beams
does not improve significantly the measurement of $\cos \dmns$.
Because these sign changes occur independently of the
$L$-dependence of the oscillation probabilities,
the physics potential of the T2KO experiment will be improved 
by dividing the total exposure time to neutrino and
anti-neutrino beams just as in T2KK \cite{HOSF}.

It is worth noting here that all the formalism presented in this
section should be useful for studying physics potential of T2K plus
NOvA \cite{nova},
whose baseline length of $L\simeq 810$~km lies between those of Tokai
to Oki Island ($L\simeq 653$~km) and to Korea ($L\gsim 1000$~km).

\section{T2K, T2KO, and T2KK baselines}
\label{sec:far}
In this section,
we study the matter profile along the baselines between 
J-PARC and Kamioka, Oki Island, and Korea
by referring to recent seismological measurements
\cite{FossaMagna}-\nocite{T2Kmat,Jmat,yamato,OKI_mat,tsushima,Kmat}
\cite{dvconv}.
We also show the relation between the off-axis angles
at SK and Oki Island, and the beam profiles
for the relevant off-axis angles \cite{website}.

\subsection{Matter profile along the baselines}

\begin{figure}[t]
\centering
 \includegraphics[scale=1.0]{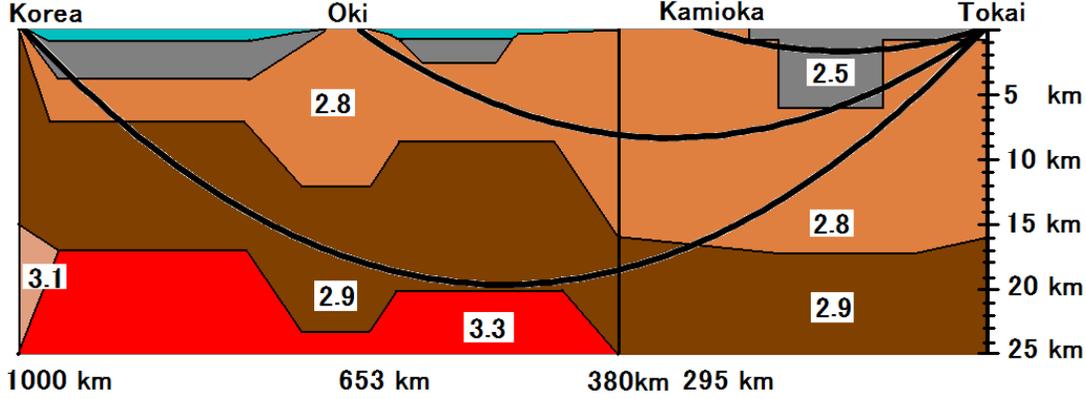}
 \caption{
The cross section view of the T2K, T2KO, and T2KK experiments
along the baselines, which are shown by the three curves.
The horizontal scale gives the distance from J-PARC along
the arc of the earth surface and the vertical scale measures the
depth of the baseline below the sea level.
The numbers in the white boxes are the average matter density 
in units of g/cm$^3$ \cite{FossaMagna}-\cite{dvconv}.
} 
\label{fig:Xsec-T2OKI}
\end{figure}

Because Oki Island is placed just at the middle point
between SK and the east shore of Korea
along the T2K beam, as shown in Fig.~\ref{fig:T2OKI1},
the cross section view of the T2K, T2KO, and T2KK experiments along the
baselines can be shown on one frame as in Fig.~\ref{fig:Xsec-T2OKI}.
The horizontal axis of Fig.~\ref{fig:Xsec-T2OKI}
gives the distance from J-PARC along the arc of
the earth surface and the vertical axis denotes the depth below the
sea level.
Three curves show the baselines from the J-PARC to SK ($L=295$Km), 
Oki Island ($L=653$km) and to the Korean east shore at $L=1000$km.
The numbers in the white boxes are the average matter density 
of the layer in units of g/cm$^3$.

If we assume that the matter density in each layer of the earth crust
as shown in Fig.~\ref{fig:Xsec-T2OKI} has the value equals to the
quoted  average,
we can estimate the average matter density $\bar{\rho}$ 
along the three baselines:
\begin{subequations}
\begin{eqnarray}
 \bar{\rho}_{\sSK} &=& 2.60~\mbox{g/cm$^3$}\,,\label{eq:matterSK}\\
 \bar{\rho}_{\sOKI}&=& 2.75~\mbox{g/cm$^3$}\,,\label{eq:matterOKI}\\
 \bar{\rho}_{\sKR} &=& 2.90~\mbox{g/cm$^3$}\,.\label{eq:matterKr}
\end{eqnarray} 
\label{eq:matter}
\end{subequations}
$\!\!\!\!$
The error of these average density can be estimated from the 
uncertainty of the matter density in each region
and the location of the boundary of each layer.
They are measured by using the velocity of the seismic
wave in most geophysical researches.
From the uncertainty in the correlation
between the matter density and the measured $p$-wave sound
velocity \cite{dvconv, HOSF},
we adopt $6\%$ overall error for the matter density.
Small fluctuation in the matter density in each layer of the earth
crust does not affect the neutrino oscillation probabilities
significantly,
because the contribution from the higher Fourier modes of the matter
density distribution is strongly suppressed \cite{koike98,HOSF}.
The locations of the boundaries are measured rather accurately from
the reflection point of the seismic wave.
From Fig.~\ref{fig:Xsec-T2OKI}, we find that
the neutrino beam for the Tokai-to-Oki baseline goes through the upper
crust layer with $\bar{\rho}=2.8$g/cm$^3$, except when it crosses
Fossa Magna filled with sediment.
The uncertainty of the boundary depth between the sediment layer
and the upper crust is only a few hundred meters.
The error from the uncertainty of the boundary
depth can hence be safely neglected for the T2KO experiment.
The traveling distance in the mantle and the crust depends on the depth 
of the boundary between the lower crust and the upper mantle for the T2KK 
experiment. 
The average matter density is sensitive to the boundary location for the 
baseline of $L \sim 1000$~km,
because the baseline almost touches the mantle as can be seen 
from Fig.~\ref{fig:Xsec-T2OKI}.
The resulting uncertainty has been estimated and is found to be 
significantly smaller than the $6\%$ overall uncertainty \cite{HOSF}.
We therefore use the average matter density along each baseline 
given in eq.~(\ref{eq:matter}) to 
estimate the oscillation probability and assign its overall
uncertainty of $6\%$ in this study.

\subsection{Off-axis angles and beam profiles}

\begin{figure}[t]
\centering
 \includegraphics[scale=0.8]{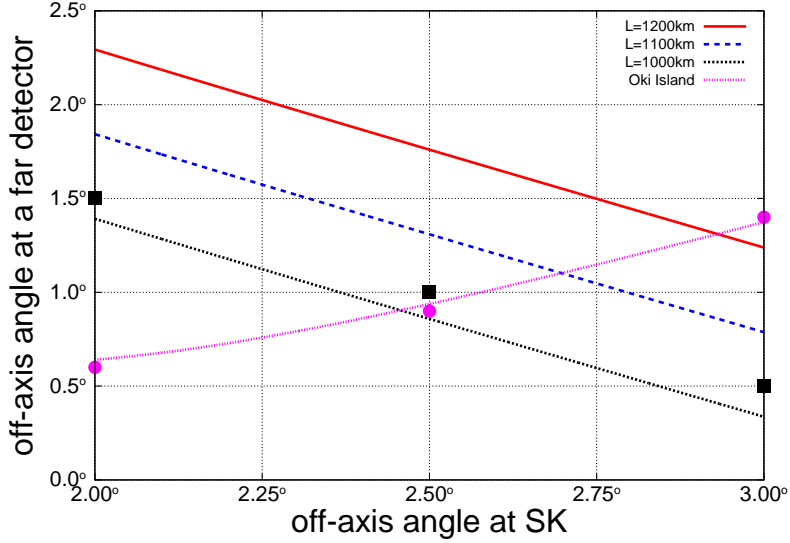}
 \caption{
The relation among the off-axis angles at SK, Oki Island,
and at a far detector in Korea ($L=1000$km, 1100km, and 1200km).
The horizontal axis gives the off-axis angle at SK
and the vertical axis is that at the far detector locations.
The purple line shows the off-axis angle at Oki Island.
The red, blue, and black lines are the smallest
off-axis angle at $L=1000$km,
1100km and 1200km, respectively.
The points corresponding to the 
purple circles and black squares are used in our numerical
analysis.
} 
\label{fig:relation}
\end{figure}

Figure~\ref{fig:relation}
shows the relation among the off-axis angles of the neutrino beam from
J-PARC observable at SK, Oki,
and a far detector in Korea ($L=1000$km, 1100km, and 1200km).
The horizontal axis gives the off-axis angle at SK
and the vertical axis gives the corresponding off-axis angles
at far detector locations.
The purple line shows the off-axis angle at Oki Island,
which grows as that at SK is increased,
because both SK and Oki Island is located in the east (upper) side of
the beam center of the T2K neutrino beam; see Fig.~\ref{fig:T2OKI1}.
The red, blue, and black lines are the smallest
off-axis angle observable in Korea 
at $L=1000$km, 1100km, and 1200km, respectively.
In contrast to the Oki Island case,
the off-axis angle at Korea decreases as that at SK grows,
because Korea is in the west (lower) side of the beam center,
also as shown in Fig.~\ref{fig:T2OKI1}.
The points corresponding to the purple circles and black squares in
the figure are used in our
numerical analysis in the following sections.

\begin{figure}[t]
\centering
\includegraphics[scale=0.65]{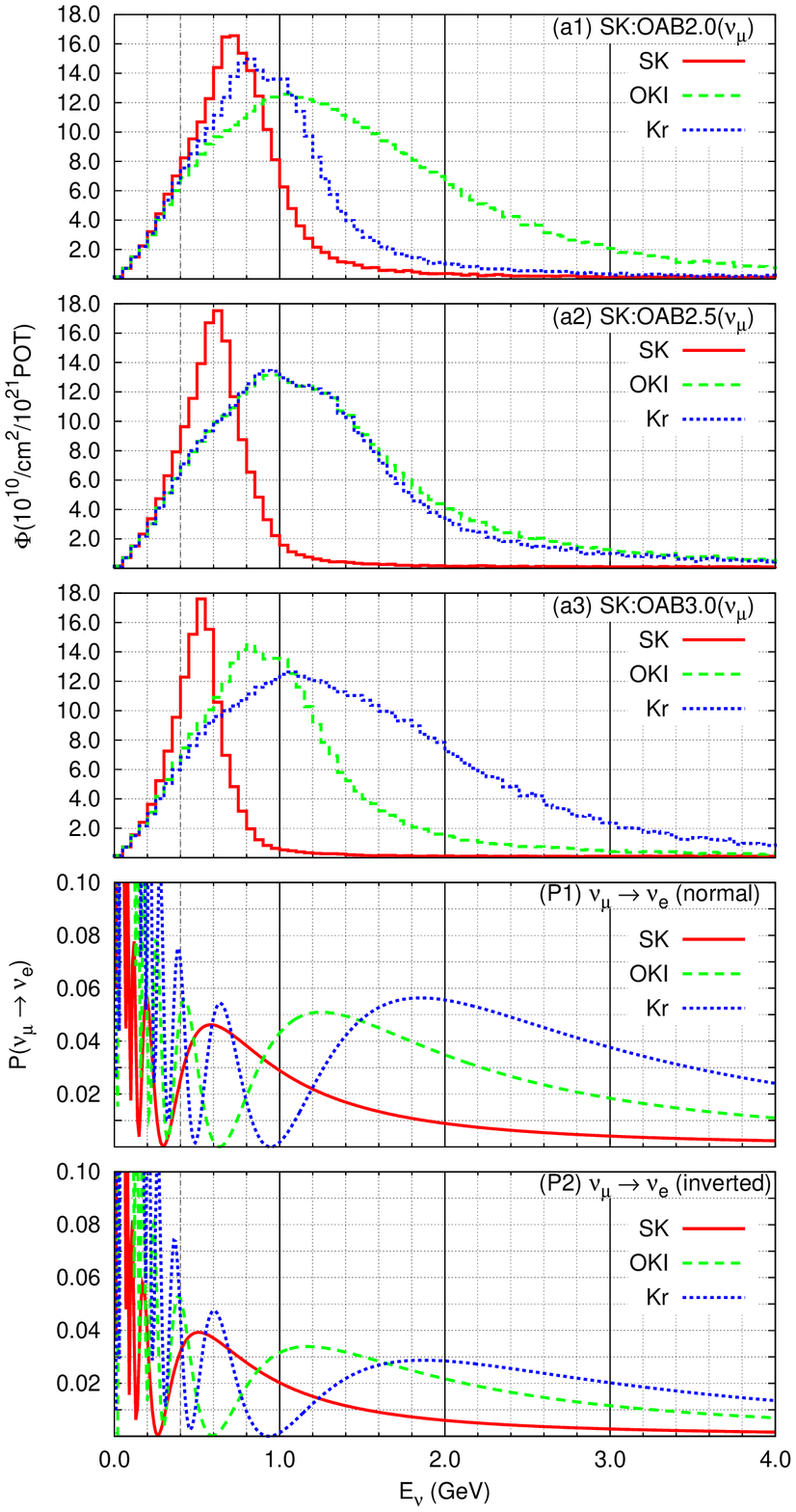}
\includegraphics[scale=0.65]{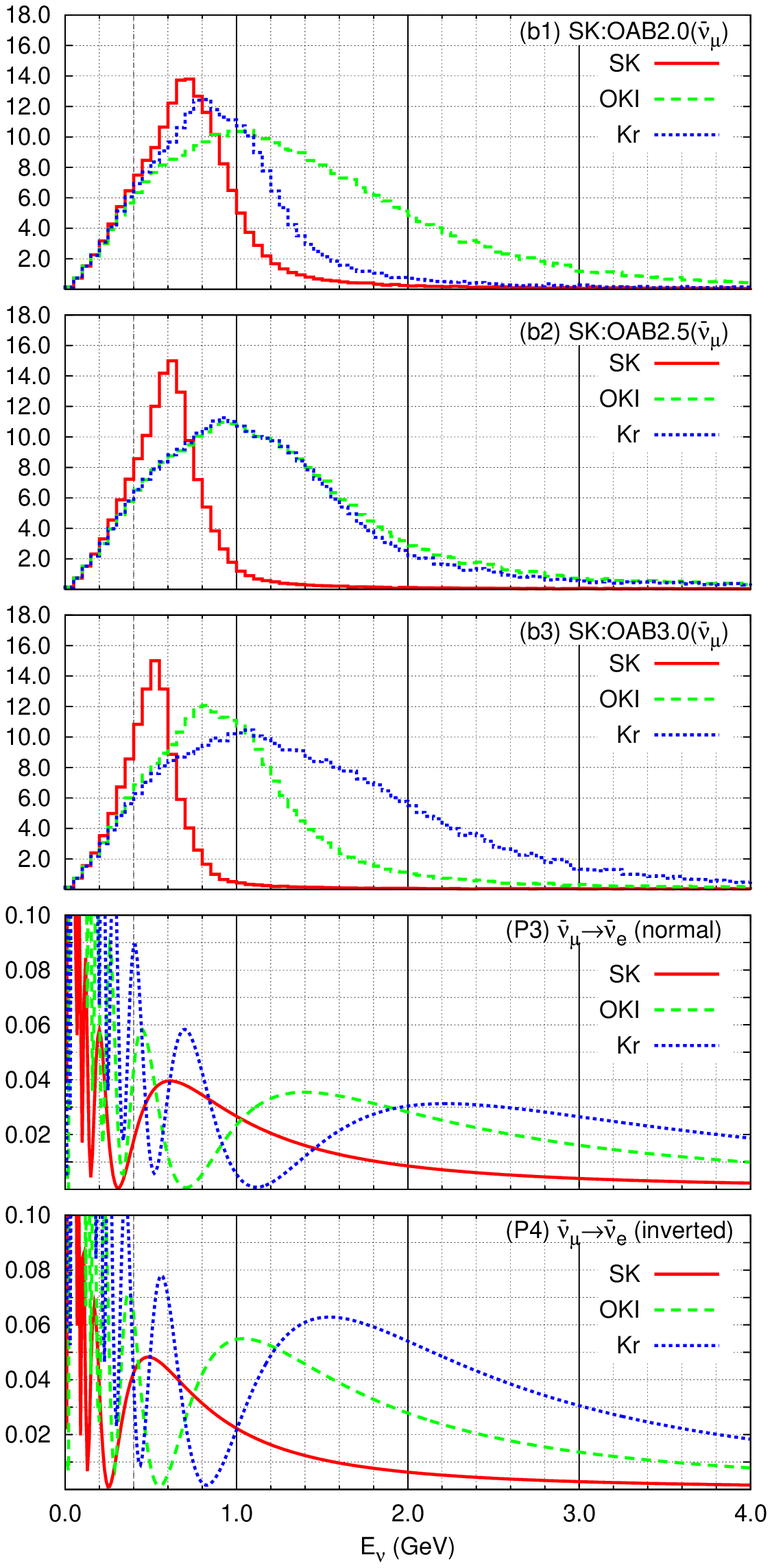}
 \caption{
The beam profiles and the oscillation probabilities as functions of
the neutrino energy.
The left column panels are for the $\nu_\mu$ focusing beam
and the right ones are for the $\bar{\nu}_\mu$ focusing beam.
The upper 6 panels, (a1)-(a3) and (b1)-(b3), show the beam
fluxes at $L=1$~km
and the bottom 4 panels, (P1)-(P4), show the $\nu_\mu \to \nu_e$
($\bar{\nu}_\mu\to\bar{\nu}_e$) oscillation probability with the input
parameters of eqs.~(\ref{eq:matter}) and (\ref{eq:input})
for $\srct{2}=0.08$ and $\dmns=0^\circ$.
The first row, (a1) and (b1) gives the beam profile at
each far detector for the $2.0^\circ$ off-axis beam (OAB) at SK,
the second row, (a2) and (b2) is for the $2.5^\circ$ OAB at SK,
and the third row, (a3) and (b3) is for the $3.0^\circ$ OAB at SK.
The fourth row, (P1) and (P3), shows the 
$\nu_\mu \to \nu_e$ and $\bar{\nu}_\mu \to \bar{\nu}_e$
oscillation probability for the normal hierarchy, respectively,
and
the bottom row, (P2) and (P4), is for the inverted hierarchy.
} 
\label{fig:profile}
\end{figure}

In Fig.~\ref{fig:profile},
the beam profiles 
of the three off-axis beams (OAB) with $2.0^\circ$, $2.5^\circ$, and
$3.0^\circ$ off-axis angles at SK are shown at SK, Oki Island (three
purple circles in Fig.~\ref{fig:relation}) and at $L=1000$km in Korea
(three black squares in Fig.~\ref{fig:relation}), together with 
the $\nu_\mu \to \nu_e$ and $\bar{\nu}_\mu \to \bar{\nu}_e$ 
oscillation probabilities as functions of the neutrino energy, 
$E_\nu$ in units of GeV.
The left column panels are for the $\nu_\mu$ focusing beam and
the right column ones are for the $\bar{\nu}_\mu$ focusing beam.
The upper three rows, (a1)-(a3) and (b1)-(b3), display the beam flux 
for the $\nu_\mu$ and $\bar{\nu}_\mu$ focusing beam, respectively,
at $L=1$ km 
in units of $10^{10}$/cm$^2$/$10^{21}$POT (protons on target).
The first row, (a1) and (b1), gives the beam profile observable at each far
detector for the $2.0^\circ$ off-axis beam (OAB) at the SK
(red solid line), 
which gives the $0.6^\circ$ OAB at Oki Island (green dashed line),
and
the $1.5^\circ$ OAB for a far detector in Korea at $L=1000$~km
(blue dotted line).
The second row, (a2) and (b2), is for the $2.5^\circ$ OAB at SK,
which gives the $0.9^\circ$ OAB at Oki Island, and the $1.0^\circ$ OAB at
$L=1000$~km.
The case for the $3.0^\circ$ OAB at SK, 
which gives the $1.4^\circ$ OAB at Oki Island 
and the $0.5^\circ$ OAB at $L=1000$~km, is shown in
the third row.
It is clearly seen that all fluxes at Oki and Korea are harder than
those at SK.
The hardest flux with $E^{\rm peak} \gsim 1$~GeV is found for the
$2.0^\circ$ at SK beam at Oki, and for the $3.0^\circ$ at SK beam in
Korea.
The $2.5^\circ$ at SK beam gives almost the same flux shape of
 $\sim 1.0^\circ$ OAB at Oki and at the $L=1000$~km
location in Korea,
as shown in Fig.~\ref{fig:T2OKI1}.
This may call upon the possibility of locating detectors both
in Oki and in Korea, where both detectors observe exactly the same
OAB around the off-axis angle of $1.0^\circ$.

The lower two rows in Fig.~\ref{fig:profile}
show the $\nu_\mu \to \nu_e$
($\bar{\nu}_\mu \to \bar{\nu}_e$) 
oscillation probability
in the left (right) column,
for the normal hierarchy, (P1) and (P3),
and
for the inverted hierarchy, (P2) and (P4).
The oscillation probabilities  are calculated for the average matter
densities of eq.~(\ref{eq:matter}),
and the three-neutrino-model parameters:
\begin{subequations} 
\begin{eqnarray}
\dm12&=&\numt{7.5}{-5}~\mbox{eV$^2$}\,, \label{eq:input_dm12} \\
\dm13&=&\pm\numt{2.35}{-3}~\mbox{eV$^2$}\,,  \label{eq:input_dm13} \\
\satm{} &=&0.5\,,\label{eq:input_satm} \\
\ssun{2}&=&0.852\,,\label{eq:input_ssun}
\end{eqnarray}
\label{eq:input}
\end{subequations}
$\!\!\!$
with
\begin{subequations} 
\begin{eqnarray}
\srct{2} &=& 0.08\,, \\
\dmns    &=& 0.0^\circ\,.
\end{eqnarray}
\label{eq:input_rct}
\end{subequations}

\noindent
The plus (minus) sign in eq.~(\ref{eq:input_dm13}) is taken for the
normal (inverted) hierarchy,
whose predictions are shown in the fourth (bottom) row
of Fig.~\ref{fig:profile}.
The red solid line in these panels is the oscillation probability at the SK,
the green dashed line is for Oki Island,
and the blue dotted line is for a far detector in Korea, 
at $L=1000$~km.

From the four plots (P1)$\sim$(P4) 
in the bottom two rows in Fig.~\ref{fig:profile},
we find that
the first oscillation peak at SK appears at $E_\nu$ around  
$0.4 \sim 0.6$~GeV for all the four cases, $\nu_\mu$ vs $\bar{\nu}_\mu$
and for both hierarchies.
Since the OAB peaks at around $E_\nu=0.7$, 0.6, and 0.5~GeV for
$2.0^\circ$, $2.5^\circ$, and $3.0^\circ$, respectively,
as shown by the red solid curves in the first three rows of
Fig.~\ref{fig:profile},
all the OAB at SK between $2.0^\circ$ and $3.0^\circ$ can observe
$\nu_\mu \to \nu_e$ and $\bar{\nu}_\mu \to \bar{\nu}_e$ oscillations 
with good efficiency.
On the other hand, the oscillation probabilities at Oki Island are
high at $E_\nu$ around $1.0 \sim 1.4$~GeV from the green dashed 
curves in the bottom two rows,
where the flux is slightly small for the $3.0^\circ$ OAB at SK,
as shown also by green dashed curves in the third row.
Finally, at $L=1000$~km, the $\nu_\mu \to \nu_e$ and $\bar{\nu}_\mu
\to \bar{\nu}_e$ oscillation probabilities are large at around
$E_\nu\sim 2$~GeV,
and hence the OAB with $3.0^\circ$ at SK is most favorable
\cite{HOS1}-\cite{T2KKbg}.
We will confirm the above observations for the whole region of the
three neutrino model parameter space in the following sections.

\section{Typical event numbers at each detector}
\label{sec:event}
In this section,
we explain how we estimate event numbers
at each far detector.
\subsection{Event calculation}

In order to compare the physics capability of the T2KO experiment
and that of T2KK experiment with a far detector at various baseline
lengths and OAB's,
we adopt the same conditions as those in Refs.~\cite{HOS1,HOS2,HOSF}:
All the detectors at Kamioka, Oki Island, and Korea 
are assumed to
have excellent detection and kinematical reconstruction capabilities for
$\nu_\mu$ and $\nu_e$ Charged Current Quasi-Elastic (CCQE) events
within the fiducial volumes of the 22.5~kton at Kamioka (SK)
and 100~kton at Oki Island or in Korea.
We use the neutrino fluxes of J-PARC beam at various off-axis
angles \cite{website} and the CCQE cross sections off 
water target \cite{Xsec}
to compute event numbers as functions of (reconstructed)
neutrino energy with the bin width of $\delta E_\nu=200$~MeV
at $E_\nu >400$~MeV.
The energy bin width of 200~MeV is chosen to take account of kinematical
reconstruction errors due to Fermi motion, resonance production and
detector resolutions \cite{T2KKbg}.
We take account of background from secondary beams, such as $\nu_e$,
$\bar{\nu}_e$ and $\bar{\nu}_\mu$ fluxes for the $\nu_\mu$ focusing
beam,
but we do not consider other backgrounds including the single $\pi^0$
production from the neutral current (NC) events \cite{T2KKbg}.
In other words, the results of our studies show what a perfect
detector of a given fiducial volume can achieve with neutrino beams
from J-PARC when it is placed along the T2K beam line.
Reconstruction efficiency and errors as well as the background
rejection capabilities depend on specific detector designs and their
studies are beyond the scope of this paper.

The number of $\nu_\alpha$ CCQE events from the $\nu_\beta$ flux in
the $\nu_\mu$-focusing beam in the $n$-th energy bin, 
$E_\nu^n=E_\nu^{\rm th}+(n-1) \times \delta E_\nu< E_\nu <E_\nu^n+\delta E_\nu$,
at each site are calculated as
\begin{equation}
 N_{\rm D}^n (\nu_\beta\to\nu_\alpha)
=
M_{\rm D} N_A \int_{E_\nu^n}^{E_\nu^n+\delta E_\nu}
\Phi_{\nu_\beta}(E) P_{\nu_\beta \to \nu_\alpha}(E)
\sigma_\alpha^{\rm CCQE}(E)~dE\,,
\label{eq:event_number}
\end{equation}
where the suffix ``D'' denotes the place of the far detector
(D=SK, Oki, Kr),
$\nu_{\alpha,\beta}$ is neutrino or anti-neutrino flavor 
$(\nu_{\alpha,\beta}=\nu_\mu, \nu_e, \bar{\nu}_\mu, \bar{\nu}_e)$,
$M_{\rm D}$ is the fiducial mass of each detector 
($M_{\rm SK}=22.5$~kton and $M_{\rm Oki}=M_{\rm Kr}=100$~kton),
$N_A=\numt{6.017}{23}$ is the Avogadro number,
$\Phi_{\nu_\beta}$ is the $\nu_\beta$ flux from J-PARC \cite{website}
which scales as $1/L^2$, 
$P_{\nu_\beta \to \nu_\alpha}$ is the neutrino or anti-neutrino
oscillation probability calculated by eq.~(\ref{eq:PPPs})
with the average matter density of eq.~(\ref{eq:matter}),
and
$\sigma_\alpha^{\rm CCQE}$ is the CCQE cross sections per nucleon in water
for each flavor type \cite{Xsec}.
We also define the event number 
for $\bar{\nu}_\mu$ focusing beam as
\begin{equation}
\overline{N}_{\rm D}^n ({\nu}_\beta\to{\nu}_\alpha)
=
M_{\rm D} N_A \int_{E_\nu^n}^{E_\nu^n+\delta E_\nu}
\overline{\Phi}_{{\nu}_\beta}(E) P_{{\nu}_\beta \to {\nu}_\alpha}(E)
\sigma_\alpha^{\rm CCQE}(E)~dE\,,
\label{eq:event_number_anti}
\end{equation}
where $\overline{\Phi}_{{\nu}_\beta}$ gives the ${\nu}_\beta$ flux
in the $\bar{\nu}_\mu$ focusing beam \cite{website}.

Because we only consider the contribution from the secondary neutrino
fluxes of the primary beam as the background in this analysis,
each $e$- and $\mu$-like event numbers
in the $n$-th bin are calculated as
\begin{equation}
 N_{\alpha,{\rm D}}^n = 
\sum_{\beta = e,\mu}
\left\{
N_{\rm D}^n(\nu_\beta \to \nu_\alpha)
+
N_{\rm D}^n(\bar{\nu}_\beta \to \bar{\nu}_\alpha)
\right\}
\,,
\label{eq:eventNU}
\end{equation}
for $\alpha = e,\mu$
at each far detector ``D'' (D=SK, Oki, Kr).
We also define the event numbers with $\bar{\nu}_\mu$ focusing beam
as 
\begin{equation}
 \overline{N}_{\alpha,{\rm D}}^n = 
\sum_{\beta = e,\mu}
\left\{
\overline{N}_{\rm D}^n(\bar{\nu}_\beta \to \bar{\nu}_\alpha)
+
\overline{N}_{\rm D}^n(\nu_\beta \to \nu_\alpha)
\right\}
\,,
\label{eq:eventANTI}
\end{equation}
for $\alpha = e,\mu$ at each far detector.

\subsection{SK}

\begin{figure}[t]
\centering
 \includegraphics[scale=0.6]{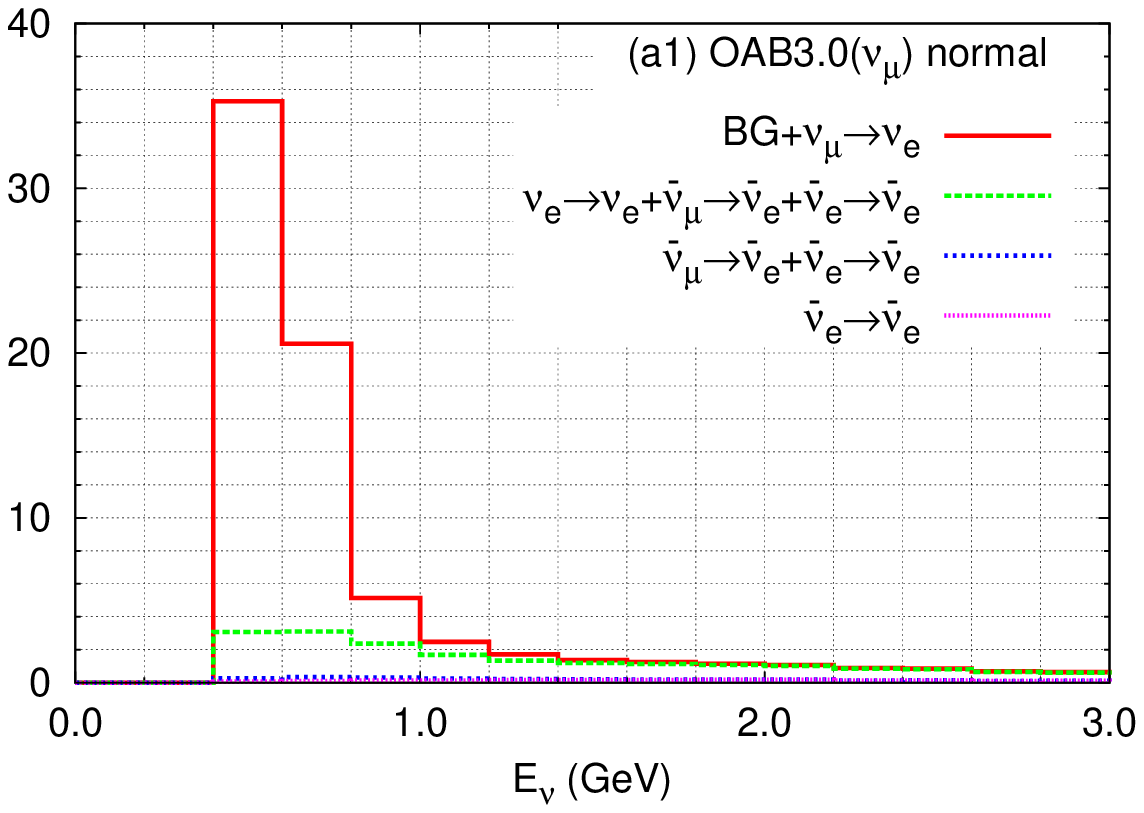}
~~~
 \includegraphics[scale=0.6]{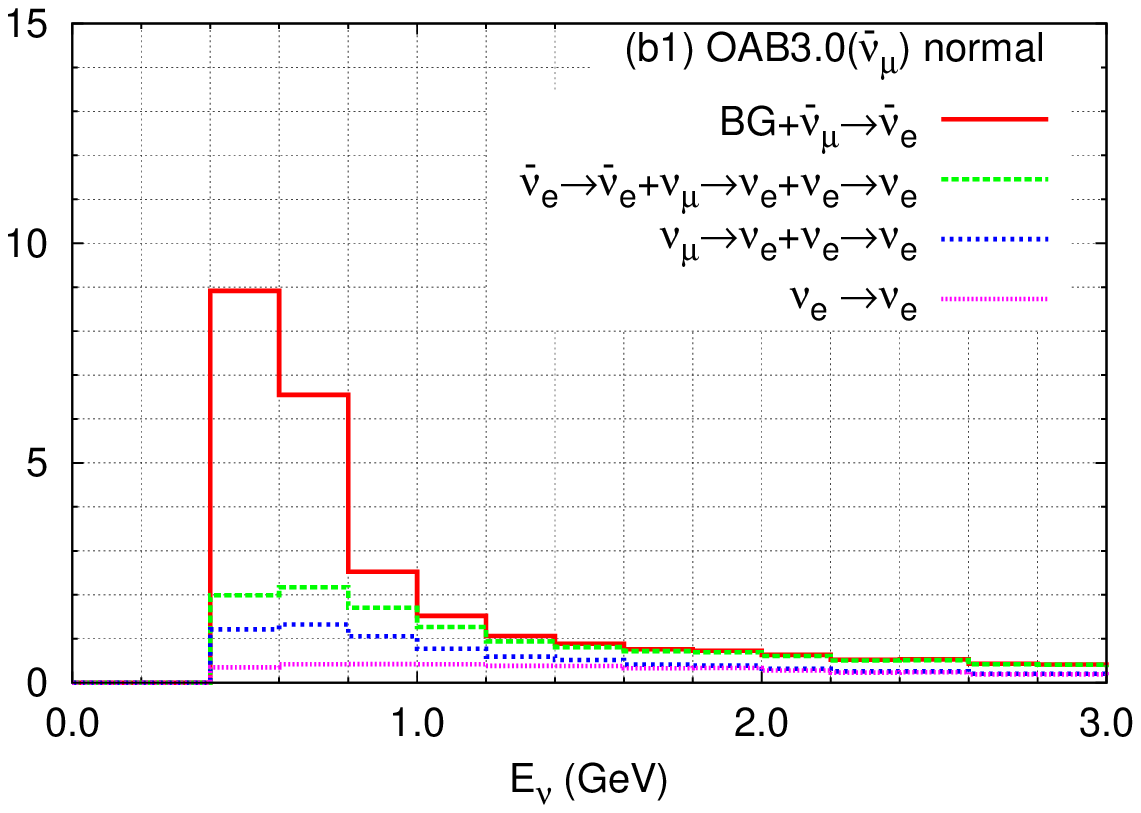}

 \includegraphics[scale=0.6]{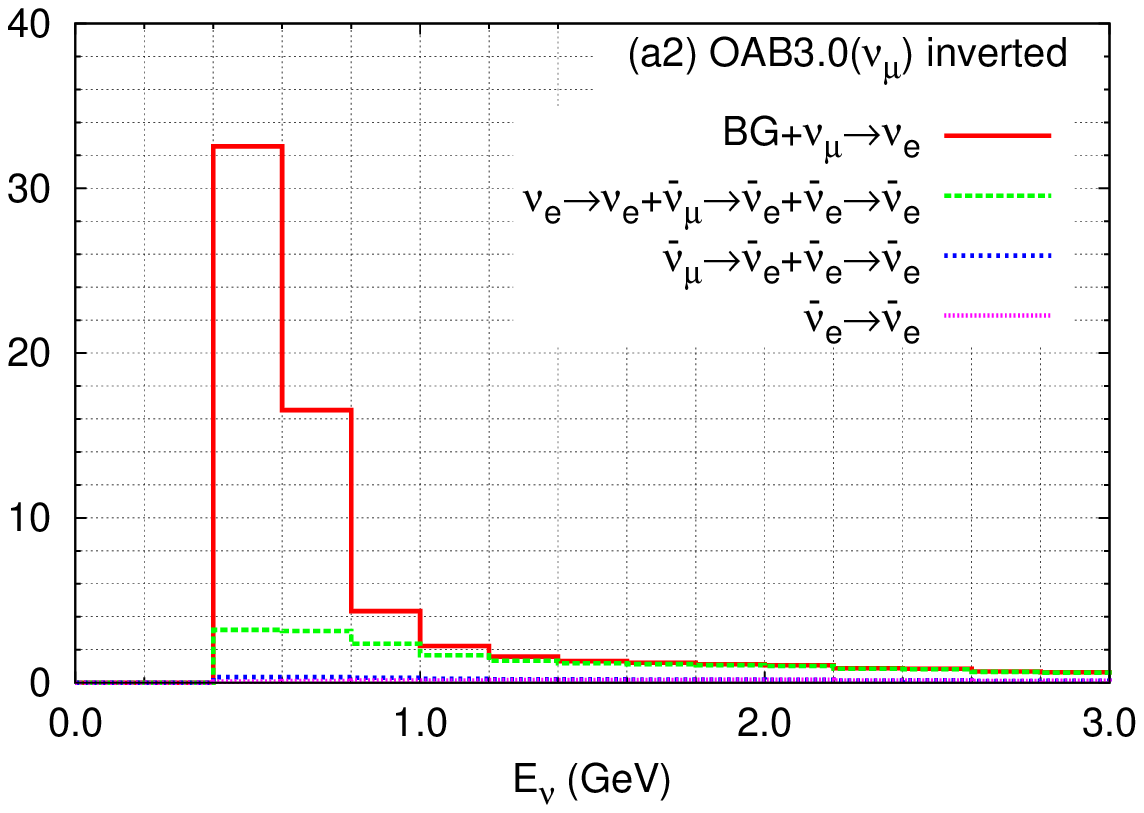}
~~~
 \includegraphics[scale=0.6]{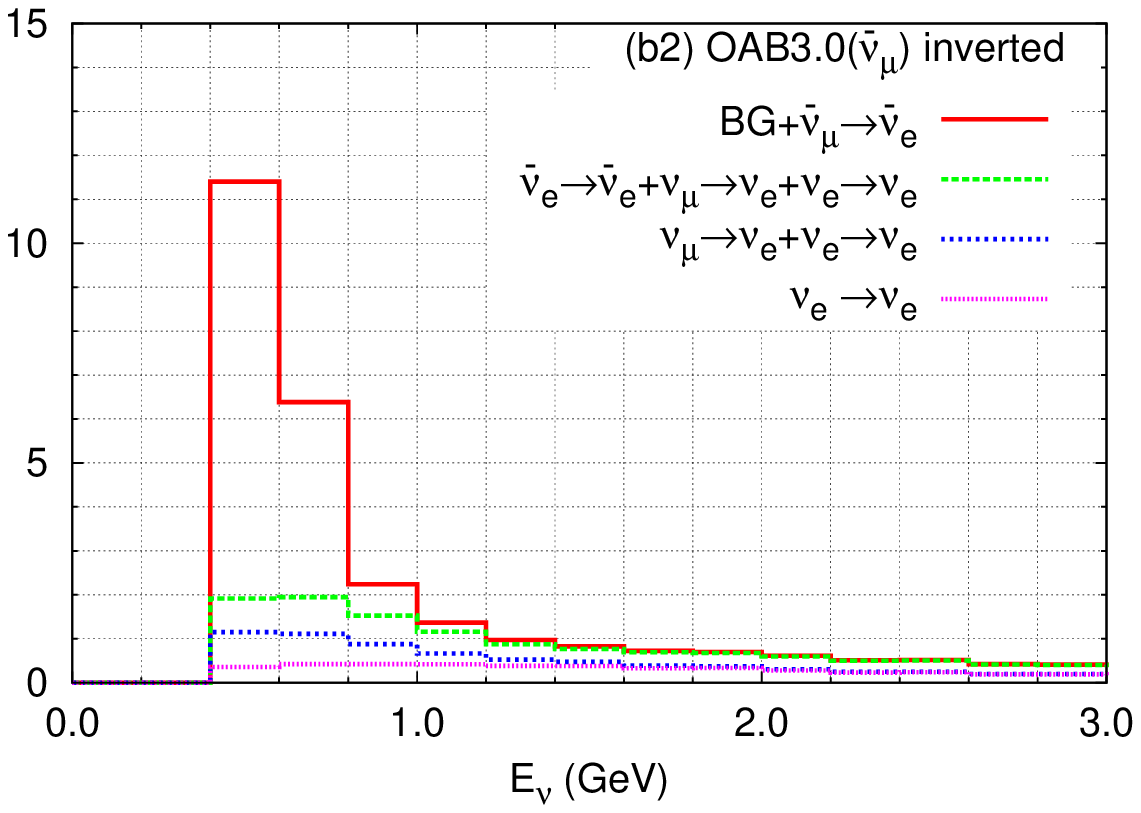}
 \caption{
Typical event number of the $e$-like CCQE events 
for $E_\nu$ ($E_{\bar{\nu}}$) $>0.4$~GeV
at SK (22.5kton)
with $3.0^\circ$ off-axis angle
and $\numt{2.5}{21}$POT exposure:
(a1) and (a2) are for $\nu_\mu$ focusing beam,
and
(b1) and (b2) are for $\bar{\nu}_\mu$ focusing beam;
whereas
(a1) and (b1) are for the normal hierarchy,
and
(a2) and (b2) are for the inverted hierarchy.
The results are for $\srct{2}=0.08$, $\dmns=0^\circ$,
and $\bar{\rho}_{\sSK} = 2.60$~g/cm$^3$
and the other input parameters of eq.~(\ref{eq:input}).
The red solid line in each panel denotes the total event numbers,
the purple dotted line shows the $\bar{\nu}_e \to \bar{\nu}_e$
contribution in (a1) and (a2),
$\nu_e \to \nu_e$ in (b1) and (b2),
the blue short dashed line gives the sum of the contributions from
$\bar{\nu}_\mu \to \bar{\nu}_e$ and $\bar{\nu}_e \to \bar{\nu}_e$
in (a1) and (a2),
that of $\nu_\mu \to \nu_e$ and $\nu_e \to \nu_e$
in (b1) and (b2),
and
the green dashed line shows the total background contribution
from the secondary beams.
}
\label{fig:T2K}
\end{figure}

The typical event numbers of the $e$-like CCQE events
for $E_\nu$ ($E_{\bar{\nu}}$) $>0.4$~GeV
at SK ($22.5$~kton)
with the $3.0^\circ$ OAB 
and $\numt{2.5}{21}$ POT exposure
is shown in Fig.~\ref{fig:T2K}:
(a1) and (a2) are for the $\nu_\mu$ focusing,
and
(b1) and (b2) are for the $\bar{\nu}_\mu$ focusing beam;
(a1) and (b1) are for the normal,
and
(a2) and (b2) are for the inverted hierarchy.
These results are for $\srct{2}=0.08$, $\dmns=0^\circ$
and $\bar{\rho}_{\sSK}=2.60$g/cm$^3$
the other input parameters of eq.~(\ref{eq:input}).
In each panel,
the red solid line denotes the total event number,
which is the sum of the signal events, $\nu_\mu \to \nu_e$ for
(a1) and (a2),
or $\bar{\nu}_\mu \to \bar{\nu}_e$ for
(b1) and (b2),
and
the total background from the secondary beams
shown by the green dashed line.
The purple dotted line shows the 
$\bar{\nu}_e \to \bar{\nu}_e$ contribution in (a1) and (a2),
${\nu}_e \to {\nu}_e$ in (b1) and (b2).
The blue short dashed line gives the sum of the contributions from
$\bar{\nu}_{\mu}\to\bar{\nu}_e$ and $\bar{\nu}_e\to\bar{\nu}_e$
in (a1) and (a2),
and those from ${\nu}_{\mu}\to\nu_e$ and ${\nu}_{e}\to\nu_e$
in (b1) and (b2).

In Fig.~\ref{fig:T2K},
both the signal (red solid minus green dash) and the total (red solid)
number of events peak in the first bin (400~MeV $<E_\nu<$ 600~MeV),
because both the $3.0^\circ$ OAB fluxes and the oscillation
probabilities are largest in the region;
see Fig.~\ref{fig:profile}, (a3), (P1), and (P2) for 
$\nu_\mu \to \nu_e$,
(b3), (P3), and (P4) for $\bar{\nu}_\mu \to \bar{\nu}_e$.
The background levels are higher for the $\bar{\nu}_\mu \to \bar{\nu}_e$
oscillation experiments than the $\nu_\mu \to \nu_e$ case for both
hierarchies.
This is partly because of the higher level of the secondary beam
fluxes in the $\bar{\nu}_\mu$ focusing beam \cite{website},
and partly because the CCQE cross sections are larger for $\nu_\ell$
than the $\bar{\nu}_\ell$ ($\ell = \mu, e$) \cite{Xsec}.

Since the ratio of the $\bar{\nu}_\mu \to \bar{\nu}_e$ to the
${\nu}_\mu \to {\nu}_e$ event numbers is significantly larger for the
inverted hierarchy than that for the normal hierarchy case
in Fig.~\ref{fig:T2K},
one may be tempted to conclude that the neutrino mass hierarchy can be
determined by using both $\nu_\mu$ and $\bar{\nu}_\mu$ focusing beams
at T2K.
This is not the case since the same trend can be expected for 
$\sin \dmns \sim -1$, as can be seen clearly from eq.~(\ref{eq:diff_Ae}).

\subsection{Oki Island}

\begin{figure}[t]
\centering
 \includegraphics[scale=0.6]{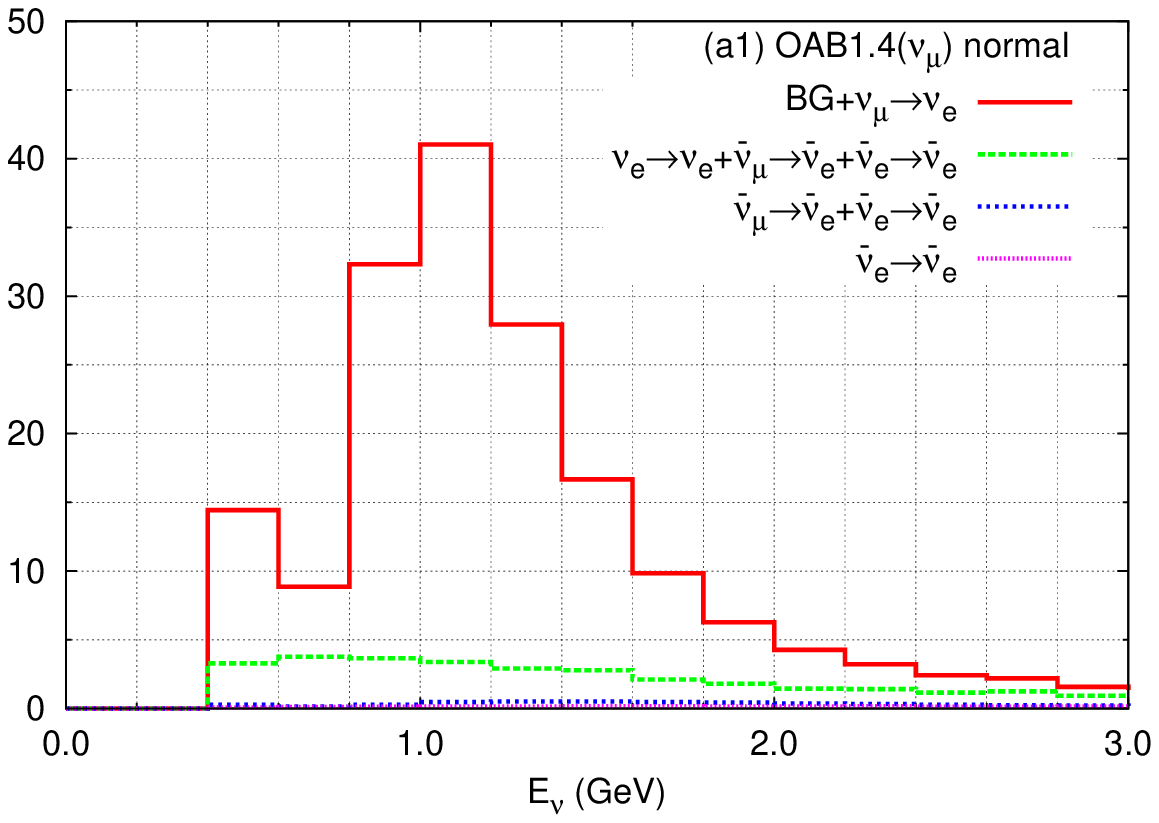}
~~~
 \includegraphics[scale=0.6]{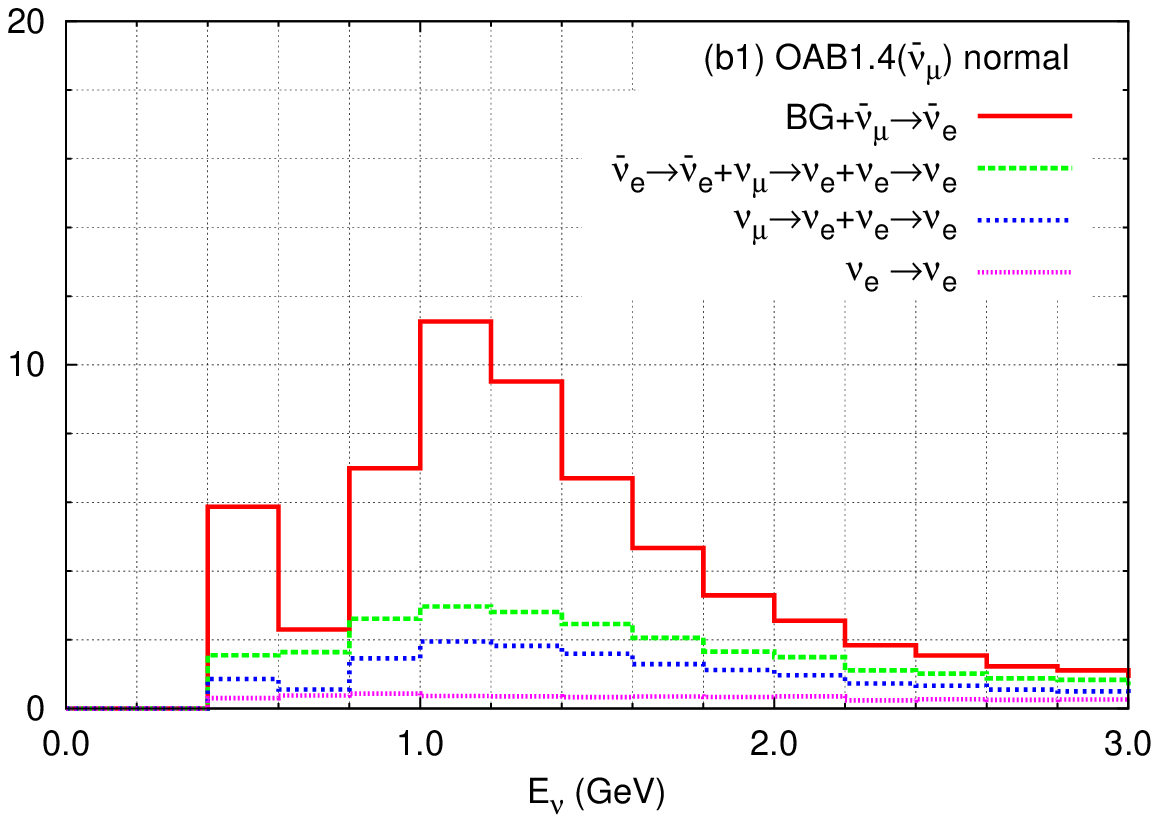}

 \includegraphics[scale=0.6]{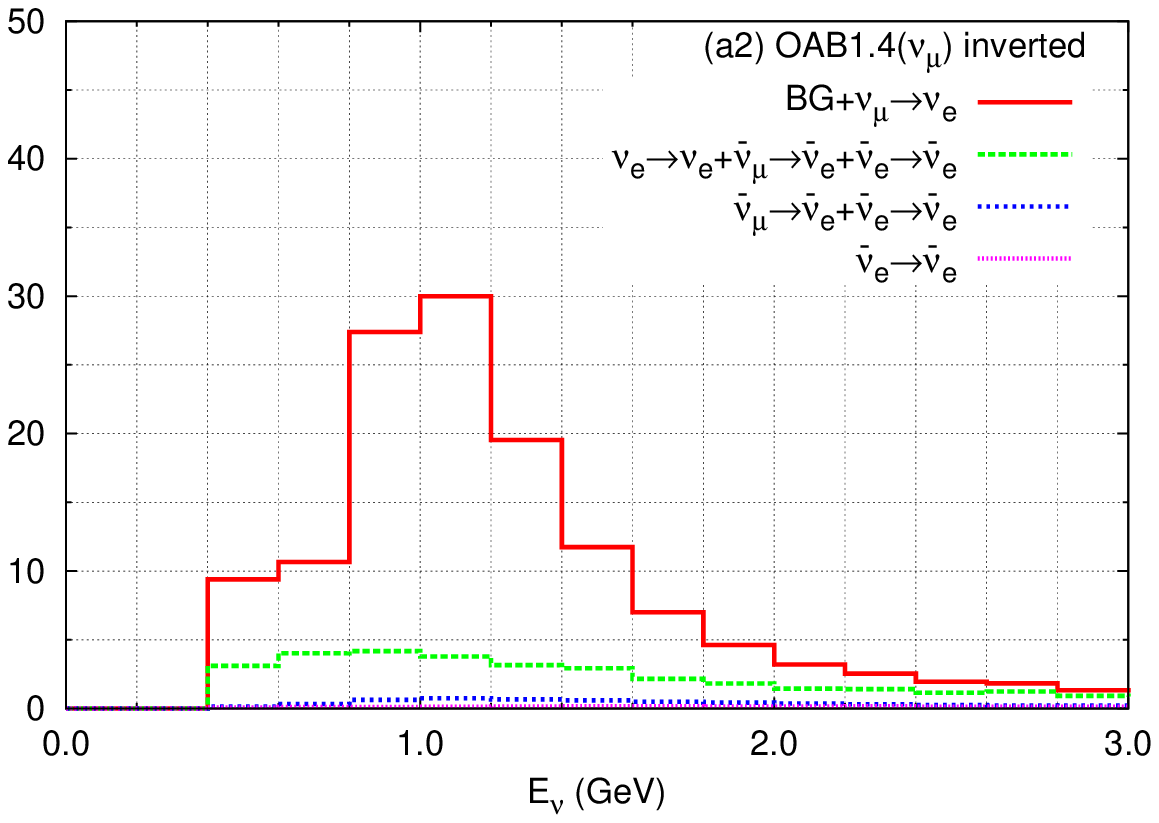}
~~~
 \includegraphics[scale=0.6]{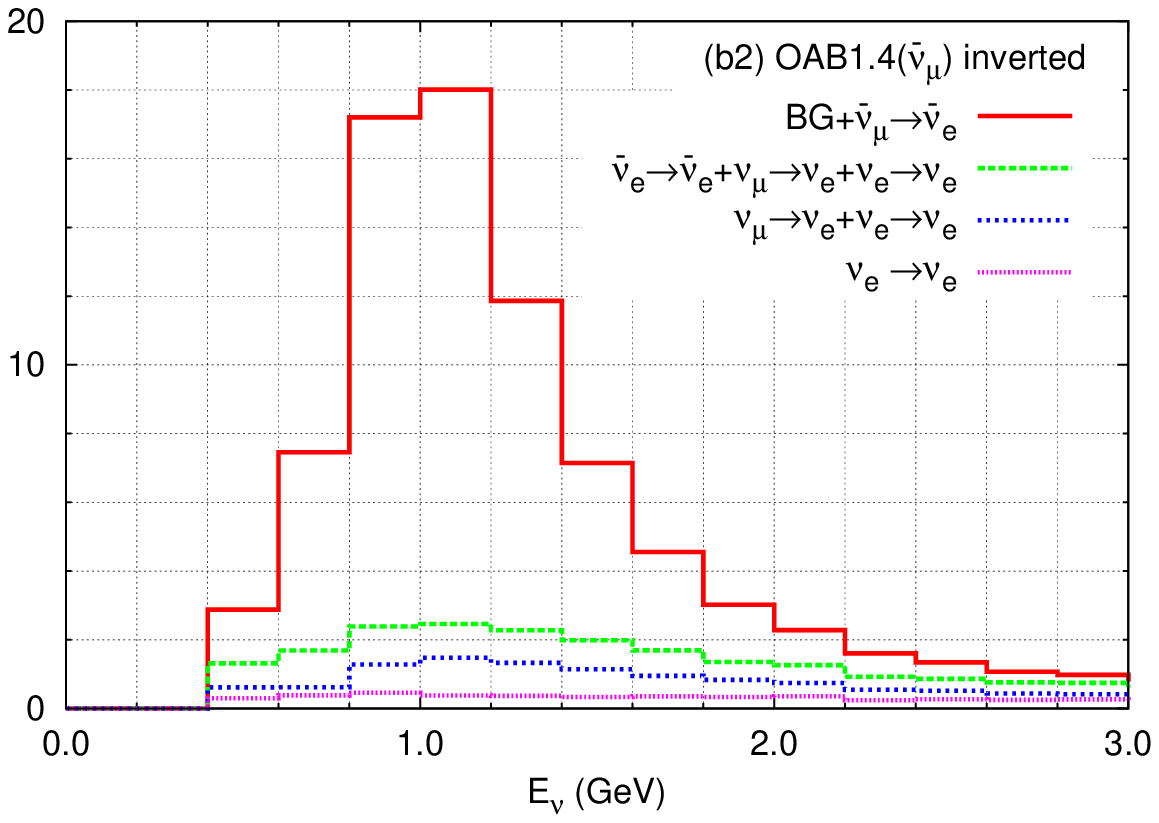}
 \caption{
The same as Fig.~\ref{fig:T2K}, 
but for Tokai-to-Oki Island ($L=653$km),
where the fiducial volume of 100~kton
and the off-axis angle of $1.4^\circ$ is assumed for the
far detector.
The results are for $\srct{2}=0.08$ and $\dmns=0^\circ$,
and $\bar{\rho}_{\sOKI}=2.75$~g/cm$^3$,
see eq.~(\ref{eq:matterOKI}).
} 
\label{fig:T2OKI}
\end{figure}

Figure \ref{fig:T2OKI} shows the typical event numbers 
for the $e$-like CCQE event at Oki Island ($L=653$~km),
where we suppose to place a 100~kton fiducial volume detector.
The results are for $\numt{2.5}{21}$ POT at J-PARC
for both $\nu_\mu$ and $\bar{\nu}_\mu$ focusing beam
at the off-axis angle of $1.4^\circ$, 
which corresponds to the $3.0^\circ$ OAB at SK, as shown in
Fig.~\ref{fig:relation}.
We use eqs.~(\ref{eq:input}) and (\ref{eq:input_rct})
for physics parameters
and eq.~(\ref{eq:matterOKI}) for the average matter density
to generate these event numbers.
The left two panels in Fig.~\ref{fig:T2OKI}, (a1) and (a2),
are for the $\nu_\mu$ focusing beam,
while the right two panels (b1) and (b2) are 
for the $\bar{\nu}_\mu$ focusing beam.
The top two panels, (a1) and (b1), are for the normal mass hierarchy,
and the bottom two panels, (a2) and (b2), are for the
inverted mass hierarchy.
The line types are the same as in Fig.~\ref{fig:T2K},
the red solid lines give the total event numbers,
the green dashed lines are the sum of all the background events
from secondary beams,
the blue short dashed lines give 
the sum of $\bar{\nu}_\mu \to \bar{\nu}_e$ and $\bar{\nu}_e \to \bar{\nu}_e$
for the $\nu_\mu$ focusing beam,
that of ${\nu}_\mu \to {\nu}_e$ and ${\nu}_e \to {\nu}_e$
for the $\bar{\nu}_\mu$ focusing beam.
The purple dotted lines shows the contribution from
$\nu_e \to \nu_e$ ($\bar{\nu}_e \to \bar{\nu}_e$)
for $\nu_\mu$ ($\bar{\nu}_\mu$) focusing beam.
As in the case of SK shown in Fig.~\ref{fig:T2K},
the background from the secondary beam contributions are higher for
the $\bar{\nu}_\mu$ focusing beam than those for the $\nu_\mu$
focusing beam.

In Fig.~\ref{fig:T2OKI},
we find that the first oscillation peak appears at
around 1.0~GeV at Oki Island for all the four cases.
The peaks at $E_\nu \sim 1$~GeV in the event numbers are obtained by
the convolution of the oscillation probability, whose first peak is
located around $E_\nu \sim 1.2$~GeV for the normal or slightly above
1~GeV for the inverted hierarchy, as shown
in Fig.~\ref{fig:profile} (P1)-(P4), and the
$1.4^\circ$ OAB fluxes that have a peak at around $0.8$~GeV in
Fig.~\ref{fig:profile} (a3) and (b3), as shown by green dashed lines.
The difference between the normal and inverted
hierarchy is larger than that of the T2K experiment.
One can observe the second peak in the 400$\sim$600~MeV bin for the normal
hierarchy case, (a1) and (b1), but not for the inverted case.
This is because the matter effect to the oscillation phase,
the term $B^e$ in eqs.~(\ref{eq:Pme}) and (\ref{eq:BeAprox}),
grows with the baseline length L,
and shifts the peaks of the oscillation maximum at 
$|\Delta_{13}+B^e|\sim \pi, 3\pi, \cdots$ in the opposite directions;
toward higher (lower) energies for the normal (inverted) hierarchy.
More accurately speaking, the above phase shift pattern applies for
$\cos \dmns \sim 1$, as can be read off from eq.~(\ref{eq:BeAprox}),
where the matter effect diminishes (enhances) the shift for the normal
(inverted) hierarchy.
The pattern reverses for $\cos \dmns \sim -1$.
Likewise, the matter effects on the oscillation 
amplitudes are also clearly seen:
we expect more (less) $\nu_\mu \to \nu_e$ events than
$\bar{\nu}_\mu \to \bar{\nu}_e$ events around the first oscillation
peak when the hierarchy is normal (inverted)
as can be seen from eq.~(\ref{eq:AeAprox}).
This pattern is enhanced when $\sin \dmns \sim -1$
whereas it is diminished when $\sin \dmns \sim 1$.

\subsection{Korea}

\begin{figure}[t]
\centering
 \includegraphics[scale=0.6]{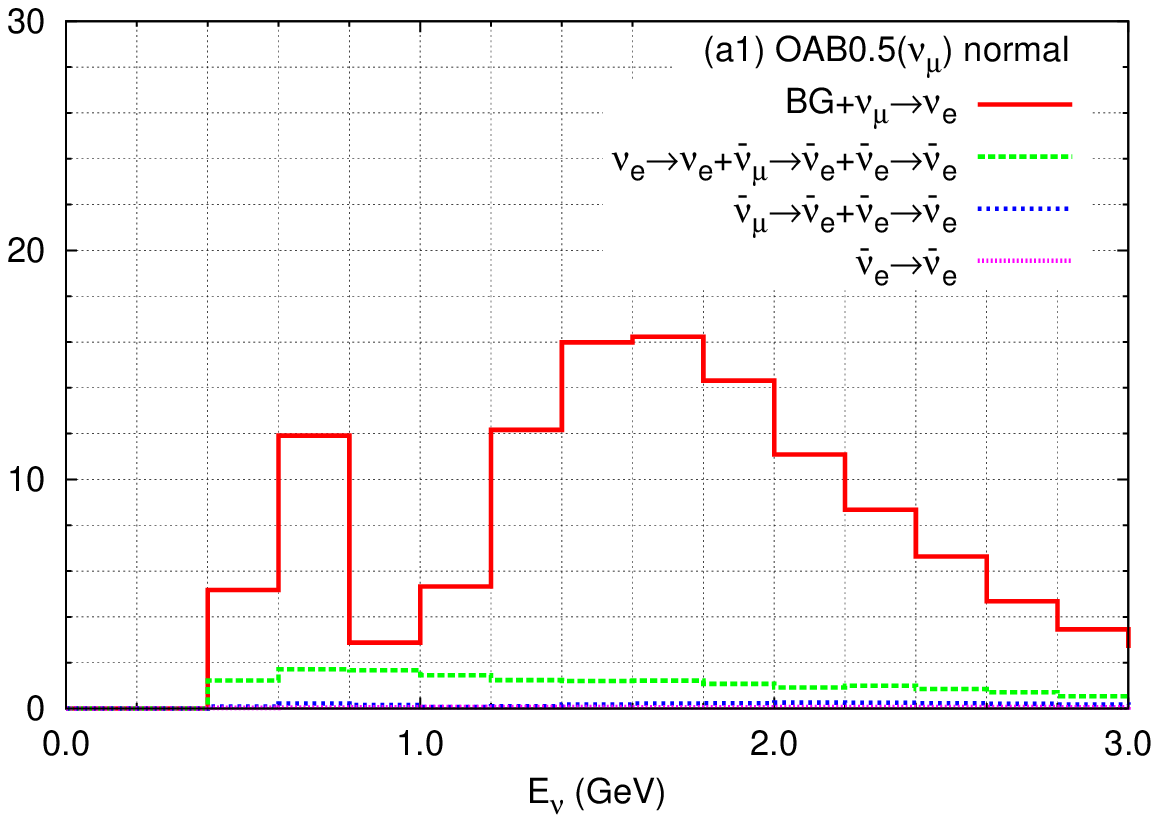}
~~~
 \includegraphics[scale=0.6]{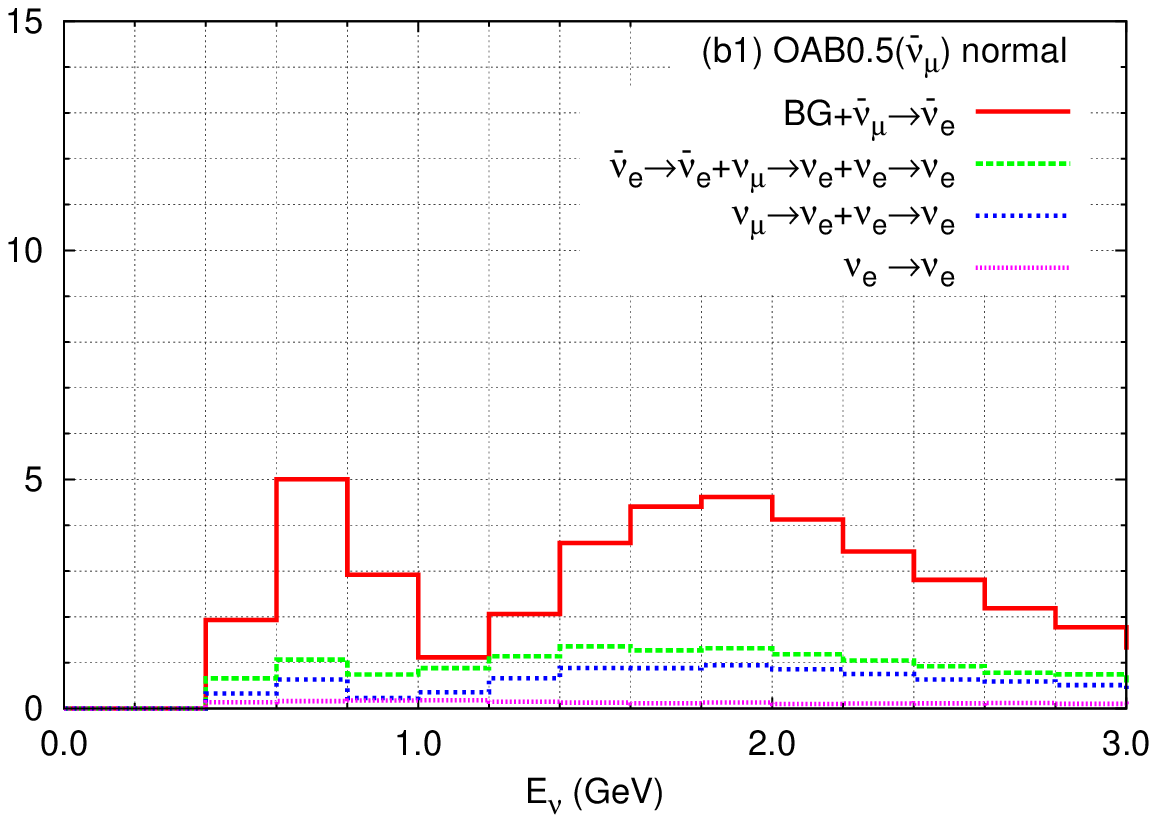}

 \includegraphics[scale=0.6]{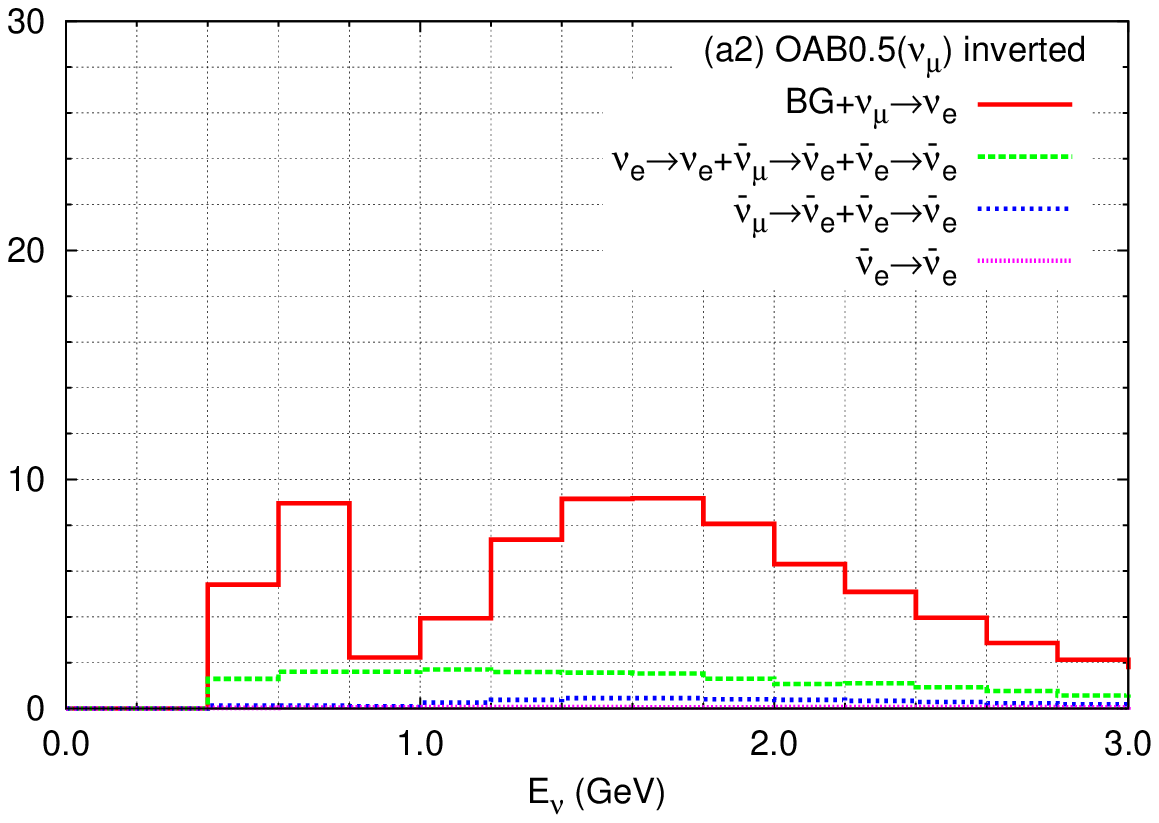}
~~~
 \includegraphics[scale=0.6]{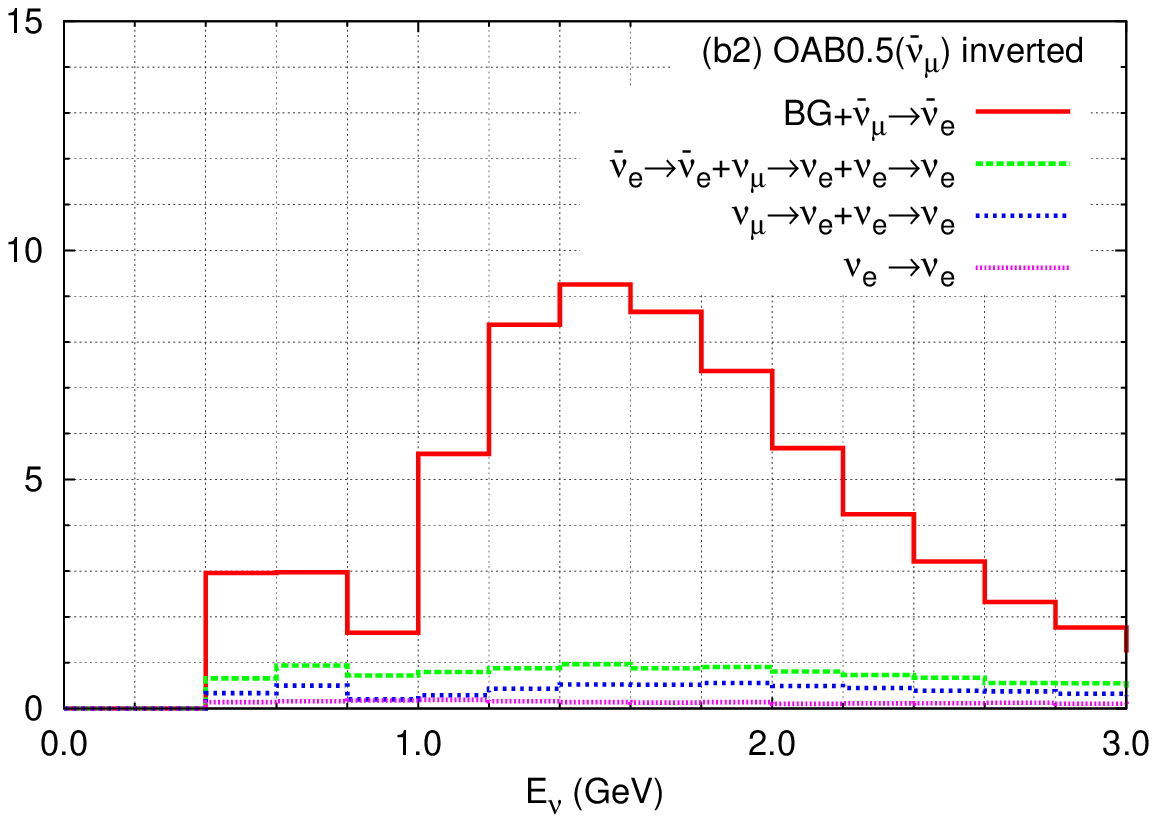}
 \caption{
The same as Fig.~\ref{fig:T2K}, 
but for Korea ($L=1000$km),
where the fiducial volume of 100~kton
and the off-axis angle of $0.5^\circ$ is assumed for the
far detector.
The results are for $\srct{2}=0.08$ and $\dmns=0^\circ$,
and $\bar{\rho}_{\sOKI}=2.9$~g/cm$^3$,
see eq.~(\ref{eq:matterKr}).
} 
\label{fig:T2KK}
\end{figure}

In Figs.~\ref{fig:T2KK},
we show the typical event numbers for $e$-like CCQE event at
a far detector in Korea ($L=1000$~km) with
a 100~kton fiducial volume detector.
The off-axis angle there is chosen to be $0.5^\circ$, 
which optimized the T2KK performance with the
$3.0^\circ$ OAB at SK \cite{HOS1,HOS2,T2KKbg}.
The results are for $\srct{2}=0.8$ and $\dmns=0^\circ$ and the
parameters of eqs.~(\ref{eq:input}) and (\ref{eq:input_rct}),
with the average matter density of 2.9 g/cm$^3$ eq.~(\ref{eq:matterKr}).
The panels and the lines types are the same as those
in Figs.~\ref{fig:T2K} and \ref{fig:T2OKI}.

The first peak of the event appears at around $1.6$~GeV 
for the $\nu_\mu$ beam for both hierarchies,
but that is around 1.8 (1.5)~GeV for the $\bar{\nu}_\mu$ beam
for the normal (inverted) hierarchy.
This is because of the matter effect contribution
to the oscillation phase,
$B^e$ in eq.~(\ref{eq:BeAprox}),
and $\bar{B}^e$ obtained from eq.~(\ref{eq:BeAprox})
by reversing the sign of $\bar{\rho}$.
The magnitude of $B^e$ is small for $\cos \dmns = 1$ in
eq.~(\ref{eq:BeAprox}),
whereas that of $\bar{B}^e$ is enhanced.
As expected, the sign of the difference $B^e - \bar{B}^e$ in
eq.~(\ref{eq:diff_Be})
reflects the neutrino mass hierarchy.

Likewise, the difference in the heights of the first oscillation peak
is more distinct in Fig.~\ref{fig:T2KK} for $L=1000$~km then
that in Fig.~\ref{fig:T2OKI} for $L=653$~km.
As expected, the oscillation amplitude for the $\nu_\mu \to \nu_e$
transition is bigger (smaller) than that for the
$\bar{\nu}_\mu \to \bar{\nu}_e$ transition when the hierarchy is
normal (inverted).
Although the magnitudes of those enhancement or suppression factor
depends on $\sin \dmns$,
as can be seen from eqs.~(\ref{eq:AeAprox}) and (\ref{eq:diff_Ae}),
the difference between different baseline lengths among 
Figs.~\ref{fig:T2K}, \ref{fig:T2OKI}, and \ref{fig:T2KK}
depend solely on the mass hierarchy pattern,
as expressed in eq.~(\ref{eq:Ae_diffs}).

\section{Analysis method}
\label{sec:chi}

We introduce a $\Delta \chi^2$ function
\begin{equation}
 \Delta \chi^2 \equiv 
 \chi^2_{\rm stat} + \chi^2_{\rm sys} + \chi^2_{\rm para}\,, 
\label{eq:def_chi1}
\end{equation}
in order to compare the physics potential of T2K, T2KO, and T2KK
experiments quantitatively on the same footing.
The first term of eq.~(\ref{eq:def_chi1})
gives statistical constraints on the model parameters from the number
of the CCQE events in each bin at each detector:
\begin{eqnarray}
 \chi^2_{\rm stat} &=& \sum_{\rm D} \sum_{n}
\left\{
\left(
\dfrac{\left(N_{\mu,{\rm D}}^n\right)^{\rm input} 
- \left(N_{\mu,{\rm D}}^n\right)^{\rm fit}}
{\sqrt{\left(N_{\mu,{\rm D}}^n\right)^{\rm input}}}
\right)^2
+
\left(
\dfrac{\left(N_{e,{\rm D}}^n\right)^{\rm input} 
- \left(N_{e,{\rm D}}^n\right)^{\rm fit}}
{\sqrt{\left(N_{e,{\rm D}}^n\right)^{\rm input}}}
\right)^2
\right.\nn\\
&&\hspace*{7ex}
+ 
\left.
\left(
\dfrac{\left(\overline{N}_{\mu,{\rm D}}^n\right)^{\rm input} 
- \left(\overline{N}_{\mu,{\rm D}}^n\right)^{\rm fit}}
{\sqrt{\left(\overline{N}_{\mu,{\rm D}}^n\right)^{\rm input}}}
\right)^2
+
\left(
\dfrac{\left(\overline{N}_{e,{\rm D}}^n\right)^{\rm input} 
- \left(\overline{N}_{e,{\rm D}}^n\right)^{\rm fit}}
{\sqrt{\left(\overline{N}_{e,{\rm D}}^n\right)^{\rm input}}}
\right)^2
\right\},
\label{eq:chi_stat}
\end{eqnarray}
where $N_{\mu,{\rm D}}^n$ and $N_{e,{\rm D}}^n$ denotes
the number of $\mu$- and $e$-like events, respectively,
for the $\nu_\mu$ focusing beam
calculated with eq.~(\ref{eq:eventNU}) in the $n$-th bin 
at each detector, D=SK, Oki, Kr,
whereas
$\overline{N}_{\mu,{\rm D}}^n$ and $\overline{N}_{e,{\rm D}}^n$
are for the $\bar{\nu}_\mu$ focusing beam.
Their square-roots give statistical errors.
The summation is over all bins from 0.4~GeV to 5.0~GeV for
$\mu$-like events at all sites,
and for $e$-like events
from 0.4~GeV to 1.2~GeV at SK,
from 0.4~GeV to 2.4~GeV at Oki,
from 0.4~GeV to 2.8~GeV at a far detector in Korea, respectively.
The input event numbers, $(N_{\alpha,{\rm D}}^n)^{\rm input}$,
are generated with eq.~(\ref{eq:input}) for both hierarchies,
and for various value of $\srct{2}$ and $\dmns$.
We use the average matter density 
along each baseline of eq.~(\ref{eq:matter})
when we calculate the input event numbers.

The event numbers in the fit, $(N_{\alpha,{\rm D}}^n)^{\rm fit}$
and $(\overline{N}_{\alpha,{\rm D}}^n)^{\rm fit}$,
are obtained by varying all the 6 parameters of
the three-neutrino model freely
and also by allowing for the systematic errors.
We consider the following systematic errors in this analysis.
We assign $6\%$ uncertainty to the overall matter density along
each baseline,
\begin{equation}
\bar{\rho}_{\rm D}^{\rm fit} = f_\rho^{\rm D} \bar{\rho}_{\rm D}\,,
\hspace*{3ex}
f_\rho^{\rm D} = 1.00 \pm 0.06\,,
\hspace*{3ex}
({\mbox{for D=SK, Oki, Kr}}).
\label{eq:rhofit}
\end{equation}
Although we expect positive correlation among $f_\rho^{D}$,
we allow them to vary independently as a conservative estimate.
We assign $3\%$ uncertainty in the flux normalization of
$\nu_\mu$ and $\bar{\nu}_\mu$ focusing beam as
\begin{subequations}
\begin{eqnarray}
&& \left(\Phi_{\nu_\alpha}^{\rm D}\right)^{\rm fit} =
  f_{\nu_\alpha}^{\rm D} \Phi_{\nu_\alpha}^{\rm D}\,,
 \hspace*{2ex}
 f_{\nu_\alpha}^{\rm D} = 1.00 \pm 0.03\,,
 \hspace*{2ex}
 ({\mbox{for }} \nu_\alpha=\nu_{\mu},\nu_e,\bar{\nu}_\mu,\nu_e)\,,
\label{eq:fluxfitN}
\\
&& \left(\bar{\Phi}_{\nu_\alpha}^{\rm D}\right)^{\rm fit} =
  \bar{f}_{\nu_\alpha}^{\rm D} \bar{\Phi}_{\nu_\alpha}^{\rm D}\,,
 \hspace*{2ex}
 \bar{f}_{\nu_\alpha}^{\rm D} = 1.00 \pm 0.03\,,
 \hspace*{2ex}
 ({\mbox{for }} \nu_\alpha=\nu_{\mu},\nu_e,\bar{\nu}_\mu,\nu_e)\,,
\label{eq:fluxfitA}
\end{eqnarray}
\label{eq:fluxfit}
\end{subequations}
$\!\!\!$
respectively, 
where D=SK, Oki, Kr.
We also ignore possible correlations among the flux errors.
For the CCQE cross sections of neutrino and anti-neutrinos,
we assume common $3\%$ error as
\begin{subequations}
 \begin{eqnarray}
\left(\sigma_{\nu_{\mu,e}}^{\rm CCQE}(E)\right)^{\rm fit}
=f_{\ell}  
\left(\sigma_{\nu_{\mu,e}}^{\rm CCQE}(E)\right)\,,
&\hspace*{1ex}&
f_{\ell} = 1.00 \pm 0.03\,,  
\label{eq:nuXsecfit}\\
\left(\sigma_{\bar{\nu}_{\mu,e}}^{\rm CCQE}(E)\right)^{\rm fit}
=f_{\bar{\ell}}  
\left(\sigma_{\bar{\nu}_{\mu,e}}^{\rm CCQE}(E)\right)\,,
&\hspace*{1ex}&
f_{\bar{\ell}} = 1.00 \pm 0.03\,,  
\label{eq:nubarXsecfit}
 \end{eqnarray}
\label{eq:Xsecfit}
\end{subequations}
$\!\!\!\!$
for neutrino and anti-neutrino events, independently,
but we take $f_{\nu_\mu}=f_{\nu_e}=f_\ell$ and
$f_{\bar{\nu}_\mu}=f_{\bar{\nu}_e}=f_{\bar{\ell}}$
because of the $e$-$\mu$ universality.
Here also, we neglect the correlation between $\nu_{\ell}$ and
$\bar{\nu}_{\ell}$ cross section errors.
The systematic error for the fiducial volume of each far detector
is also assigned as
\begin{equation}
 M_{\rm D}^{\rm fit} = f_{\rm D} M_{\rm D}\,,
 \hspace*{3ex}
 f_{\rm D} = 1.00 \pm 0.03~~~
 (\mbox{for } {\rm D}={\rm SK}, {\rm Oki}, {\rm Kr})\,.
\label{eq:Volfit}
\end{equation}
Summing up,
we take account of 32 systematic uncertainties in terms of which
$\chi^2_{\rm sys}$ in eq.~(\ref{eq:def_chi1}) is expressed as
\begin{eqnarray}
 \chi^2_{\rm sys}&=&
\sum_{\rm D}
\left[
\left(\dfrac{1-f_{\rm D}}{0.03}\right)^2
+
\left(\dfrac{1-f_{\rho}^{\rm D}}{0.06}\right)^2
+
\sum_{\nu_\alpha}
\left\{
\left(\dfrac{1-f_{\nu_\alpha}^{\rm D}}{0.03}\right)^2
+
\left(\dfrac{1-\bar{f}_{\nu_\alpha}^{\rm D}}{0.03}\right)^2
\right\}
\right]\nn\\
&&+
\sum_{\beta=\ell,\bar{\ell}}
\left(\dfrac{1-f_\beta}{0.03}\right)^2\,.
\label{eq:def_chi_sys}
\end{eqnarray}
There are 32 normalization factors since
for each detector ``D'' the fiducial volume $(f_{\rm D})$,
the average matter density $(f_{\rho}^{\rm D})$,
and the 4 fluxes each for $\nu_\mu$ $(f_{\nu_\alpha}^{\rm D})$ and
$\bar{\nu}_\mu$ $(\bar{f}_{\nu_\alpha}^{\rm D})$ focusing beam
are accounted for,
in addition to the overall theoretical uncertainties for 
$\nu_\ell$ $(f_\ell)$ and $\bar{\nu}_\ell$ $(f_{\bar{\ell}})$
CCQE cross sections.

The last term of eq.~(\ref{eq:def_chi1}), $\chi^2_{\rm para}$,
accounts for the external constraints on the model parameters:
\begin{eqnarray}
 \chi^2_{\rm para}
&=&
\left(
\dfrac{\numt{7.5}{-5}\mbox{eV}^2-(\dm12)^{\rm fit}}{\numt{0.2}{-5}}
\right)^2
+
\left(
\dfrac{0.852-\ssun{2}^{\rm fit}}{0.025}
\right)\nn\\
&&+
\left(
\dfrac{\srct{2}^{\rm input}-\srct{2}^{\rm fit}}{0.01}
\right)^2\,.
\end{eqnarray}
The first two terms are from the present constraints from the KamLAND
experiment \cite{KamLAND} listed in eq.~(\ref{eq:sun_data}).
In the last term, we assume that the new reactor experiments
\cite{DoubleCHOOZ,DayaBay,RENO}
will measure $\srct{2}$ with an uncertainty of 0.01
in the near future.
We do not impose the present constraints on $|\dm13|$ and $\satm{2}$
given in eq.~(\ref{eq:atm_data}),
since the experiments studied in this report will measure them more
accurately.

\section{mass hierarchy}
\label{sec:mass}

In this section,
we show physics capability of the T2KO experiment
to determine the mass hierarchy
and compare it with that of the T2KK \cite{HOS1}-\cite{HOSF}
and the \T2Kn experiment.
Here by \T2Kn, we examine the option where the
additional 100~kton detector is placed in the Kamioka site
to make the total fiducial volume 122~kton at $L=295$~km,
which may be regarded as a small scale version of 
Hyper-Kamiokande \cite{HK}.

\begin{figure}[t]
\centering
 \includegraphics[scale=1.0]{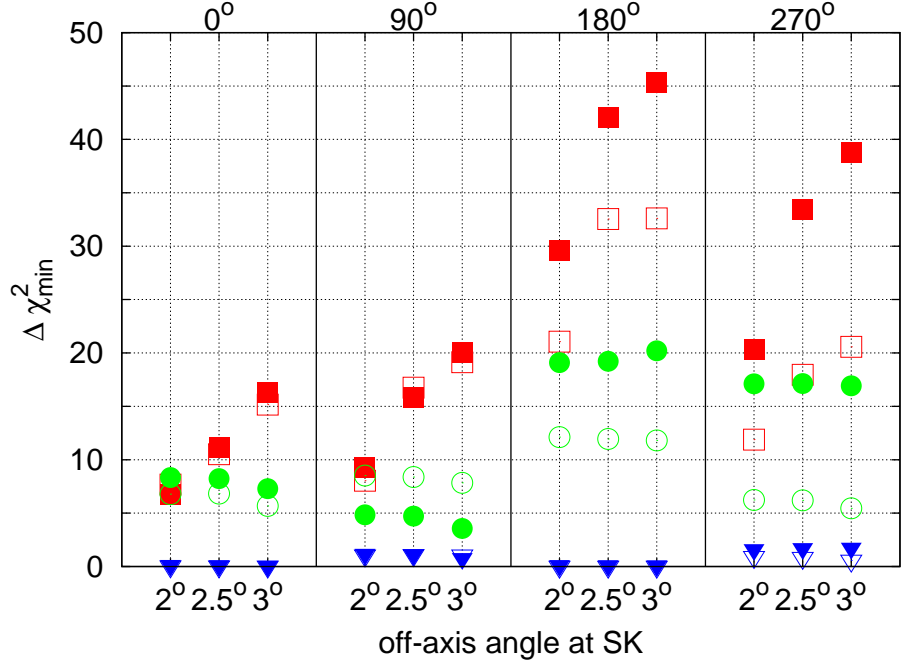}
 \caption{
Minimum $\Delta \chi^2$ of the T2KK , T2KO, and \T2Kn experiment
to exclude the wrong mass hierarchy
when only the $\nu_\mu$ focusing beam of $\numt{5.0}{21}$ POT
is used.
Four columns give results for
$\dmns=0^\circ$, $90^\circ$, $180^\circ$, and $270^\circ$,
respectively,
and the horizontal axis in each column
gives the off-axis angle at SK.
The solid (open) square, circle, and triangle denotes,
respectively,
the T2KK, T2KO and \T2Kn results for the normal 
(inverted) hierarchy.
The results are for $\srct{2}=0.08$
and the other input model parameters in
eq.~(\ref{eq:input}),
as well as the average matter density of eq.~(\ref{eq:matter})
along the three baselines.
} 
\label{fig:mass1}
\end{figure}

In Fig.~\ref{fig:mass1},
we show the minimum $\Delta \chi^2$ of T2KK, T2KO, and \T2Kn
experiment for the $\nu_\mu$ focusing beam with $\numt{5.0}{21}$ POT.
Four columns of Fig.~\ref{fig:mass1} give results for
$\dmns=0^\circ$, $90^\circ$, $180^\circ$, and $270^\circ$,
respectively,
and the horizontal axis in each column gives the off-axis angle at
the SK.
The solid (open) square, circle, and triangle denotes,
respectively,
the sensitivity of the T2KK, T2KO, and \T2Kn experiments
for the normal (inverted) hierarchy.
The results are for $\srct{2}=0.08$ and the other input model
parameters in eq.~(\ref{eq:input}),
as well as the average matter density of eq.~(\ref{eq:matter})
along the three baselines.

From Fig.~\ref{fig:mass1},
we can tell that
the physics potential for the mass hierarchy determination
of the T2KK experiment is far better than the other experiments
when the combination of $3.0^\circ$ OAB at SK
and $0.5^\circ$ OAB at a far detector in Korea
is taken, where the mass hierarchy can be
determined by more than 4$\sigma$ level for all the 8 cases 
(4 values of $\dmns{}$ and both hierarchies).
The sensitivity of the T2KK on the mass hierarchy reduces
significantly
as the off-axis angle at SK is reduced.
This is because OAB with small off-axis angle cannot reach Korea when
off-axis angle at SK is below $\sim 2.5^\circ$ \cite{HOS1,HOS2};
see Fig.~\ref{fig:relation}.

For the T2KO experiment,
the mass hierarchy can be determined at the 2$\sigma$ level or higher.
This capability does not depend strongly on the off-axis angle at SK,
because the beam intensity around the first peak 
($E_\nu \sim 1$GeV) at Oki Island does not change much
with the OAB at SK, see Fig.~\ref{fig:profile}.

Furthermore, it is clearly shown in Fig.~\ref{fig:mass1} that
the \T2Kn experiment does not have any sensitivity to the 
neutrino mass hierarchy pattern
for any combinations of $\dmns$ and the mass hierarchy.
This is essentially because the small differences in the 
oscillation probabilities between the normal and the inverted
hierarchy can easily be compensated by small shifts in the model parameters,
such as $|\dm13|$, $\satm{}$, and $\dmns$.

\begin{figure}[t]
\centering
 \includegraphics[scale=1.0]{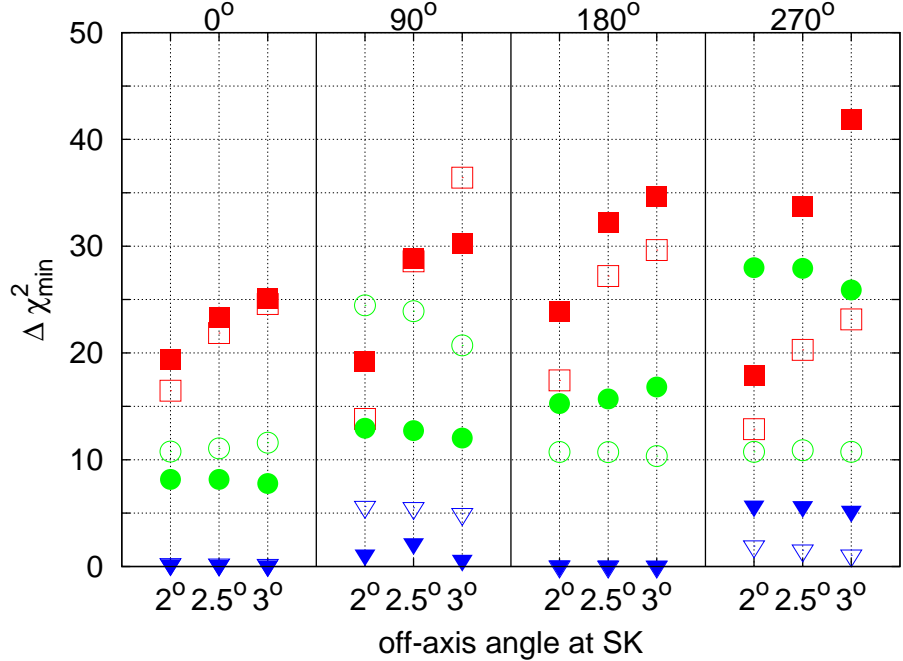}
 \caption{
The same as Fig.~\ref{fig:mass1},
but with both $\nu_\mu$ and $\bar{\nu}_\mu$ focusing beams
each with $\numt{2.5}{21}$ POT.
} 
\label{fig:mass2}
\end{figure}

In Fig.~\ref{fig:mass2}, we show the
minimum $\Delta \chi^2$ of the T2KK, T2KO, and \T2Kn experiment
to exclude the wrong mass hierarchy when
both $\nu_\mu$ and $\bar{\nu}_\mu$ focusing beams are used,
each with $\numt{2.5}{21}$ POT
to keep the total exposure the same.

As for the T2KK experiment, shown by the red squares,
the improvement is significant especially when
$\dmns\simeq 0^\circ$ and $90^\circ$,
making $\Delta \chi^2_{\rm min}$
greater than 20 for all the 8 combinations of
$\dmns$ and sgn$(\dm13)$,
not only for the $3.0^\circ$ OAB but also for the
$2.5^\circ$ OAB at SK \cite{HOSF}.
Likewise for the T2KO experiment, shown by green circles,
the improvement is most significant at $\dmns=90^\circ$
where the smallest $\Delta \chi^2_{\rm min}$ grows from
$\sim 4$ in Fig.~\ref{fig:mass1} to $\sim 12$ in Fig.~\ref{fig:mass2}.
This is essentially because the $\sin \dmns$ term in the oscillation
amplitude shift $A^e$ in eq.~(\ref{eq:AeAprox}) changes sign for the
$\bar{\nu}_\mu \to \bar{\nu}_e$ oscillation as shown in
eq.~(\ref{eq:diff_Ae}) for the difference $A^e - \bar{A}^e$.
This helps resolving the entanglement between the $\sin \dmns$
dependence ad the matter effect in the oscillation amplitudes.
Significant improvements are also found for $\dmns=270^\circ$
by the same reason.
Moderate improvements are found for the $\dmns=0^\circ$ case,
allowing the T2KO experiment with half-and-half $\nu_\mu$ and
$\bar{\nu}_\mu$ beams to resolve the mass hierarchy at $3\sigma$
level for the worst case
($\dmns=0$ and normal hierarchy).
No significant dependence on the OAB at SK is found.

The capability of determining the mass hierarchy pattern
by the \T2Kn experiment 
does not appear even by using both $\nu_\mu$ and $\bar{\nu}_\mu$
beams.
The value of $\Delta \chi^2_{\rm min}$ stays almost zero,
except for just two special combinations;
$\dmns=270^\circ$ for the normal hierarchy
and
$\dmns=90^\circ$ for the inverted hierarchy.
These are the two cases, where the difference between the
$\nu_\mu \to \nu_e$ and $\bar{\nu}_\mu \to \bar{\nu}_e$
oscillation amplitudes,
$A^e-\bar{A}^e$ in eq.~(\ref{eq:diff_Ae}) is largest or smallest,
respectively,
such that variation of the other model parameters cannot account for
the difference if the wrong mass hierarchy is assumed.
With the same token, $\Delta \chi^2_{\rm min}$ exceeds 20 for
T2KO or 30 for T2KK (with $\gsim 2.5^\circ$ OAB at SK)
for these two particular combinations.

\begin{figure}[t]
\centering
 \includegraphics[scale=0.70]{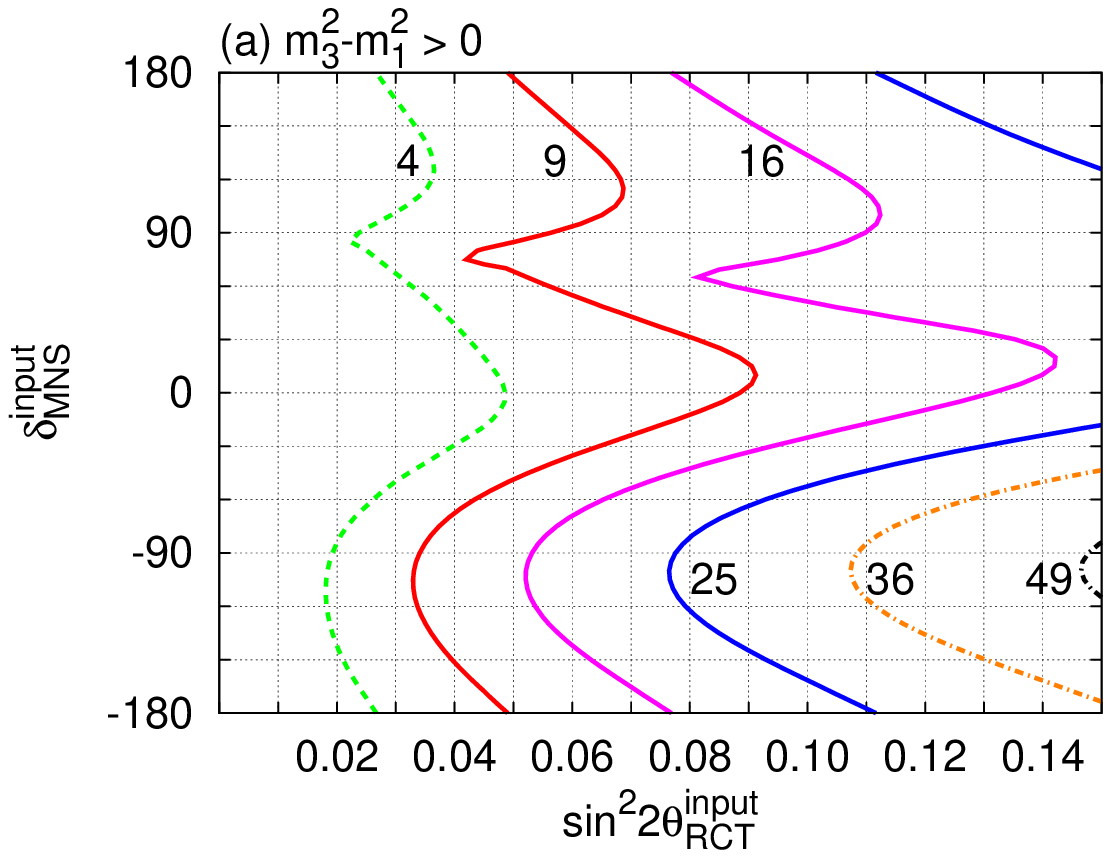}
~~
 \includegraphics[scale=0.70]{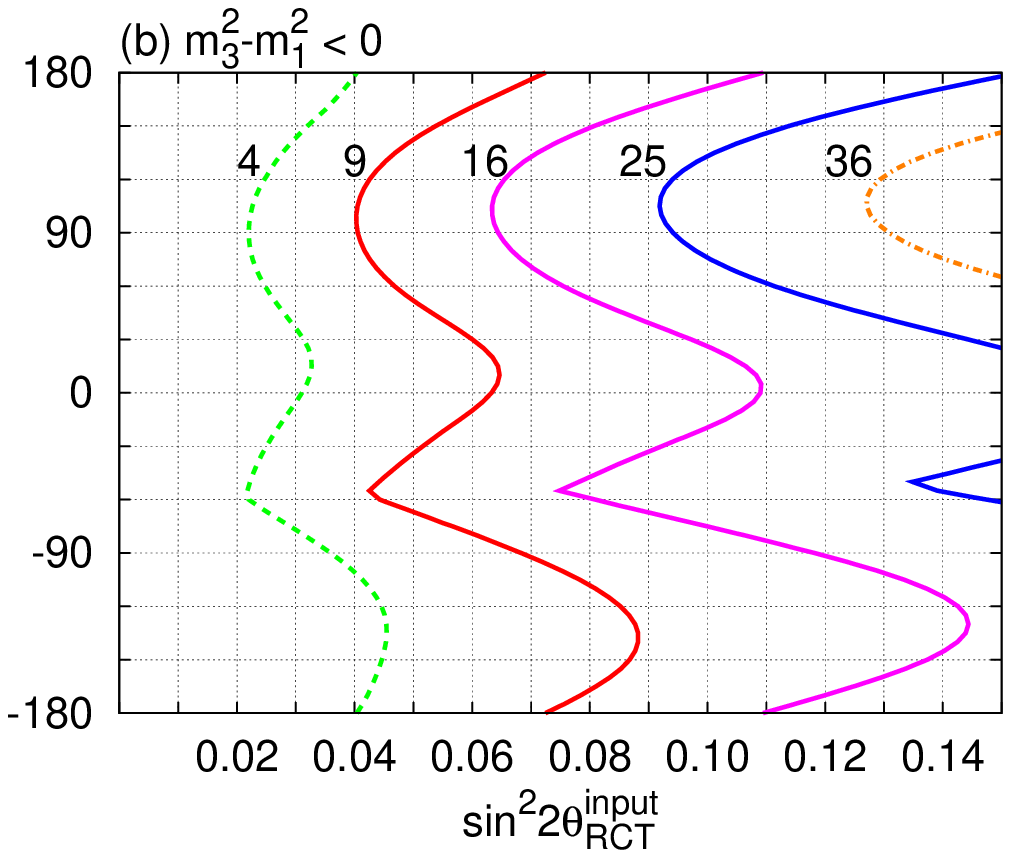}
 \caption{
The $\Delta \chi^2_{\rm min}$ contour plot 
for the T2KO experiment 
to exclude the wrong mass hierarchy in the plane of
$\srct{2}$ and $\dmns$.
The left figure is for the normal hierarchy
and the right one is for the inverted hierarchy.
The OAB combination for both figures is
$3.0^\circ$ at SK and $1.4^\circ$ at Oki Island
with $\numt{2.5}{21}$ POT 
for both $\nu_\mu$ and $\bar{\nu}_\mu$ focusing beams.
Contours for $\Delta \chi^2_{\rm min}=4$, 9, 16, 25, 36,
49 are shown.
All the input parameters other than $\srct{2}$ and $\dmns$
are shown in eqs.~(\ref{eq:matter}) and (\ref{eq:input}).
} 
\label{fig:cnt-T2OKI}
\end{figure}

So far, we have shown results for $\srct{2}=0.8$ and
4 representative values of $\dmns$;
$0^\circ$, $180^\circ$, and $\pm 90^\circ$.
In Fig.~\ref{fig:cnt-T2OKI},
we show the contour plot of the $\Delta \chi^2_{\rm min}$ 
for the T2KO experiment to exclude the 
wrong mass hierarchy in the whole plane of $\srct{2}$ and $\dmns$:
The left figure (a) is for the normal hierarchy,
whereas the right one (b) is for the inverted mass hierarchy.
Since no strong dependence of $\Delta \chi^2_{\rm min}$ on the OAB
angle
is found for T2KO potential in Fig.~\ref{fig:cnt-T2OKI},
we choose $3.0^\circ$ OAB at SK, which gives $1.4^\circ$ OAB
at Oki, as shown in Fig.~\ref{fig:relation}.
The results are for $\numt{2.5}{21}$ POT each for
$\nu_\mu$ and $\bar{\nu}_\mu$ focusing beam.
Contours in each figure are for
$\Delta \chi^2_{\rm min}=4$, 9, 16, 25, 36, and 49.
The input parameters other than $\srct{2}$ and $\dmns$
are shown in eqs.~(\ref{eq:matter}) and (\ref{eq:input}),
exactly the same as those adopted in Fig.~\ref{fig:mass2}.
Accordingly the $\Delta \chi^2_{\rm min}$ values at
$\srct{2}=0.08$ agree exactly with those presented in
Fig.~\ref{fig:mass2}
for T2KO with $3.0^\circ$ OAB at SK, for the 4
representative $\dmns$ values.

It is clearly seen from Fig.~\ref{fig:cnt-T2OKI} that 
the mass hierarchy pattern can be distinguished at $3\sigma$
if $\srct{2}^{\rm input}\gsim 0.09$ for any value of
$\dmns^{\rm input}$ and for both hierarchies.
The most difficult case is found for $\dmns \sim 0^\circ$
for the normal hierarchy,
while $\dmns \sim -135^\circ$ for the inverted hierarchy.
On the other hand, the discrimination is easiest at
$\dmns \sim -90^\circ$ for the normal and 
at $\dmns \sim 90^\circ$ for the inverted hierarchy,
in accordance with the argument presented above
for Fig.~\ref{fig:mass2}.

In addition, the contour plots Fig.~\ref{fig:cnt-T2OKI} identify
another case at $\dmns \simeq 60^\circ$ ($\simeq -60^\circ$)
for the normal (inverted) hierarchy,
where the difference between the right and the wrong hierarchy is
large,
giving high $\Delta \chi^2_{\rm min}$.
The spikes of the contours around
these $\dmns$ values appear as a consequence of the conspiracy
between the mass hierarchy dependences in
the oscillation amplitude $A^e$ and
in the phase shift term $B^e$.
When $\dmns^{\rm input}\simeq 60^\circ$,
the $\nu_\mu \to \nu_e$ oscillation amplitude shift $A^e$ cancels
between the matter effect term and the $\sin \dmns$ term for
the normal (inverted) hierarchy at around the Tokai-to-Oki baseline;
see eq.~(\ref{eq:AeAprox}) at $L=653$~km.
The cancellation is not significant at $L=295$~km for T2K,
and the two effects add up constructively for
$\bar{\nu}_\mu \to \bar{\nu}_e$ oscillation.
When the wrong hierarchy is assumed the best fit is found for
$\sin \dmns^{\rm fit}\sim 0$, for which there are
two solutions, $\cos \dmns^{\rm fit} \sim 1$ and
$\cos \dmns^{\rm fit} \sim -1$,
which have significantly different phase-shift;
see eq.~(\ref{eq:BeAprox}).
We find that the spike around $\dmns \sim \pm 60^\circ$
in Fig.~\ref{fig:cnt-T2OKI} occurs
when the $\Delta \chi^2_{\rm min}$ solution of the wrong hierarchy
model jumps from $\cos \dmns^{\rm fit} \sim 1$
to $\cos \dmns^{\rm fit} \sim -1$.

\begin{figure}[t]
\centering
 \includegraphics[scale=0.7]{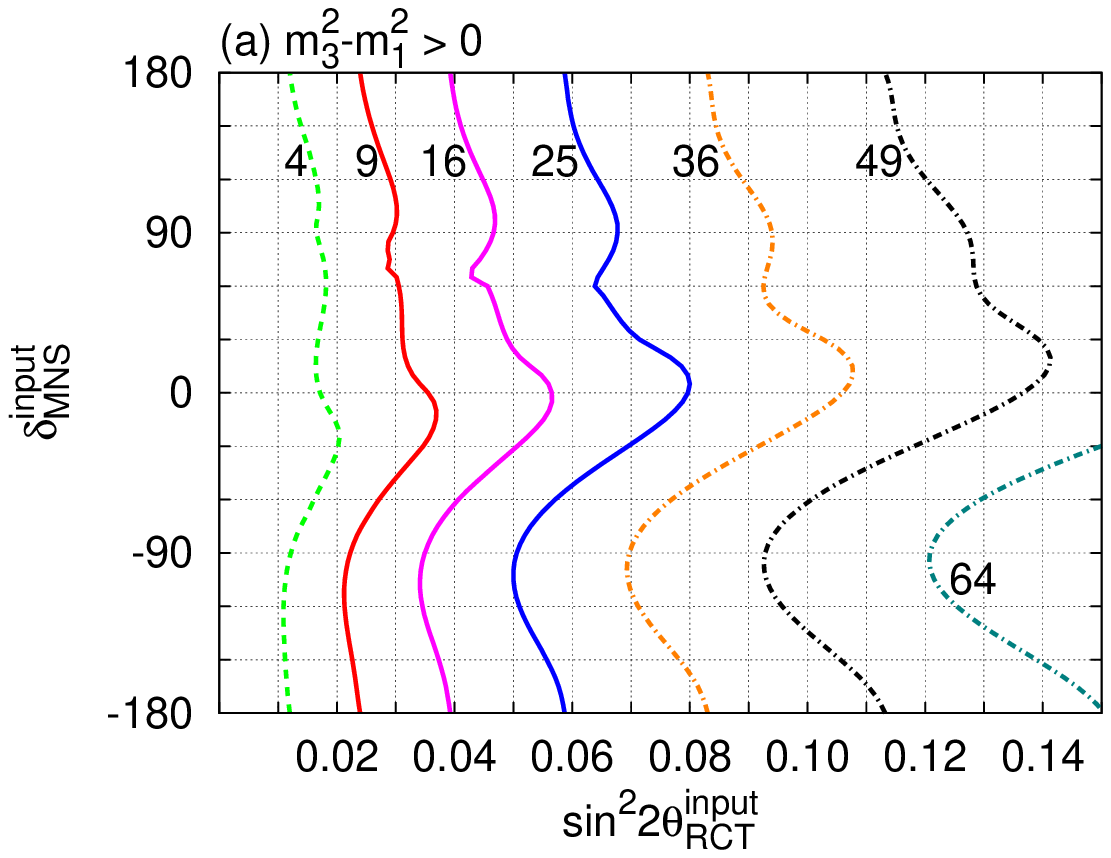}
~~~
 \includegraphics[scale=0.7]{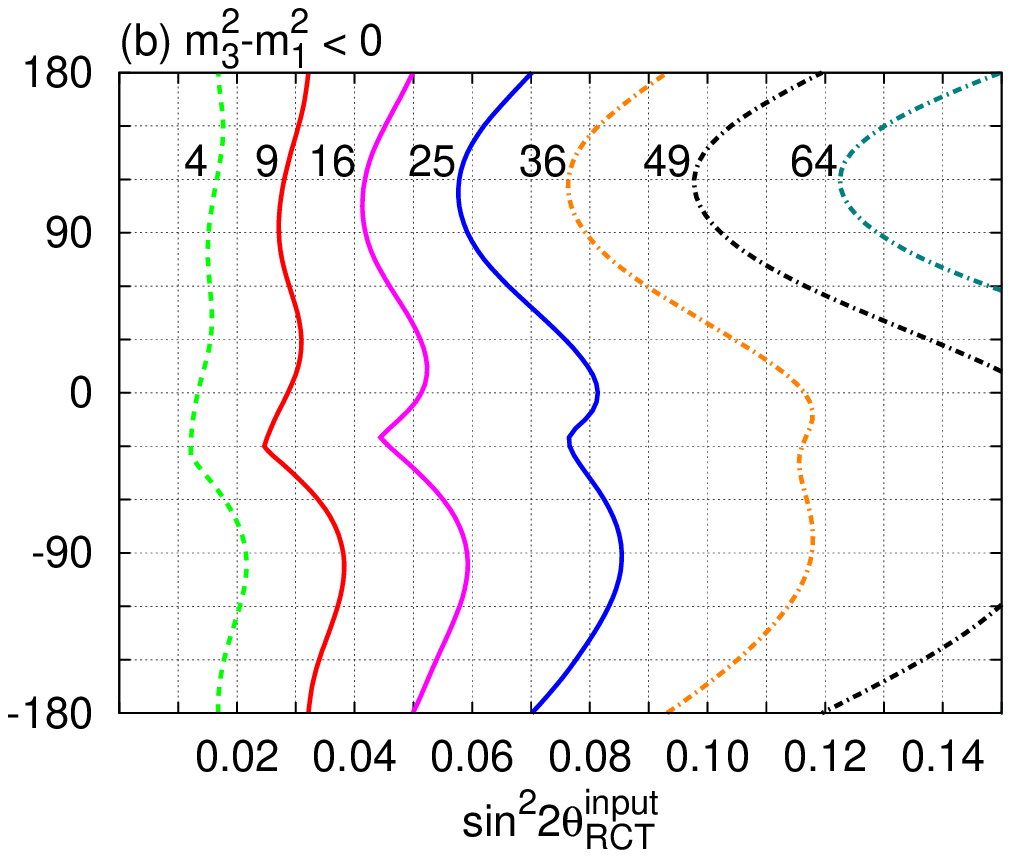}
 \caption{
The same as Fig.~\ref{fig:cnt-T2OKI}, but for T2KK experiment
with the optimum OAB combination,
$3.0^\circ$ OAB at SK and
$0.5^\circ$ OAB at $L=1000$km.
$\Delta \chi^2_{\rm min}$ values are given along the contours.
} 
\label{fig:cnt-T2KK}
\end{figure}

Figure \ref{fig:cnt-T2KK} shows the same contour plots as
Fig.~\ref{fig:cnt-T2OKI},
but for T2KK with $3.0^\circ$ OAB at SK and
$0.5^\circ$ OAB at $L=1000$km.
Significant increase in the $\Delta \chi^2_{\rm min}$ values
at T2KK is clearly seen against those in Fig.~\ref{fig:cnt-T2OKI}
for T2KO.
Now,
the wrong mass hierarchy can be excluded at $5\sigma$ level
for $\srct{2} > 0.08$ (0.09) if the mass hierarchy is
normal (inverted).
Because the measurement error is dominated by statistics,
we find that the $3\sigma$ sensitivity of T2KO with 100~kton
detector can be archived with a 40~kton detector for T2KK.
The CP phase dependence of the T2KK contours is much weaker than
that of the T2KO contours, especially for smaller
$\srct{2}$.
This is simply because the matter effect terms
at $L \gsim 1000$~km dominate
over the $\sin \dmns$ and $\cos \dmns$ terms,
in the correction terms $A^e$ and $B^e$;
see eq.~(\ref{eq:ABeApprox}).

\section{CP phase}
\label{sec:CP}

\begin{figure}[t]
\centering
 \includegraphics[scale=0.21]{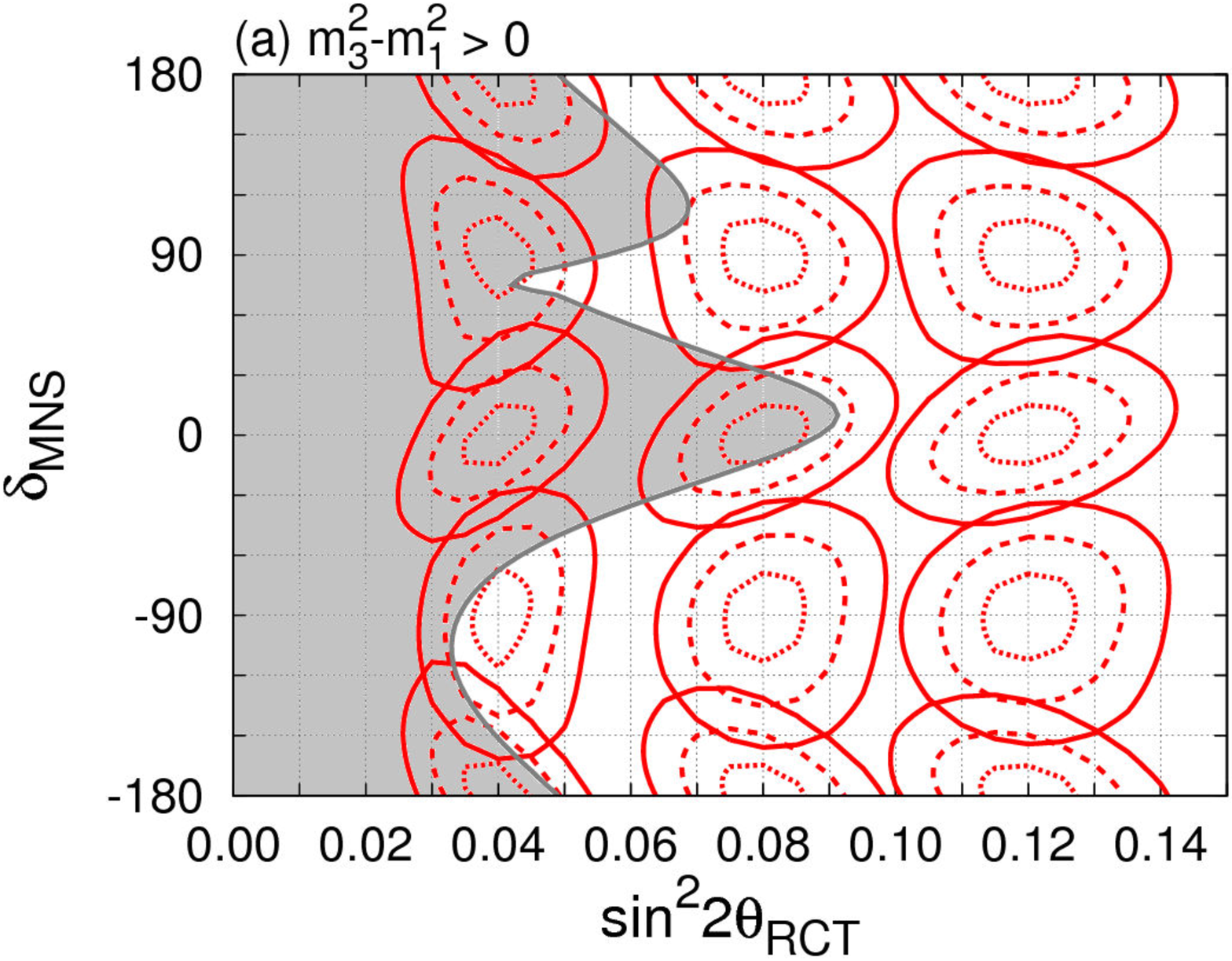}
~~
 \includegraphics[scale=0.21]{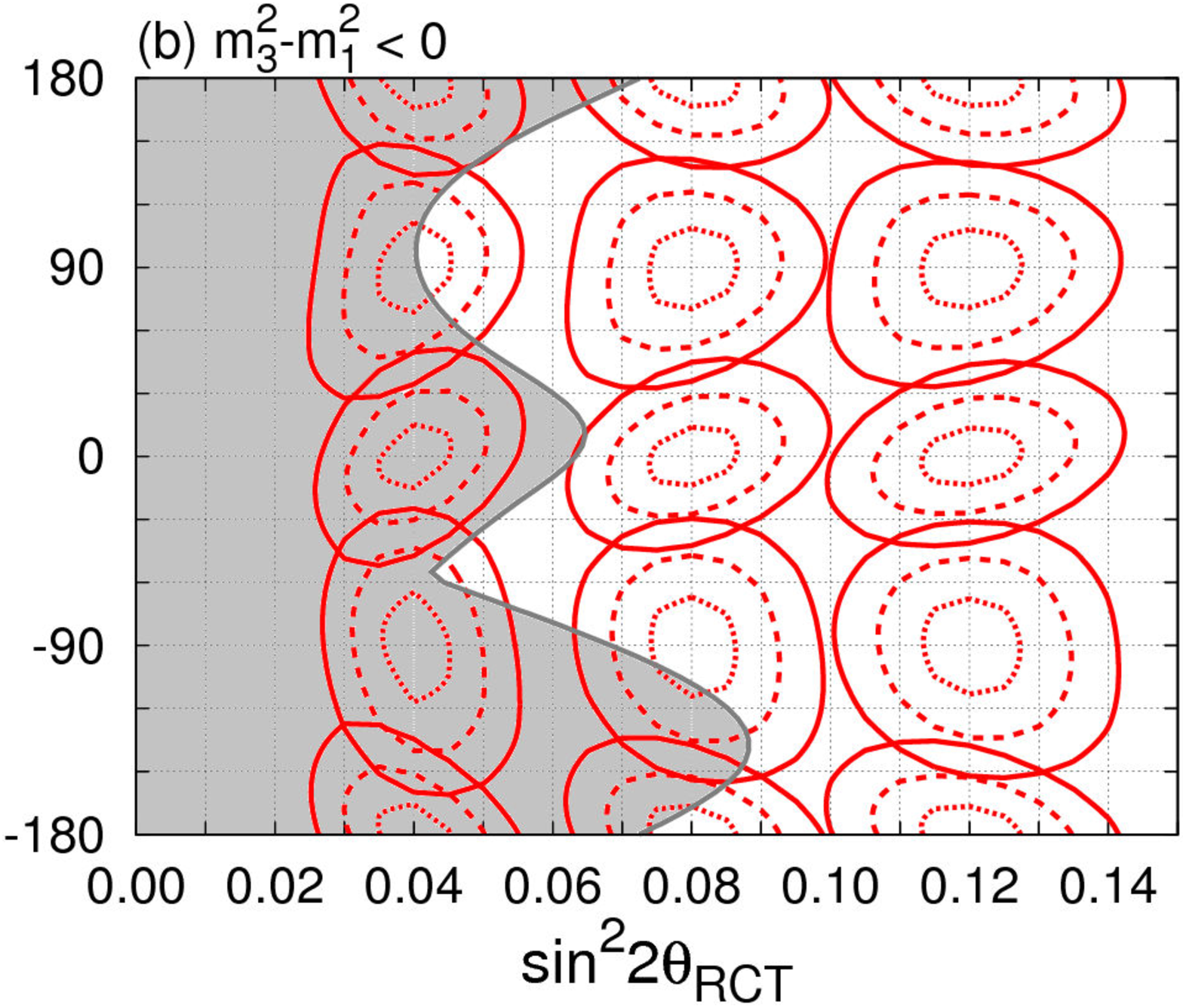}
 \caption{
The $\Delta \chi^2_{\rm min}$ contour plot for the T2KO experiment in
the plane of $\srct{2}$ and $\dmns$
when the mass hierarchy is assumed to be normal (left) or 
inverted (right).
Allowed regions in the plane of $\srct{2}$ and $\dmns$ are shown for
the combination of $3.0^\circ$ OAB at SK and $1.4^\circ$ at Oki Island
with $\numt{2.5}{21}$ POT each for $\nu_\mu$ and $\bar{\nu}_\mu$ 
focusing beams.
The input values of $\srct{2}$ is 0.04, 0.08, and 0.12
and $\dmns$ is $0^\circ$, $90^\circ$, $180^\circ$, and
$270^\circ$.
The other input parameters are given in eqs.~(\ref{eq:matter}) and
(\ref{eq:input}).
The dotted-lines, dashed-lines, and solid-lines show
$\Delta \chi^2_{\rm min}=1$, 4, and 9 respectively.
The shaded region has ``mirror'' solutions for the wrong mass
hierarchy giving $\Delta \chi_{\rm min}^2<9$.
} 
\label{fig:CP-T2KO}
\end{figure}

In this section, we investigate the measurement of
CP phase $\dmns$ 
in the T2KO experiment, as compared to the T2KK
and \T2Kn options.
In all the cases we adopt $3.0^\circ$ OAB at SK,
which makes the SK contribution to the measurements exactly the same,
and assume
$\numt{2.5}{21}$ POT each for $\nu_\mu$ and $\bar{\nu}_\mu$ 
focusing beam.

We show in Fig.~\ref{fig:CP-T2KO}
the $\Delta \chi^2_{\rm min}$ contour plots for the T2KO experiment 
in the plane of $\srct{2}$ and $\dmns$,
when the mass hierarchy is assumed to be normal (left) or
inverted (right).
The 12 cases are shown in each figure for
$\dmns=0^\circ$, $90^\circ$, $180^\circ$, $270^\circ$,
and for $\srct{2}=0.04$, 0.08, and 0.12.
The other input parameters are given in eqs.~(\ref{eq:matter})
and (\ref{eq:input}).
The allowed regions in the plane of $\srct{2}$ and $\dmns$ are
inside of the dotted-, dashed-, and solid-contours at
$\Delta \chi^2_{\rm min}=1$, 4, and 9, respectively.
The shaded region has ``mirror'' solutions for the wrong mass
hierarchy giving $\Delta \chi^2_{\rm min}<9$,
as shown by the red solid contours in Fig.~\ref{fig:cnt-T2OKI}.
Since the $\srct{2}=0.04$ input cases are no longer relevant after
the measurements eqs.~(\ref{eq:DayaBay_s13}) and (\ref{eq:RENO_s13})
by DayaBay \cite{DayaBay} and RENO\cite{RENO}, respectively, the only
parameter regions where we should worry about the mirror solution with
the wrong hierarchy are around $\dmns\simeq 0^\circ$ at
$\srct{2}\simeq 0.08$ for the normal hierarchy,
and
around $\dmns \simeq -135^\circ$ at $\srct{2}\simeq 0.08$
for the inverted hierarchy.
Since these regions are near the $3\sigma$ boundary, the mirror
solutions may be excluded by extending the experimental period
or by enhancing the beam power.
The $\srct{2}=0.04$ input cases are kept, since they show the
independence of the $\dmns$ measurement error on $\srct{2}$ clearly.
As explained in Ref.\cite{HOS1,HOS2}, this independence is a
consequence of the $1/\sqrt{\srct{2}}$ enhancement of the $\sin \dmns$
and $\cos \dmns$ dependencies in $A^e$ and $B^e$,
respectively,
in eqs.~(\ref{eq:AeAprox}) and (\ref{eq:BeAprox}),
which cancels precisely the statistical error
which is proportional to $\sqrt{\srct{2}}$,
or the square-root of the
$\nu_\mu \to \nu_e$ and $\bar{\nu}_\mu \to \bar{\nu}_e$ event numbers.

It is clearly seen that $\dmns$ can be measured with $\pm 20^\circ$
error for all the 24 cases presented in 
Fig.~\ref{fig:CP-T2KO} (a) and (b).
This is essentially because the magnitude of the coefficient of
$\sin \dmns$ in the amplitude shift in eq.~(\ref{eq:AeAprox})
and that of $\cos \dmns$ in the phase shift in eq.~(\ref{eq:BeAprox})
are approximately equal.
It should be noted that the uncertainty in the $\srct{2}$ is dictated
by the external constraint with the error of $\pm 0.01$ on the
$\chi^2$ function eq.~(\ref{eq:def_chi1}).
Because of the nearly zero correlation between the errors of
$\dmns$ and $\srct{2}$ in Fig.~\ref{fig:CP-T2KO},
further improvements in the precise measurements of
$\srct{2}$ will not reduce the errors of $\dmns$ significantly.

\begin{figure}[t]
\centering
 \includegraphics[scale=0.21]{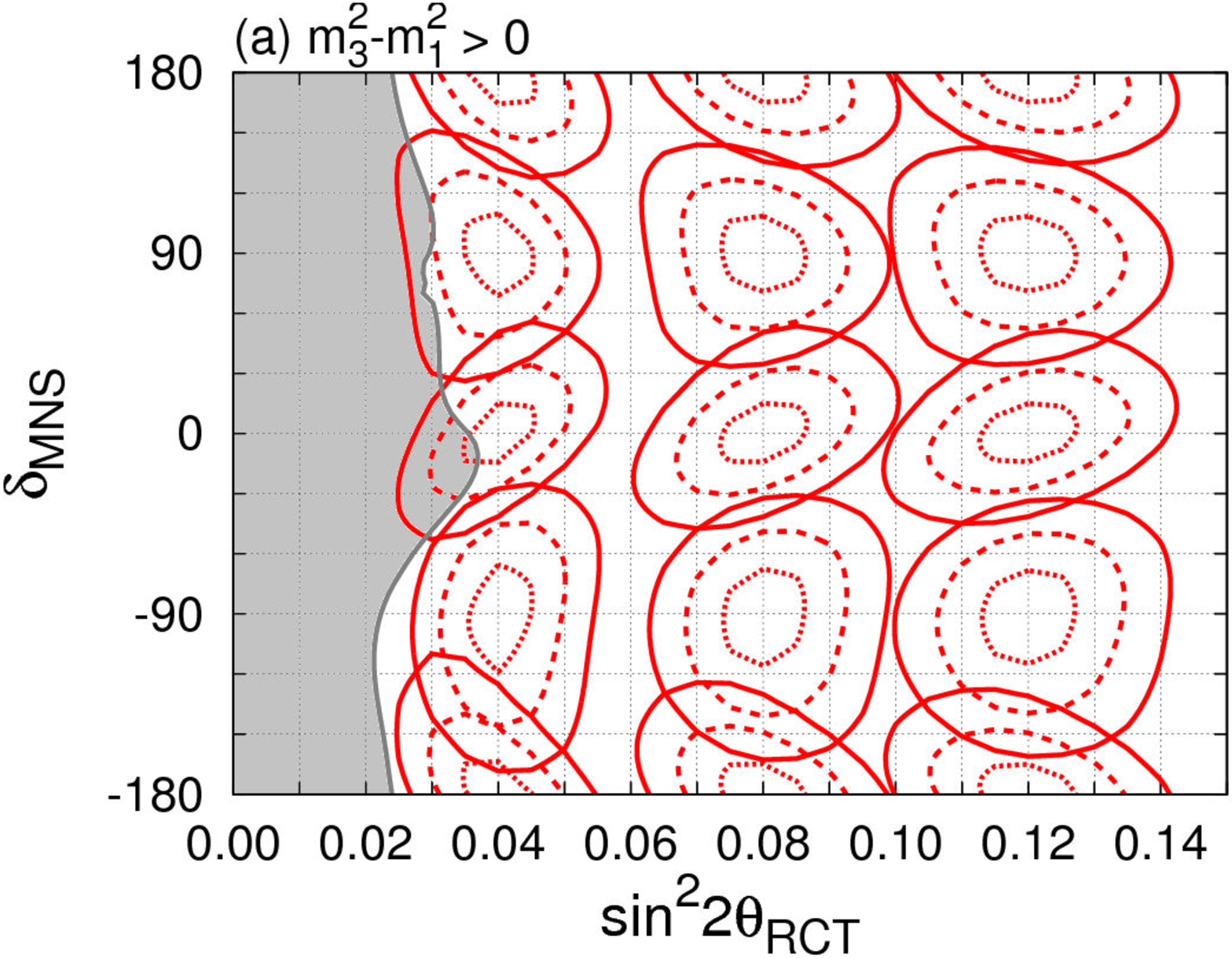}
~~
 \includegraphics[scale=0.21]{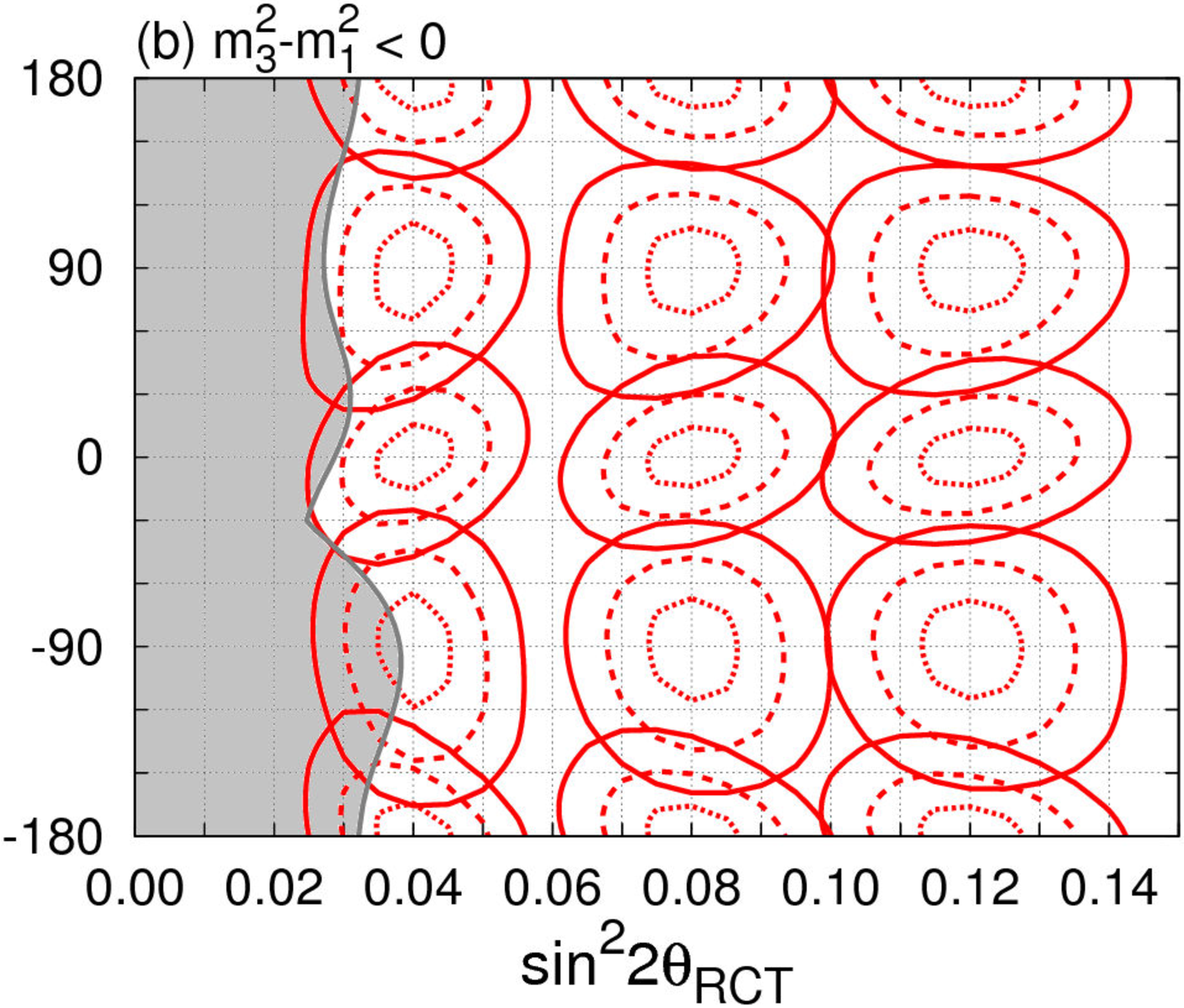}
 \caption{
The same as Fig.~\ref{fig:CP-T2KO}, but for T2KK experiment
with $3.0^\circ$ OAB at SK and
$0.5^\circ$ OAB at $L=1000$km.
} 
\label{fig:CP-T2KK}
\end{figure}

In order to compare the sensitivity of the $\dmns$ measurement
between the T2KO and the T2KK experiments,
we show in Fig.~\ref{fig:CP-T2KK}
the same contours for the T2KK experiment with
$3.0^\circ$ OAB at SK and
$0.5^\circ$ OAB at $L=1000$km.

It is clearly seen from the 12 sets of contours,
each for normal (left) and inverted (right) hierarchy, that the
expected error of $\dmns$ is $\pm 20^\circ$ for all the combinations,
just as for the T2KO experiment shown in Fig.~\ref{fig:CP-T2KO}.
This is remarkable since the event number at a far detector in Korea at
$L\simeq 1000$~km is significantly smaller than that in Oki at
$L=653$~km
because of the flux which decreases as $1/L^2$ at long distances.
This decrease the overall flux is compensated by the wide-band
structure of the $0.5^\circ$ OAB as shown by blue dotted lines 
in Fig.~\ref{fig:profile} (a3) and (b3), which enables the far
detector in Korea to observe not only the first oscillation peak but
also the second one as in Fig.~\ref{fig:T2KK}.
Around the second peak, $|\Delta_{13}/\pi| \sim 3$, and the sensitivity
to $\sin \dmns$ and $\cos \dmns$ can be three times higher than the
first peak with $|\Delta_{13}/\pi| \sim 1$; see eqs.~(\ref{eq:ABeApprox})
and (\ref{eq:diff_Ae_And_Be}).
In addition, the extended energy range covered by the $0.5^\circ$ OAB
allows the T2KK experiment to measure the phase-shifts $B^e$ and
$\bar{B}^e$ accurately,
and hence $\cos \dmns$; see eqs.~(\ref{eq:BeAprox}) and (\ref{eq:diff_Be}).
Indeed, we notice in Fig.~\ref{fig:CP-T2KK} (a) and (b) that the error
of $\dmns$ can be as small as $\pm 15^\circ$ when
$\dmns=0^\circ$ or $180^\circ$.

\begin{figure}[t]
\centering
 \includegraphics[scale=0.7]{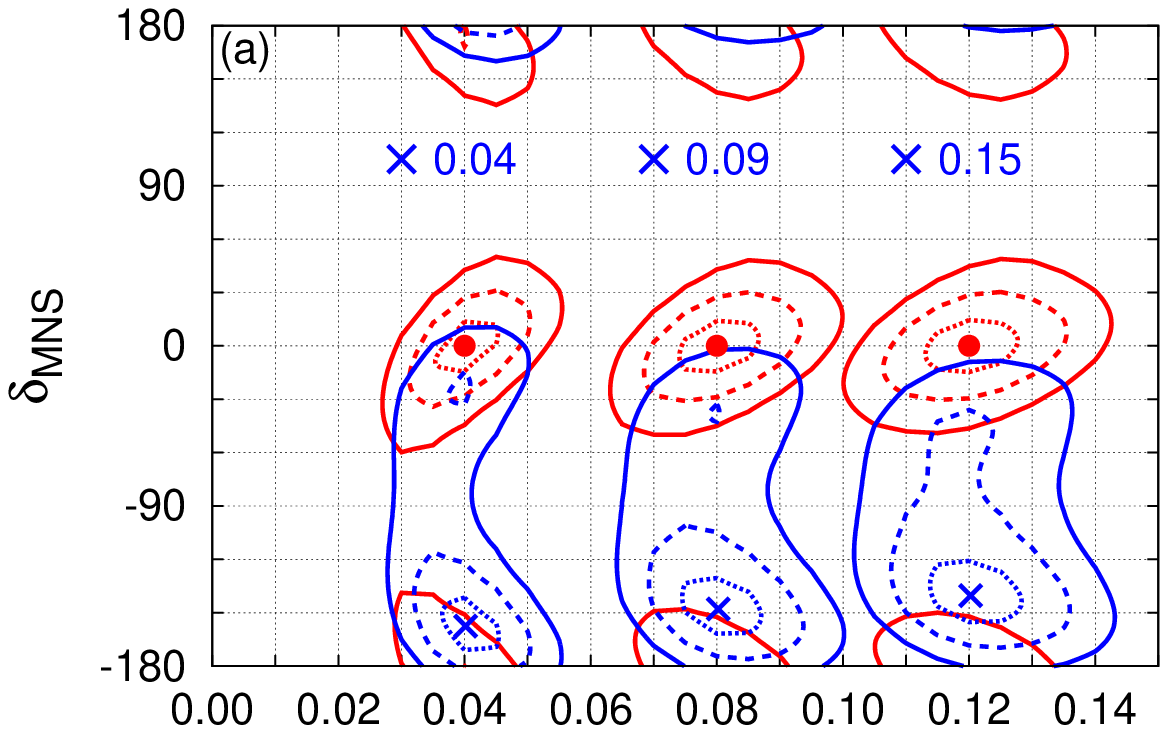}
~~
 \includegraphics[scale=0.7]{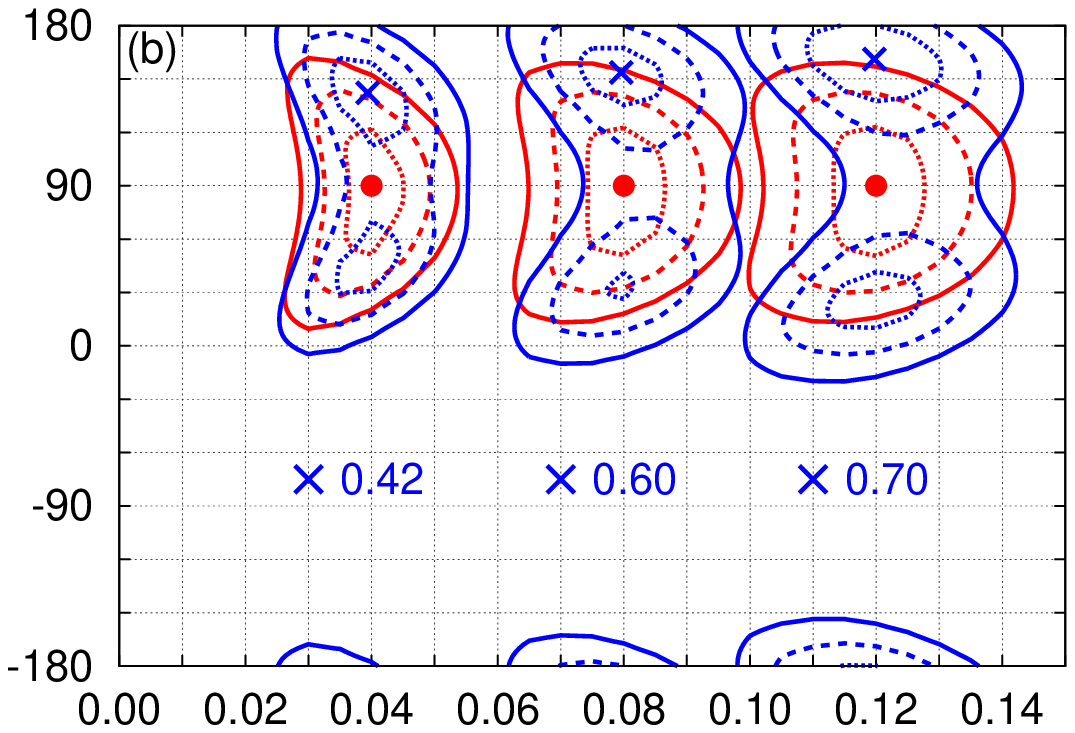}

 \includegraphics[scale=0.7]{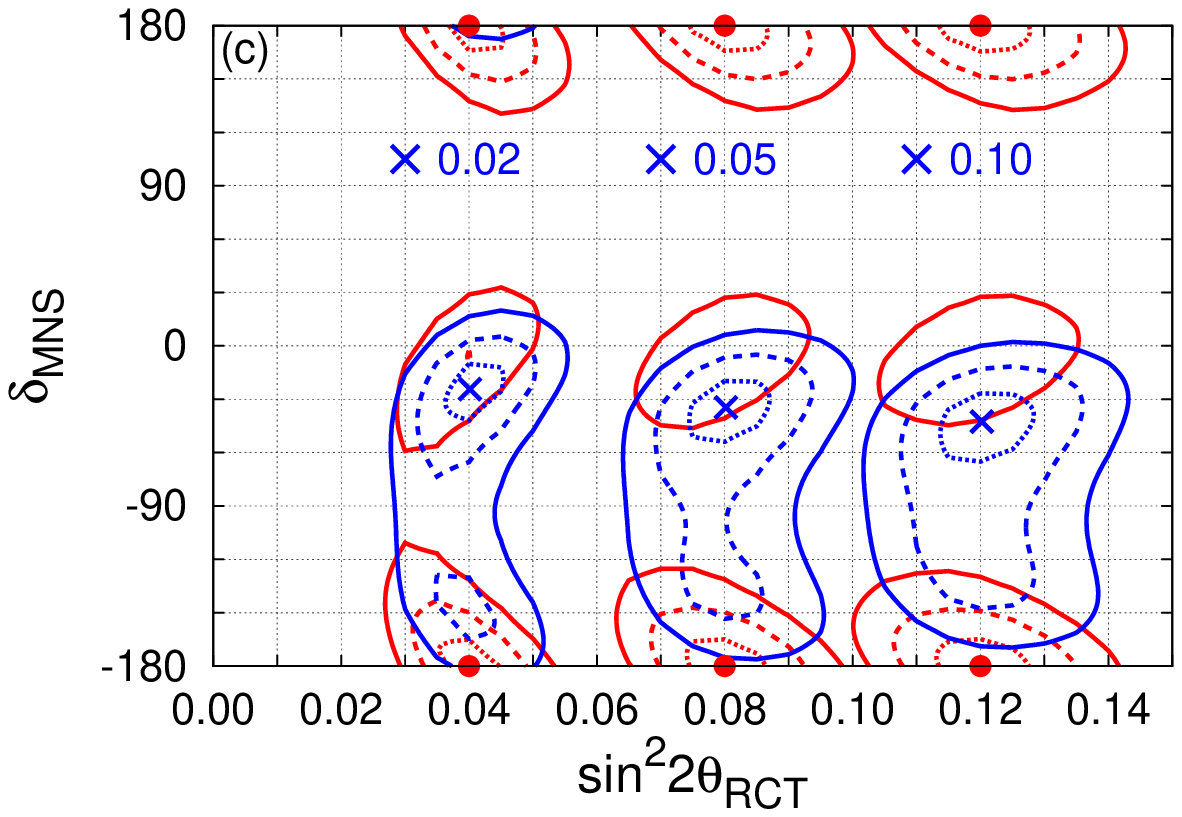}
~~
 \includegraphics[scale=0.7]{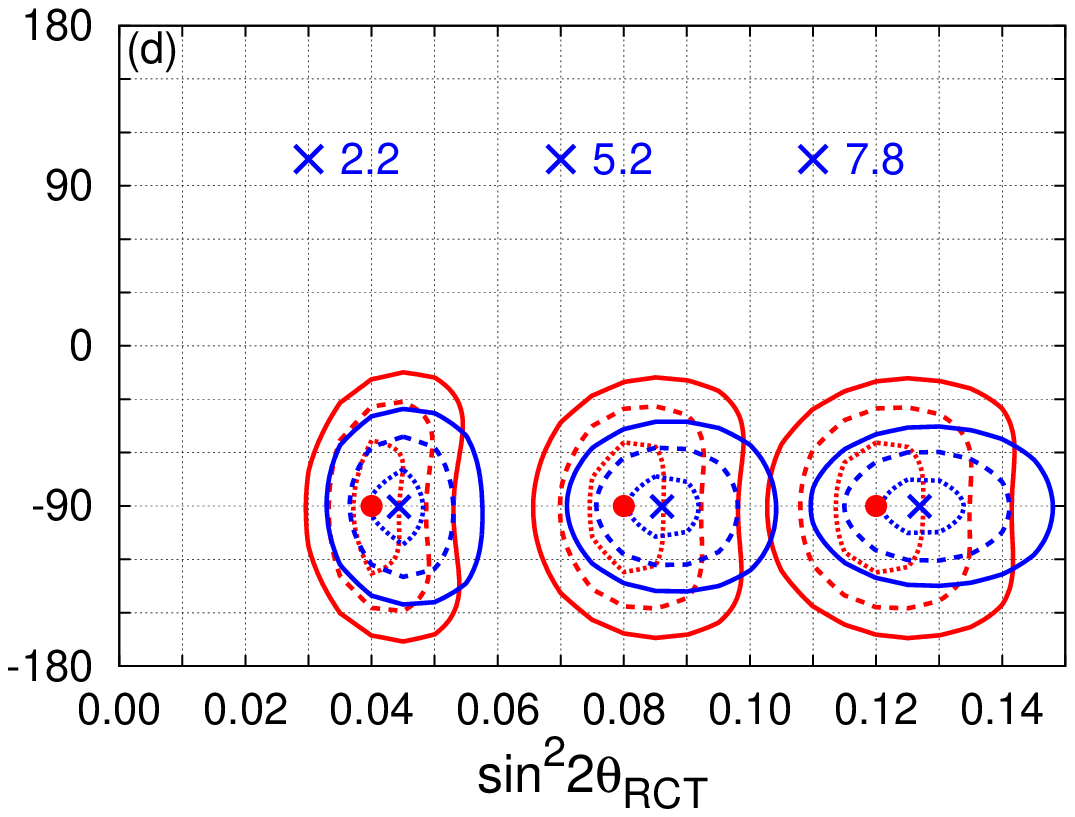}

 \caption{
The $\Delta \chi^2_{\rm min}$ contour plot for the \T2Kn experiment
in the plane of $\srct{2}$ and $\dmns$
when the mass hierarchy is assumed to be normal ($m_3^2-m_1^2 > 0$).
Allowed regions in the plane of $\srct{2}$ and $\dmns$ are shown
for experiments with $\numt{2.5}{21}$ POT 
each for $\nu_\mu$ and $\bar{\nu}_\mu$ focusing beam at 
$3.0^\circ$ off-axis angle.
The input values of $\srct{2}$ are 0.04, 0.08, and 0.12
and $\dmns$ are $0^\circ$ (a), $90^\circ$ (b),
$180^\circ$ (c), and $270^\circ$ (d).
The other input parameters are listed in eqs.~(\ref{eq:input})
and (\ref{eq:matter}).
The red dotted-lines, dashed-lines, and solid-lines show
$\Delta \chi^2_{\rm min}=1$, 4, and 9 contours, respectively,
when the right mass hierarchy is assumed in the fit,
whereas the blue contours give $\Delta \chi^2_{\rm min}$
measured from the local minimum value (shown besides the 
$\times$ symbol) at the cross point
when the wrong hierarchy is assumed in the fit.
} 
\label{fig:CP-T2Kn}
\end{figure}
\begin{figure}[t]
\centering
 \includegraphics[scale=0.7]{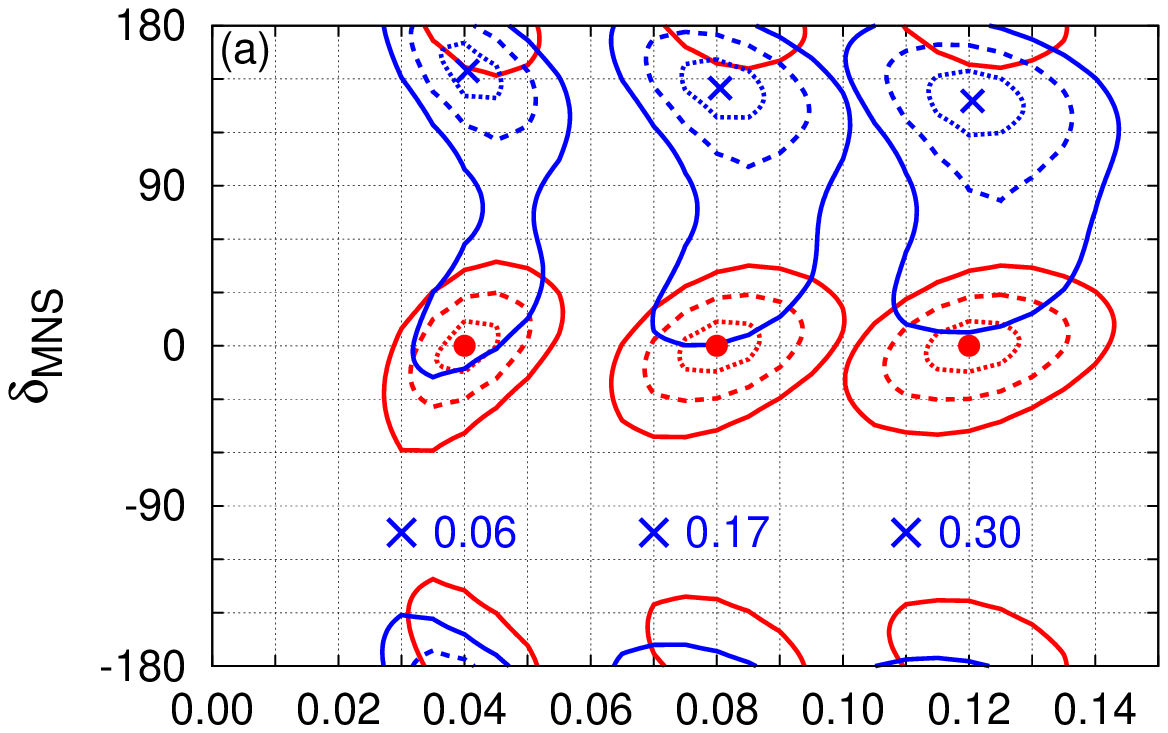}
~~
 \includegraphics[scale=0.7]{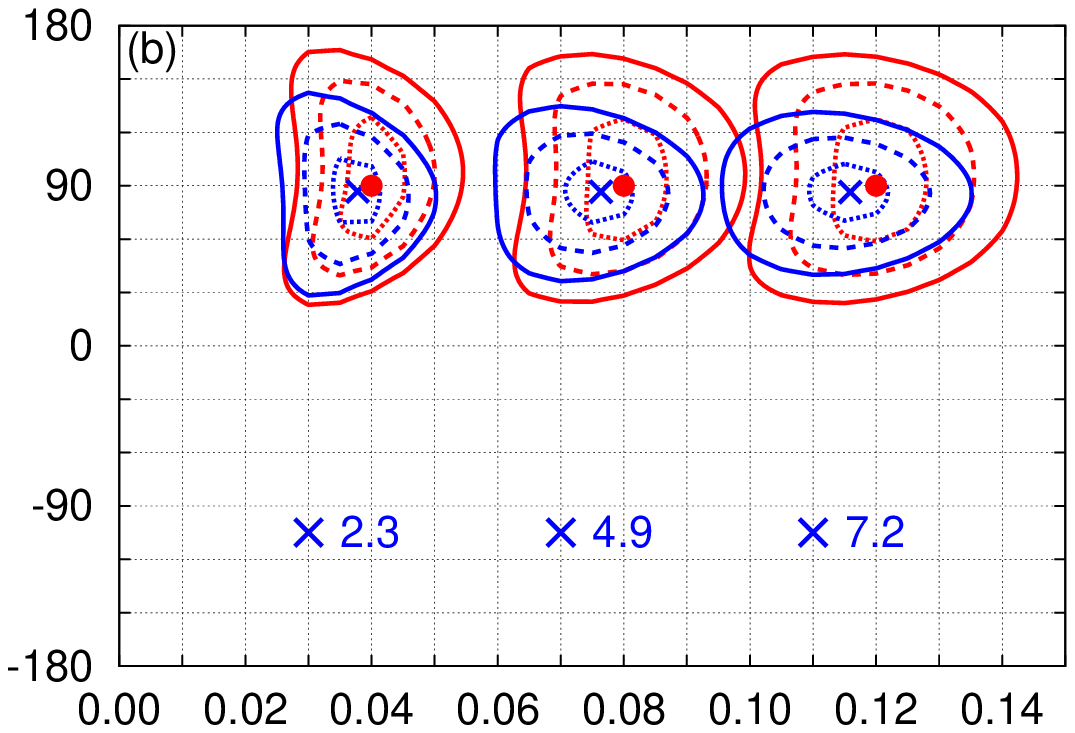}

 \includegraphics[scale=0.7]{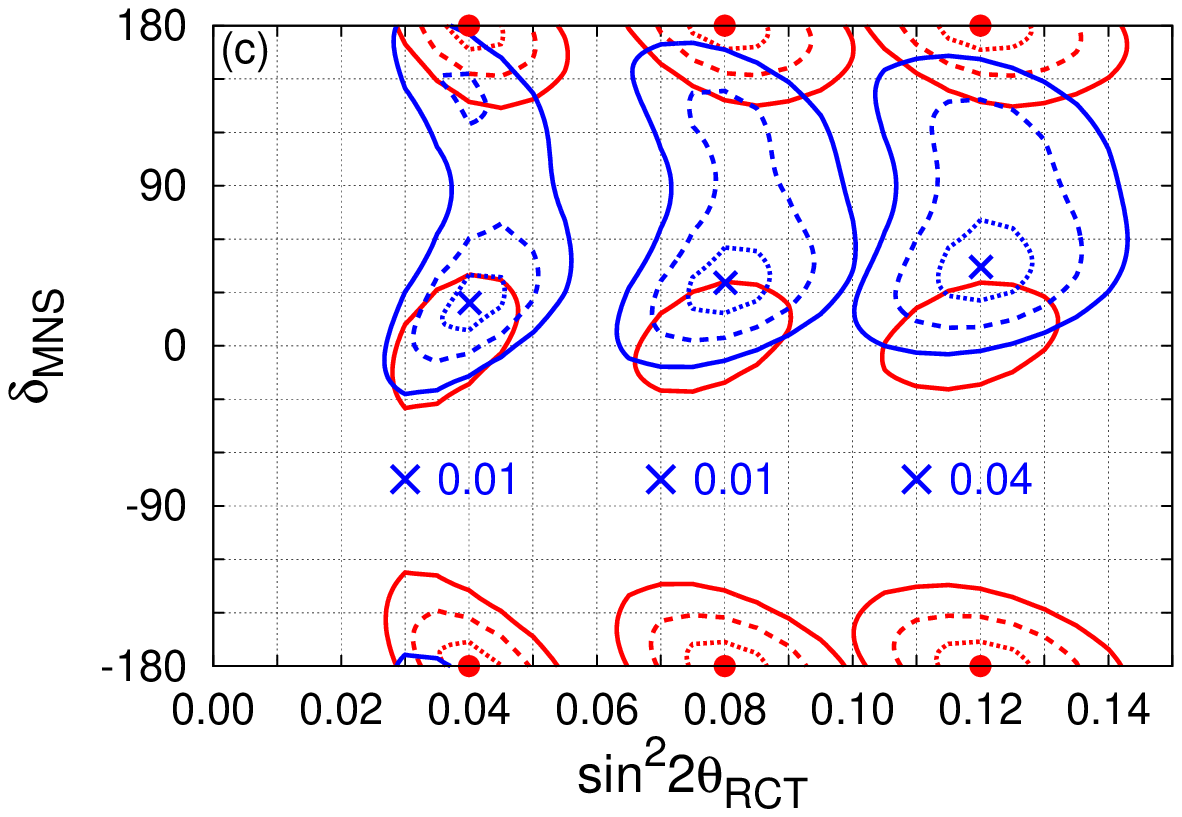}
~~
 \includegraphics[scale=0.7]{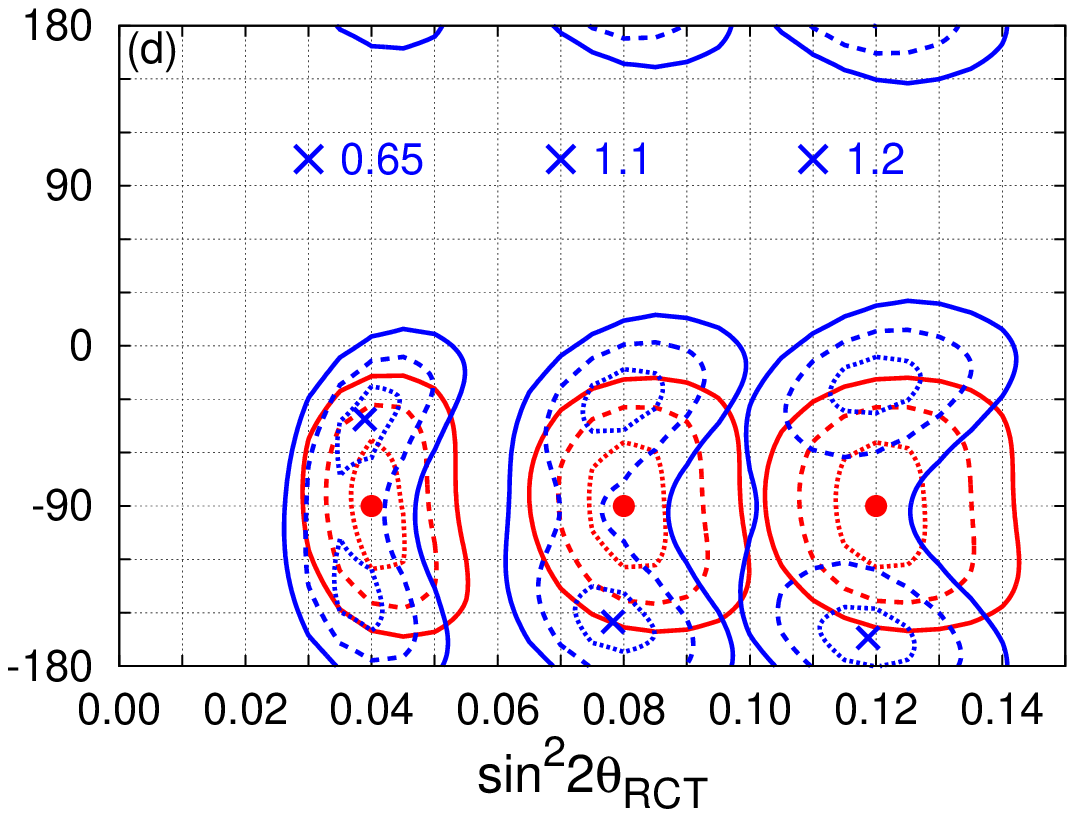}

 \caption{
The same as Fig.~\ref{fig:CP-T2Kn}, but for
the inverted mass hierarchy ($m^2_3-m^2_1 < 0$).
} 
\label{fig:CP-T2Ki}
\end{figure}

Finally in Figs.~\ref{fig:CP-T2Kn} and \ref{fig:CP-T2Ki},
we show the allowed regions by the \T2Kn experiment
in the plane of $\srct{2}$ and $\dmns$
for the normal (Fig.~\ref{fig:CP-T2Kn})
and the inverted (Fig.~\ref{fig:CP-T2Ki})
mass hierarchy,
when both $\nu_\mu$ and $\bar{\nu}_\mu$ focusing beam 
at $3.0^\circ$ off-axis angle
are used
each with $\numt{2.5}{21}$ POT.
The four $\dmns$ cases of
$0^\circ$(a), $90^\circ$(b), $180^\circ$(c), $270^\circ$(d)
are examined for the three
$\srct{2}^{\rm input} = 0.04$, 0.08, 0.12,
just as in 
Fig.~\ref{fig:CP-T2KO} for T2KO and 
Fig.~\ref{fig:CP-T2KK} for T2KK.
The other input parameters are also the same,
taken from eqs.~(\ref{eq:matter}) and (\ref{eq:input}).
The red dotted-lines, dashed-lines, and solid-lines show
$\Delta \chi^2_{\rm min}=1$, 4, 9 contours, respectively,
when the right mass hierarchy is assumed in the fit,
whereas the blue contours gives $\Delta \chi^2_{\rm min}$
measured from the local minimum at the blue cross point
when the wrong hierarchy is assumed in the fit.

The local minimal value of $\Delta \chi^2_{\rm min}$
at the blue cross point is given besides the cross mark in the
corresponding input $\srct{2}$ column.
As has been explained in section~\ref{sec:mass} and shown in
Fig.~\ref{fig:mass2},
the local $\Delta \chi^2_{\rm min}$ values are significant only for
$\dmns\simeq -90^\circ$ when the hierarchy is normal
(Fig.~\ref{fig:CP-T2Kn} (d)),
and for $\dmns\simeq 90^\circ$ when the hierarchy is inverted
(Fig.~\ref{fig:CP-T2Ki} (b)).
In order to show the location of the input parameters clearly,
we show the global minimal point by the red solid blob for each input
parameter case.
The global minimum gives $\Delta \chi^2_{\rm min}=0$ in our
analysis which ignores fluctuations in the mean number of events in
each bin.

We first note that $\dmns$ can be constrained uniquely around the
above two specific cases ($\dmns \simeq -90^\circ$ for $m^2_3-m^2_1>0$,
or $\dmns \simeq 90^\circ$ for $m^2_3-m^2_1<0$),
since not only the wrong mass hierarchy assumption gives
non-negligible local $\Delta \chi^2_{\rm min}$ as given in
Fig.~\ref{fig:CP-T2Kn} (d) and Fig.~\ref{fig:CP-T2Ki} (b),
but also the wrong hierarchy assumption favors the right $\dmns$,
with slightly larger (smaller) fitted $\srct{2}$ that compensate for
the matter effect for the normal (inverted) hierarchy.
The $1\sigma$ error shown by the red dotted contour is rather large,
however,
about $\pm 35^\circ$.
In all the other cases,
the presence of the wrong hierarchy solutions as shown by blue
contours significantly reduce the capability of measuring $\dmns$
with \T2Kn.
More importantly,
even if we can remove the wrong hierarchy by other experiments such as
Nova \cite{nova},
the next generation reactor neutrino oscillation experiments
\cite{ngRct},
or by an atmospheric neutrino observation with a huge detector
\cite{PINGU,SmironovICE},
the \T2Kn experiment with only one baseline length cannot measure $\dmns$
with high accuracy when $\dmns \simeq \pm 90^\circ$,
or suffers from the $\dmns^{\rm fit} = 180^\circ-\dmns^{\rm input}$
solution
when $\dmns \simeq 0^\circ$ or $180^\circ$ as can be seen from the
separate red contour islands on the (a) and (c) plots in
Figs.~\ref{fig:CP-T2Kn} and \ref{fig:CP-T2Ki}.

Throughout the analysis of this and the previous sections
we fix the OAB at SK at $3.0^\circ$ in order to make the SK
contributions to all the three experiments T2KO, T2KK and \T2Kn
identical.
We find that the performance of \T2Kn slightly improves if the 
$2.5^\circ \sim 2.0^\circ$ OAB is adopted instead, mainly because
these fluxes are slightly wider (harder) than the $3.0^\circ$ OAB
as shown by red solid curves in Figs.~\ref{fig:profile}
(a1)-(a3), (b1)-(b3).

\section{summary}
\label{sec:summary}

In this paper,
we examine physics potential of 
a one-beam two-detectors neutrino oscillation experiment
with an additional 100~kton water \cerenkov detector
in Oki Island, which is located along the T2K 
beam line at the baseline length of
$L=653$~km.
Together with Super-Kamiokande (SK) at $L=295$~km,
we can measure neutrino oscillations at two different energies
for the same oscillation phase proportional to $L/E$.
We may call this proposal as T2KO (Tokai-to-Kamioka-and-Oki),
whose capability has been compared with T2KK
(Tokai-to-Kamioka-and-Korea) 
with $L\simeq 1000$~km for the far detector in Korea,
and also with \T2Kn
where the same 100~kton detector 
is placed at the SK site ($L=295$~km).

As shown in Fig.~\ref{fig:T2OKI1},
since the Oki Island is located in the east side of the
T2K beam center, just like the SK,
the off-axis angle at Oki Island increases as that at SK
increases, as shown in Fig.~\ref{fig:relation}.
The off-axis beam (OAB) with $1.4^\circ$, $0,9^\circ$, and $0.6^\circ$
from J-PARC can be observed at Oki Island,
when the $3.0^\circ$, $2.5^\circ$, and $2.0^\circ$ OAB
reaches at SK, respectively.
The neutrino energy of the first oscillation maximum for the
$P(\nu_\mu \to \nu_e)$ and $P(\bar{\nu}_\mu \to \bar{\nu}_e)$
is between 1.0~GeV and 1.5~GeV,
which depend on the CP phase and the mass hierarchy pattern,
at $L = 653$~km.
Since 
the $\nu_\mu$ ($\bar{\nu}_\mu$) beam between $0.6^\circ$ and
$1.4^\circ$ off-axis angles has significant intensity around
these energies, 
as shown by green dashed lines in the upper six panels
in Fig.~\ref{fig:profile},
we expect that the T2KO experiment 
can be sensitive to the neutrino mass hierarchy and the CP phase,
just like the T2KK experiment 
\cite{HOSetc}, \cite{HOS1}-\cite{HOSF}.

For a detector of 100~kton fiducial volume and $\numt{2.5}{21}$ POT
exposure each for both $\nu_\mu$ and $\bar{\nu}_\mu$ beams,
we find that the T2KO experiment can determine the
mass hierarchy pattern at $3\sigma$ level if 
$\srct{2} = 4|U_{e3}|^2(1-|U_{e3}|^2)$ is larger than 0.09,
by observing the $e$-like CCQE (Charged-Current Quasi Elastic) events;
see Fig.~\ref{fig:cnt-T2OKI}.
This result does not strongly depend on the off-axis angle of the
$\nu_\mu$ ($\bar{\nu}_\mu$) beam at Oki Island,
because the neutrino intensity at the first oscillation maximum
does not strongly depend on the off-axis angle at Oki Island.
The T2KO sensitivity to the mass hierarchy is about $1/3$
(in $\Delta \chi^2_{\rm min}$) of the T2KK experiment
with the optimum OAB combination of
$3.0^\circ$ at SK and $0.5^\circ$ at a far detector in Korea
with the baseline length around 1000~km.
This is because the factor of two higher sensitivity of the T2KK over
the T2KO experiment as shown by eqs.~(\ref{eq:Ae_diffs}) and
(\ref{eq:Be_diffs}),
which should give a factor of 4 in $\Delta \chi^2_{\rm min}$
is partially compensated by the smaller average
flux by a factor of (635~km/1000~km)$^2\simeq 0.4$
at a far detector in Korea.
The sensitivity of the mass hierarchy pattern
of the \T2Kn experiment,
where a 100~Kton detector is added at the SK location,
is almost zero,
except around 
$\dmns \sim -90^\circ$ ($90^\circ$) for the normal (inverted)
hierarchy;
see Fig.~\ref{fig:mass2}.

The CP phase in the MNS (Maki-Nakagawa-Sakata) lepton flavor mixing
matrix \cite{MNS}, $\dmns$ can be measured with $\pm 20^\circ$ error
for all the four cases at $\dmns=0^\circ$, $\pm90^\circ$, and $180^\circ$,
almost independent of the $\srct{2}$ values \cite{HOS1,HOS2}
as long as the neutrino mass hierarchy is determined.
This is because $\sin \dmns$ can be constrained by
the difference between the magnitudes of the oscillation probabilities
$P(\nu_\mu \to \nu_e)$ and $P(\bar{\nu}_\mu \to \bar{\nu}_e)$
around the oscillation maximum,
whereas $\cos \dmns$ can be determined by the oscillation
phase around the first oscillation maximum,
or the location of the oscillation peak(s);
see eqs.~(\ref{eq:AeAprox}) and (\ref{eq:diff_Ae}).
The sensitivity to the CP phase, $\dmns$, of the T2KO experiment
is similar to that of the T2KK experiment,
mainly because the smallness of the flux at T2KK is compensated by its
capability to measure the second oscillation peak at
$|\Delta_{13}/\pi|\sim 3$ when the sensitivity to both $\cos \dmns$
and $\sin \dmns$ is a factor of 3 higher than that around the first
peak; see eqs.~(\ref{eq:ABeApprox}).

The \T2Kn option, which may be regarded as a first step toward
the Hyper-Kamiokande \cite{HK},
cannot generally determine $\dmns$, mainly because it cannot resolve
mass hierarchy by itself.
Only when $\dmns \simeq - 90^\circ$ for the normal hierarchy
(see, Fig.~\ref{fig:CP-T2Kn} (d))
and when  $\dmns \simeq 90^\circ$ for the inverted hierarchy
(see, Fig.~\ref{fig:CP-T2Ki} (b)),
the constraints for both hierarchy assumptions overlap,
and the CP phase can be determined uniquely.
Even if the mass hierarchy is determined by other experiments
\cite{nova}, \cite{ngRct}-\cite{SmironovICE}
the sensitivity to $\dmns$ is rather poor at \T2Kn as shown 
in Figs.~\ref{fig:CP-T2Kn} and \ref{fig:CP-T2Ki}.
This is essentially because of the parameter degeneracy unavoidable in
experiments with only one baseline length, such as those between
$\dmns$ and $180^\circ-\dmns$ when $\sin \dmns \simeq 0$.

\bigskip
\bigskip
\noindent
{\it Acknowledgments}

We would like to thank our experimentalist colleagues
Y.~Hayato, A.K.~Ichikawa, T.~Kobayashi, and T.~Nakaya,
from whom we learn about the K2K and T2K experiments.
We thank N.~Isezaki and M.~Komazawa for
teaching us about the geophysics measurements in
the Sea of Japan, or the East Sea of Korea.
We are also grateful to Japanese Coast Guard
for showing the detailed geological information around Oki Island.
The numerical calculations have been carried out on KEKCC at KEK.

\end{document}